\documentclass[11pt,a4paper]{article}
\usepackage{jheppub,amsmath,  amssymb,slashed,url,bm,textgreek,upgreek}
\usepackage{graphicx}
\usepackage{epstopdf}
\def\t{{ \sf t}} 
\def\J{{\mathcal J}}
\def\K{{\mathcal K}}

\def\X{{\mathcal X}}

\def\p{{\eurm p}}

\def\g{\mathfrak g}

\def\be{\begin{equation}}
\def\ee{\end{equation}}

\def\hat{\widehat}

\def\tilde{\widetilde}

\def\h{\widehat}

\def\S{{\mathcal S}}

\def\V{{\mathcal V}}
\def\O{{\mathcal O}}

\def\A{{\mathcal A}}
\def\d{{\mathrm d}}

\def\HH{{\mathbb H}}

\def\R{{\mathbb R}}
\def\C{{\mathbb C}}
\def\U{{\mathcal U}}

\def\[{\bigl [}

\def\]{\bigr ]}

\def\CP{{\mathbb{CP}}}
\def\N{{\mathcal N}}

\def\tr{{\mathrm {tr}}}
\def\Z{{\mathbb Z}}

\def\ad{{\mathrm{ad}}}

\def\L{{\mathcal  L}}
\def\t{\widetilde }
\def\h{\widehat}

\def\V{{\mathcal V}}
\def\I{{\mathcal I}}
\def\G{{\mathcal G}}
\def\B{{\mathcal B}}

\def\M{{\mathcal M}}

\def\H{{\mathcal H}}

\def\tilde{\widetilde}

\def\bar{\overline}

\font\teneurm=eurm10 \font\seveneurm=eurm7  \font\fiveeurm=eurm5
\newfam\eurmfam
\textfont\eurmfam=\teneurm \scriptfont\eurmfam=\seveneurm
\scriptscriptfont\eurmfam=\fiveeurm
\def\eurm#1{{\fam\eurmfam\relax#1}}
\font\teneusm=eusm10 \font\seveneusm=eusm7 \font\fiveeusm=eusm5
\newfam\eusmfam
\textfont\eusmfam=\teneusm \scriptfont\eusmfam=\seveneusm
\scriptscriptfont\eusmfam=\fiveeusm
\def\eusm#1{{\fam\eusmfam\relax#1}}
\font\tencmmib=cmmib10 \skewchar\tencmmib='177
\font\sevencmmib=cmmib7 \skewchar\sevencmmib='177
\font\fivecmmib=cmmib5 \skewchar\fivecmmib='177
\newfam\cmmibfam
\textfont\cmmibfam=\tencmmib \scriptfont\cmmibfam=\sevencmmib
\scriptscriptfont\cmmibfam=\fivecmmib

\def\AdS{{\mathrm{AdS}}}
\def\S{{\mathrm S}}
\def\T{{\mathrm T}}
\def\GG{{\mathrm G}}
\def\SU{{\mathrm{SU}}}
\def\U{{\mathrm U}}
\def\SO{{\mathrm{SO}}}
\def\CS{{\mathrm{CS}}}
\def\tr{{\mathrm{tr}}}
\def\dim{{\mathrm{dim}}}
\def\i{{\mathrm i}}
\def\Sym{{\mathrm{Sym}}}
\def\PU{{\mathrm{PU}}}
\def\PSU{\PU}
\def\ad{{\mathrm{ad}}}
\def\OO{{\mathrm O}}
\def\Sp{{\mathrm {Sp}}}
\def\K{{\mathcal K}}
\def\KK{{\mathrm{K3}}}
\def\eX{{\eusm X}}
\def\ASD{{\rm ASD}}

\title{Instantons and the Large $\N=4$ Algebra}

 \author{Edward Witten}
\affiliation{School of Natural Sciences, Institute for Advanced Study,\\  1 Einstein Drive, Princeton, NJ 08540 USA \\ witten@ias.edu}

\abstract{We investigate the differential geometry of the moduli space of instantons on $\S^3 \times \S^1$. Extending previous results, we show
that a sigma-model with this target space can be expected to possess a large $\N=4$ superconformal symmetry, supporting speculations that this
sigma-model may be dual to Type IIB superstring theory on $\AdS_3\times \S^3\times \S^3\times \S^1$.   The sigma-model is parametrized by three integers -- the rank of the gauge
group, the instanton number, and a ``level'' (the integer coefficient of a topologically nontrivial $B$-field, analogous to a WZW level).    
These integers are expected to correspond to two five-brane charges and
a one-brane charge.   The sigma-model is weakly coupled when the level,  conjecturally corresponding to one of the five-brane changes, becomes very large, keeping the
other parameters fixed.  The central charges of the large $\N=4$ algebra agree, at least semiclassically, with expectations from the duality.}

\begin{document}
\maketitle
\section{Introduction}\label{intro} 

\subsection{Overview}\label{overview} 

Among the original examples of AdS/CFT duality \cite{malda} were Type IIB superstring theory on $\AdS_3 \times \S^3\times \T^4$ and $\AdS_3\times \S^3\times \KK$, which are
believed to be dual to two-dimensional sigma-models in which the target space is the moduli space of instantons on $\T^4$ or $\KK$, respectively. Some of the arguments were recently
assessed and extended in  \cite{AU}, where one can also find detailed references. These examples have
$\N=4$ supersymmetry, which greatly facilitates their understanding.

However, the superficially similar example of Type IIB superstring theory on $\AdS_3\times \S^3\times \S^3\times \S^1$ has been less well understood, despite having an
even larger superconformal symmetry.    This model and a related one based on $\AdS_3\times \S^3\times \S^3\times \R$ have been studied extensively
 \cite{Sk,EFGT,BPS,GMMS,EGGL,Tong,EGL,EG} and in particular it is known that a dual conformal field theory should possess a ``large'' $\N=4$ superconformal algebra
 (in fact, two copies of this algebra, for chiral and antichiral modes, respectively), 
 as opposed to the more familiar ``small'' $\N=4$ superconformal algebra relevant to strings on $\AdS_3\times \S^3\times \T^4$ or $\AdS_3\times \S^3\times \KK$.   The large $\N=4$ superconformal algebra is an extension of the small one with
an additional $\SU(2)$  $R$-symmetry and  some additional free fields \cite{STP,Sch}.  
The simplest indication that $\AdS_3\times \S^3\times \S^3\times \S^1$ leads to a large superconformal
algebra is simply that it has many $\SU(2)$ symmetries, acting on the left and right on the two copies of
$\S^3=\SU(2)$ \cite{Sk}.  These all turn out to be $R$-symmetries of the left or right superconformal algebra.  The detailed analysis demonstrating the existence of a large
$\N=4$ algebra was made in \cite{EFGT} from a worldsheet point of view
(for the case that the fluxes on $\AdS_3\times \S^3\times \S^3\times \S^1$ are  of Neveu-Schwarz type) and in \cite{BPS} in supergravity.  
A simple example of a two-dimensional conformal field theory with the ``large'' $\N=4$ symmetry
is an $\SU(2)_k\times \U(1)$ supersymmetric WZW model, which is the same thing as a supersymmetric sigma-model with target $\S^3\times \S^1$ and $k$ units of flux 
of  $H=\d B$ (where $B$ is the sigma-model $B$-field).   

The Type IIB supergravity solutions on $\AdS_3\times \S^3\times \S^3\times \S^1$, which were analyzed in \cite{EFGT,BPS}, depend on three parameters, namely the three-form
fluxes on $\AdS_3$ and on the two $\S^3$'s.   Supersymmetry requires that all fluxes are of the same type, Neveu-Schwarz or Ramond-Ramond or a mixture. 
The three fluxes can be parametrized by integers 
$Q_1,Q_5,Q_5'$, where $Q_5$ and $Q_5'$ have been interpreted as the numbers of fivebranes wrapped in two different ways  that produce the  fluxes on the two
$\S^3$'s, and $Q_1$ is similarly interpreted\footnote{The integrality of $Q_1$ is not visible in the supergravity solution.}  
as the number of one-branes, related to the flux on $\AdS_3$.

Proposals for a dual of superstring theory on $\AdS_3\times \S^3\times \S^3\times \S^1$ have  been mainly of two types.  One idea is that, at least for some values of $Q_1$, $Q_5$, and
$Q_5'$,  the dual might be a symmetric product
of $N$ copies of the $\SU(2)_k\times \U(1)$ model, for some $N$ \cite{EFGT,BPS}.   One motivation for this proposal is simply that this symmetric product is one of the relatively few known examples
of a model with the large $\N=4$ superconformal symmetry.  Another motivation is that the duals of $\AdS_3\times \S^3\times \T^4$ and $\AdS_3\times \S^3\times \KK$
are related to similar symmetric products (if the integers $Q_1$ and $Q_5$ that characterize those models are relatively prime).   An obvious limitation of this proposal is that the symmetric product only depends on two integers, $k$ and $N$, while strings on
$\AdS_3\times \S^3\times \S^3\times \S^1$ depend on three integers $Q_1,Q_5,Q_5'$.  
It is believed that strings on $\AdS_3\times \S^3\times \S^3\times \S^1$ do not have
dualities that would make one of the three integers irrelevant.
  However, if $Q_5=1$ (or $Q_5'=1$), there is reasonably strong evidence that strings on $\AdS_3\times \S^3 \times \S^3\times \S^1$  are indeed dual to a symmetric product of
  many copies of the $\SU(2)_k\times \U(1)$ WZW model \cite{EGL,EG}.
  
  A second idea has been that the dual of strings on $\AdS_3\times \S^3\times \S^3\times \S^1$ might be a sigma-model with target the moduli space $\M$ of instantons on $\S^3\times \S^1$
  \cite{GMMS}.
  A question about this idea has been whether it is true that this sigma-model does have the large superconformal symmetry.    There are important 
  partial results in this direction 
  \cite{book,Hitchin,MV,BCG}.   Roughly, by combining previous results, 
  it is known that $\M$ is a generalized hyper-Kahler manifold, also known as a bi-HKT or (4,4) manifold. 
Assuming conformal invariance,
   this is enough to show that the model has $\N=4$ supersymmetry with the small $\N=4$ algebra.   The main goal of the present article is to complete this story
  and show that this sigma-model  actually  possesses  large $\N=4$ superconformal symmetry.   For this,  one must show
  that Killing vector fields on $\M$ that come from the symmetries of $\S^3\times \S^1$ are covariantly constant for  appropriate connections on $\M$ with torsion.
It is also necessary to show that the sigma-model with target $\M$ is conformally-invariant and not just scale-invariant, something that  is non-trivial  for sigma-models with
  $\N=4$ supersymmetry \cite{HullTownsend}. We address this point by
  applying arguments of \cite{Polchinski,PW}.  
  
  If $Q_1$ and $Q_5$ are relatively prime, the
   moduli space of instantons on $\T^4$ or $\KK$ is a deformation of a symmetric product of copies of $\T^4$ or $\KK$.    If a similar statement were true for $\S^3\times \S^1$,
  then potentially the two proposals about the dual of string theory on $\AdS_3\times \S^3\times \S^3\times \S^1$ could both be correct.   However, generically (except for
  gauge group $\U(1)$) it is not true that the moduli space $\M$ of instantons on $\S^3\times \S^1$ is deformation equivalent to a symmetric product.   This will be explained
  in section \ref{notso}.
  
  \subsection{Motivation For The Conjecture}\label{motivation}
  
  The idea that a sigma-model with target $\M$ is dual to Type IIB string theory on $\AdS_3\times \S^3\times \S^3\times \S^1$ can be motivated by a simple brane construction.
   In describing this, we slightly amplify the discussion in \cite{GMMS} (see scenario 3 in section 3 of that article) as well as  \cite{EGL}.  We also assume that the fluxes considered
   are of Ramond-Ramond (RR) type, so the corresponding branes are D-branes.  We denote the RR two-form field as $C_2$ and its three-form field strength as $G_3=\d C_2$.  
   For the starting point, we consider
   Type IIB superstring theory on $X=\R^2\times \S^1\times \T^*\S^3\times \R$, where $\T^*\S^3$ is a noncompact Calabi-Yau manifold (the deformed conifold) with $Q'_5$ units
   of RR flux  on $\S^3\subset \T^*\S^3$.   The deformed conifold with this flux is a supersymmetric configuration, studied originally in \cite{GV,KT}. Then we wrap $Q_5$ D5-branes on $\R^2\times \S^1 \times \S^3\times p \subset X $, where $p$ is a point in the last factor of $X=\R^2\times \S^1\times \T^*\S^3\times \R$.   These D5-branes support a $\U(Q_5)$ gauge theory,
   with gauge connection $A$ and  field strength  
   $F=\d A+A\wedge A$. Because of the assumed $G_3$ flux, the effective action of the
   gauge theory (in Euclidean signature) 
   contains a coupling\footnote{With $G_3=\d C_2$, integration by parts puts this  in the form $\frac{\i}{8\pi^2}\int C_2\wedge \tr \, F\wedge F$, which may be more familiar.}
   \be\label{effcouple}-\i \int_{\R^2\times \S^1\times \S^3}\CS(A) \wedge\frac{G_3}{2\pi}\ee
involving the  Chern-Simons three-form $\CS(A)=\frac{1}{4\pi}\tr\,\left(A\d A+\frac{2}{3} A^2\right)$. 
   This configuration, in which $Q_1$ has not yet been introduced, has actually been discussed previously  \cite{LM}. (For related examples, see for instance
   \cite{AGK,ST,GKMW}.)  In everything that we have said up to this point,  $\R^2\times \S^1$ could be replaced with any three-manifold $W$, or simply with $W=\R^3$.
        As explained in \cite{LM}, the low energy physics on $W$ is a three-dimensional topological field theory, a  $\U(Q_5)$ Chern-Simons theory at level $Q_5'$ (where the Chern-Simons coupling comes
   from (\ref{effcouple})).
   We will return to this Chern-Simons theory momentarily. 
 
 We can now add $Q_1$ D1-branes wrapped on $\R^2\times p_i \subset \R^2\times \S^3\times \S^1$, where $p_i$ are points in $\S^1\times \S^3$.   However, we assume that these
 D1-branes ``dissolve'' into instantons in the $\U(Q_5)$ gauge theory.   Generically, a $\U(Q_5)$ instanton on $\S^3\times \S^1$ completely breaks the $\U(Q_5)$ gauge symmetry
 down to the center $\U(1)$. (This is true for all positive values of the instanton number.)   The  $\U(1)$ gauge field  becomes massive because of the coupling (\ref{effcouple}), and is described at low energies by a $\U(1)$ $BF$ theory at level
 $Q_5'$ (which can be understood as the reduction to $\R^2$ of a $\U(1)_{Q_5'}$ Chern-Simons theory on $W=\R^2\times \S^1$).   Being a topological field theory, this level $Q_1$ $BF$ theory
 plays no role in the analysis of the large $\N=4$ algebra, though  it is undoubtedly important in some subtle aspects of the story.\footnote{Since $\U(Q_5)=(\U(1)\times \SU(Q_5))/\Z_{Q_5}$
 is only locally a product group,
 there will be a subtle coupling between the $\U(1)$ $BF$ theory and the sigma model associated to $\SU(Q_5)$ instantons, which we concentrate on in this
 article.}
 The rest of the low energy physics is described, by standard arguments, by a supersymmetric sigma-model with target the moduli space $\M$ of instantons on $\S^3\times \S^1$, except that
 we must understand the role of the interaction (\ref{effcouple}).  As this interaction depends on an integer $Q_1$, is odd under reflection of $\R^2$, and cannot be written
 as the integral of a gauge-invariant local density, it should come as no surprise that in the sigma-model this interaction becomes the coupling to a topologically non-trivial $B$-field,
 that is, a $B$-field whose field strength $H=\d B$ has nonzero periods that must satisfy a Dirac quantization condition.   This will be explained in section \ref{topology}.
 
 So far we have arrived at a sigma-model with target space the moduli space $\M$ of instantons on $\S^3\times \S^1$.   But what does this have to do with string theory on
 $\AdS_3\times \S^3\times \S^3\times \S^1$?    To answer this question, we just follow the original analysis of holographic duality of the D1-D5 system \cite{malda}.   If the flux
 $Q'_5$ is small compared to $Q_1$ and $Q_5$, so that its local effects are small, then we can simply borrow the original analysis.   The normal bundle to $\R^2\times \S^1\times \S^3\times p\subset \R^2\times \S^1
 \times \T^*\S^3\times \R$ is, of course, locally a copy of $\R^4$.    When we take the near horizon geometry, the zero-section of the normal bundle (that is,
 the origin in this $\R^4$) is omitted, the radial direction in the normal bundle
 combines with the first factor of  $\R^2\times \S^1
 \times \T^*\S^3\times \R$ to make a copy of $\AdS_3$, and the angular directions in the normal bundle simply survive in the near horizon geometry as a factor of $\S^3$.   In general,
 this $\S^3$ would be fibered over the worldvolume of the D5-branes, which in our present discussion is $\R^2\times \S^1\times \S^3$. But as $\S^3$ is parallelizable, the fibration 
 is trivial and the angular variables just give another factor of $\S^3$.   Thus the near horizon geometry is $\AdS_3\times \S^3\times \S^3\times \S^1$.
 
Particularly if $Q_5'$ is not assumed to be small compared to $Q_1$ and $Q_5$, it is not entirely clear from this analysis
 that the near horizon geometry will be the standard maximally symmetric $\AdS_3\times \S^3\times \S^3\times \S^1$ 
 geometry.  If, however, this is the case (which is plausible but will not be proved here), then we get a reasonable basis for expecting that Type IIB superstring theory on  $\AdS_3\times \S^3\times \S^3\times \S^1$ is
 dual to a sigma-model with target the instanton moduli space $\M$.   We also learn the dictionary in this relationship: $Q_5$ maps to the rank of a $\U(Q_5)$ gauge group;
 $Q_1$ is the instanton number; and $Q_5'$ is the ``level,'' that is, the coefficient of a topologically non-trivial $B$-field in the sigma-model.  
 
 A puzzle here is that the $\AdS_3\times \S^3\times \S^3\times \S^1$ geometry, with $Q_5$ and $Q_5'$ understood as the flux of $G_3/2\pi$ over the two $\S^3$'s, has a manifest
 symmetry between $Q_5$ and $Q_5'$.   It is not at all clear why the low energy limit of the D-brane system would have that property.   However, an encouraging observation was made in
 \cite{LM}.  As remarked earlier, for $Q_1=0$, the low energy limit is  a $\U(Q_5)$ Chern-Simons theory at level $Q'_5$, which has a symmetry
 $Q_5\leftrightarrow Q_5'$, usually called level-rank duality \cite{NT}.   If it is true that string theory on $\AdS_3\times \S^3\times \S^3\times \S^1$ is dual to the sigma-model,
 then for any value of $Q_1$, the low energy limit must have the same $Q_5\leftrightarrow Q_5'$ symmetry.   The symmetry is only predicted in the low energy limit because
 on the gravity side, it only emerges in the near horizon limit of the geometry.  Since level-rank duality is rather subtle, this example suggests that understanding the 
 $Q_5\leftrightarrow Q_5'$ symmetry of the sigma-model may not be easy.
 
 A final remark is that conformal invariance of the sigma-model depends crucially on a renormalization group flow.    The sigma-model whose target is the instanton moduli space $\M$
 is somewhat analogous to a sigma-model with target a compact Lie group $\GG$: it is constructed from a target space metric  as well as a $B$-field.   Conformal invariance
 will hold only if these are properly related.   In the case of the group manifold, if the target space radius is large compared compared to the ``level,'' 
 then the model is asymptotically free but flows in the infrared to a conformally invariant fixed point at which the radius is determined in terms of the level \cite{EW}.  We anticipate a
 similar behavior for the sigma-model with target $\M$.   In the preceding discussion, we assumed no relation between the radius of the 
 $\S^3\subset \T^*\S^3$ on which the D5-branes were wrapped and the sigma-model level $Q'_5$.  If the radius is too large, 
 the sigma-model is definitely not conformally invariant; it has a target space metric that is large
 compared to the level.   What we will argue in the body of this article is that with a correctly adjusted radius, 
 the sigma-model has a conformally invariant fixed point with large $\N=4$ superconformal symmetry.  This will be an infrared stable fixed point, since a theory with large $\N=4$ 
 symmetry does not have relevant couplings \cite{GMMS}.   Hopefully, for any (or perhaps any sufficiently large)
  initially assumed radius, there is a renormalization group flow to this fixed point.
  
  Just as in the case of the WZW model, the metric of the target space at the critical point is proportional to the level, which here is $Q_5'$.  Hence at its critical point, the sigma-model
becomes weakly coupled if $Q'_5$ is taken to be large, for fixed $Q_1$ and $Q_5$.  By contrast, the supergravity description becomes reliable when $Q_1\gg Q_5,Q_5'\gg1$.

By wrapping an orientifold plane on the D5-brane world-volume, one can as usual 
replace the $\U(Q_5)$  gauge group of the D5-branes by an orthogonal or symplectic group.  The duality conjecture considered in this article likely has an analog
for those cases, but we will not discuss this in detail.   Most considerations regarding instanton
moduli space in this article are valid for any compact gauge group, although exceptional gauge groups have no obvious application in $\AdS_3\times S^3\times S^3\times \U(1)$ duality.

 \subsection{Differential Geometry Of A Four-Manifold and Its Instanton Moduli Space}\label{diffgeom}
 
In many cases,  if a four-manifold $M$ has a differential geometric structure which ensures that a sigma-model with target $M$ possesses a certain
 supersymmetry algebra, then the moduli space $\M$ of instantons on $M$ has the same differential geometric structure, so a sigma-model with target $\M$
 possesses the same supersymmetry algebra.
 
 The most familiar results of this kind arise in the absence of a  $B$-field.   If $M$ is  Kahler, so that a sigma-model with target $M$ has
  global $(2,2)$ supersymmetry, then $\M$ is also  Kahler, and a sigma-model with target $\M$ also has global $(2,2)$ supersymmetry.
  If $M$ is hyper-Kahler, so that a sigma-model with target $M$ has 
  $\N=4$ superconformal symmetry with the small $\N=4$ algebra for both left- and right-movers, 
  then $\M$ is also hyper-Kahler, again leading to small $\N=4$ superconformal symmetry for both chiralities.

Some results along these lines that are relevant to sigma-models with a $B$-field are as follows: 

\begin{enumerate}
\item If $M$ satisfies the conditions for $(0,2)$ supersymmetry -- it is a complex manifold with a hermitian metric whose torsion is closed in a sense reviewed in section \ref{sigmab} --
then $\M$ is also a complex manifold\footnote{The theorem of \cite{Buchdahl,LiYau} actually identifies $\M$ as a moduli space of
stable bundles on $M$.}  \cite{Buchdahl,LiYau}, with a natural hermitian metric that also has closed torsion \cite{book}, so the sigma-model with target
$\M$ also has $(0,2)$ supersymmetry,
\item If $M$ is a generalized Kahler manifold (the geometry that leads to $(2,2)$ supersymmetry with a $B$-field) then so is $\M$ \cite{Hitchin,BCG}.
\item  If $M$ is an HKT manifold (the geometry that leads to $(0,4)$ supersymmetry, with a small $\N=4$ algebra), then so is $\M$ \cite{MV}.
\item If $M$ is generalized hyper-Kahler  or bi-HKT (leading to $(4,4)$ supersymmetry with the small $\N=4$ algebra), then so is $\M$.   This follows on combining results in
\cite{Hitchin} and \cite{MV}; see section \ref{hkt}.
\item Finally, if $M$ has the properties that lead to invariance under the large $\N=4$ algebra, then so does $\M$.   This is shown in  section \ref{large}.
\end{enumerate}

 \subsection{Organization Of The Paper}\label{organization}
 
 This article is organized as follows.
 
 In section \ref{topology}, we describe basic aspects  of the moduli space $\M$ that will be important in this article.   In section \ref{review},
 we review the geometry required for extended supersymmetry in a two-dimensional sigma-model and the relevant geometry of $\S^3\times \S^1$.  
 In section \ref{hypercomplex}, we explain how to prove that the moduli space $\M$ is a hypercomplex manifold, and in section \ref{hkt}
 we extend that and prove that it has the geometry  associated with small $\N=4$ symmetry.   These two sections are 
 primarily based on previous results \cite{book,Hitchin,MV,BCG}, with some details added.
  In section  \ref{large}, we show that the sigma-model
 with target $\M$ actually has the geometry associated to the large $\N=4$ algebra, not just the small one.  For this, one has to show that the Killing vector
 fields on $\M$ associated to the symmetries of $\S^3\times \S^1$ are covariantly constant for appropriate connections with torsion.  We also determine the central
 charges of the $\N=4$ algebra.  
 In section \ref{conformal}, we complete the story by arguing that the sigma-model with target $\M$, with appropriate metric and $B$-field,
  is conformally-invariant and not just scale-invariant.   This is argued in several
 ways, using considerations in  \cite{Polchinski,PW}.   Putting all this together, it follows
 that the sigma-model with target $\M$ is a conformal field theory with large $\N=4$ superconformal symmetry for both chiralities.

 In section \ref{motivation}, in the starting point, we could have replaced $\T^* \S^3$ with $\T^*(\S^3/\Z_n)$  for some integer $n\geq 2$, where $\Z_n$ acts on $\S^3=\SU(2)$ on,
 say, the left.   Then the same logic as before
 would motivate the idea that Type IIB superstring theory on $\AdS_3\times \S^3\times \S^3/\Z_n\times \S^1$ is dual to instantons on $\S^3/\Z_n\times \S^1$.  Most
 of our considerations carry over directly to that case, but there is an interesting novelty, discussed in section \ref{novel}, and related to the possibility of turning on a discrete NS $B$-field
 (assuming the background is of Ramond type).
A gauge bundle on $\S^3/\Z_n\times \S^1$ has a $\Z_n$-valued discrete
 topological invariant that is absent for $\S^3\times \S^1$, and this makes it possible to construct examples that may be interesting  purely from a geometrical point of view.
  Hyper-Kahler manifolds can be generalized to include
 a non-flat $B$-field, leading to the concept of a strong HKT manifold, a notion that we will review in section \ref{review}.  
   In general, a $\sigma$-model with target a compact  strong HKT manifold will  have a large $\N=4$ superconformal algebra for one chirality.
  (To get large $\N=4$ algebras for both chiralities, one needs a pair of strong HKT structures with equal and opposite torsion.)  However,
 known examples of manifolds of this type   are very limited; apart from  hyper-Kahler manifolds, one has only  homogeneous examples \cite{Spindel,OP}, of which the
 simplest is  $\S^3\times \S^1$, and products of hyper-Kahler and homogeneous manifolds.      As we explain in section \ref{novel}, for certain values of the instanton number and the
 discrete topological invariant, the instanton moduli spaces on $\S^3/\Z_n\times \S^1$ are smooth and compact.   These may potentially give the first examples of compact strong HKT
 manifolds that are not merely products of hyper-Kahler manifolds and homogeneous spaces.

We conclude this introduction with a note on notation and terminology.   Concerning notation, we generally denote a four-manifold on which we study the instanton equation as $M$
and the corresponding instanton moduli space as $\M$.  The metric on $M$ is denoted as $g$ and tangent space indices to $M$ are denoted $i,j,k$ or $k,l,m$; the metric on $\M$ is
denoted $G$ and tangent space indices to $\M$ are denoted $\alpha,\beta,\gamma$.   Given a geometric structure on $M$, a corresponding structure on $\M$ is denoted with a hat.  For example, if $\I$ is a complex structure and $V$ is a vector field
on $M$, then the corresponding complex structure and vector field on $\M$ are denoted as $\h \I$ and $\h V$.
 The target
space of a general sigma-model is denoted $\eX$; the metric on $\eX$ is denoted as $G$ and tangent space indices are denoted $I,J,K$.   Lie algebra indices of $\SU(2)$ and
(related to this) tangent space indices of $\S^3$ are denoted $a,b,c$.
At some points, it is hard to be completely consistent with these conventions; for example,  although we usually consider $\S^3\times \S^1$  as a four-manifold on which we study the instanton
equation, it is also considered in section \ref{largefour} as the target space of a sigma-model.

Concerning terminology, the geometry that leads to a small $\N=4$ algebra (for both left-movers and right-movers) has been
called twisted generalized hyper-Kahler geometry, where ``twisting'' means that  the three-form $H$ is topologically nontrivial (closed but not exact).  
Similarly, the analog with $\N=2$ supersymmetry has been called twisted generalized Kahler geometry \cite{Gualtieri}.
 The generalized
hyper-Kahler manifolds of primary interest in the present article are  twisted in that sense\footnote{There are, however, interesting complete but not compact examples
in which $H$ is non-zero but exact.  See \cite{Hitchin} for examples.  The analysis in this article should apply to those examples, with minor changes.}    
and we will take the liberty of sometimes omitting the word ``twisted.''  Somewhat
similarly, the geometry that leads to a small $\N=4$ algebra for, say, right-movers only (or left- and right-movers both) has been called strong HKT geometry (or strong bi-HKT
geometry), where HKT stands for a generalization of hyper-Kahler
geometry to allow torsion,  ``strong'' means that $\d H=0$, and ``bi'' means that there are two separate HKT structures with opposite torsion. 
Likewise, the analogs  for  $(0,2)$ (or $(2,2)$) have been called strong KT
(or strong bi-KT) geometry.
  In this article, we only consider geometries with $\d H=0$, appropriate to Type II superstrings, and we sometimes
omit the word ``strong.''   
 
 \section{Some Properties of the Instanton Moduli Space}\label{topology}

In discussing general properties of the moduli spaces described in the introduction, we will take the gauge group to be $\SU(Q_5)$, since the center of $\U(Q_5)$ leads to a $BF$ topological field theory
that generally decouples.    We view the Lie algebra of $\SU(n)$ as the algebra of traceless antihermitian $n\times n$ matrices and we define the invariant
quadratic form
\be\label{qform}   (a,b)=-\tr\,a b, \ee
where the minus sign is needed for positivity.

  Most of what we will say will apply with minor modifications for other compact semi-simple gauge groups, such as orthogonal and symplectic groups that one would
encounter in an orientifold construction. Sometimes we consider a general compact simple Lie group $\GG$.
 
 \subsection{Cohomology: Some Simple Observations}
 
 Some low-dimensional cohomology classes of the instanton moduli space $\M$ will be important.    One thing that will emerge in the following analysis is that some statements
 are only true, or are only known to be true, if $Q_1$ and/or $Q_5$ is sufficiently large.  For small values of the charges, there may be exceptional behavior not seen in supergravity.
 
 \subsubsection{More on the $B$-Field}\label{moreb}
 
 First let us explain in more detail the statement that the interaction
    \be\label{neffcouple}- \i\int_{\R^2\times \S^3\times \S^1}\CS(A) \wedge \frac{G_3}{2\pi}\ee
    can be interpreted at low energies in terms of a $B$-field on $\M$.      The fact that the model is at low energies a sigma-model with target $\M$
    means that, if $m^\alpha$, $\alpha=1,\cdots, \dim\,\M$ are local coordinates on $\M$, then at low energies the model can be expressed in terms of fields $m^\alpha(x)$,  $x\in\R^2$,
    along with their supersymmetric partners.   This implies in particular that after integrating out massive fields,
     the gauge field $A$ -- both its components $A_\mu$, $\mu=1,2$ along $\R^2$ and its components
    along $\S^3\times \S^1$-- can be expressed in terms of the $m_\alpha$.  The $A_\mu$  vanish if the $m$'s are constant along $\R^2$ (a constant set of $m^\alpha$'s simply
    describes an instanton on $\S^3\times \S^1$ with no dependence on $\R^2$).  So the $A_\mu$ are proportional to the derivatives of the $m$'s.  The general form, modulo
    irrelevant terms of higher order,  
   is $A_\mu(x,y) =\sum_\alpha f_\alpha(m(x),y) \frac{\partial}{\partial x^\mu} m^\alpha   $, where  $y\in \S^3\times \S^1$  and
 $f_\alpha(m,y)$ are Lie algebra valued functions of  $m$ and $y$.   So we have $\partial_\mu A_\nu-\partial_\nu A_\mu =
    \sum_{\alpha,\beta} (\partial_\alpha f_\beta-\partial_\beta f_\alpha) \partial_\mu m^\alpha
    \partial_\nu m^\beta$, and $[A_\mu,A_\nu]=\sum_{\alpha,\beta}[f_\alpha,f_\beta]\partial_\mu m^\alpha  \partial_\nu  m^\beta$.   Since $\CS(A)$ depends on $A_\mu$ through $\partial_\mu A_\nu-\partial_\nu A_\mu$ and $[A_\mu,A_\nu]$,
    the form of these expressions shows that the interaction (\ref{neffcouple}) reduces at low energies, after integrating over $\S^3\times \S^1$,  to something of the  general form
    \be\label{bfield}\i\int_{\R^2}\d^2 x\,\epsilon^{\mu\nu} \,B_{\alpha\beta}(m)\partial_\mu m^\alpha \partial_\nu m^\beta,\ee
    with $B_{\alpha\beta}=-B_{\beta\alpha}$.   This is the standard form of the contribution of a $B$-field to the sigma-model action.
    
    If the $m^\alpha$ are constant at infinity along $\R^2$, so that $A_\mu$ vanishes there, then in evaluating the expression (\ref{neffcouple}) we can compactify $\R^2$ to $\S^2$.
    If we then view $\S^2$ as the boundary of a ball $U$ and extend the gauge field from $\S^2\times \S^3\times \S^1$ over $U\times \S^3\times \S^1$, then we can replace
    eqn. (\ref{neffcouple}) with
    \be\label{betterform}-\i\int_{U\times \S^3\times \S^1}\frac{\tr F\wedge F}{4\pi} \wedge \frac{G_3}{2\pi}.\ee
    This is an improved formula because the integrand is gauge-invariant, but it does potentially depend on the choice of $U$ and of the extension of the gauge field over $U$.  To compare
    two different choices with different extensions over possibly different manifolds $U$ and $U'$, we glue together $U$ and $U'$ along their boundaries to make a closed oriented manifold
    $W$ and  learn that the difference in the two evaluations of the action is
    \be\label{etterform}-\i \int_{W\times \S^3\times \S^1} \frac{\tr F\wedge F}{4\pi} \wedge \frac{G_3}{2\pi}.\ee
    We can explicitly do the integral over $\S^3$, using the fact that in the setup described in section (\ref{motivation}), 
    $G_3$ is a pullback from $\S^3$ and $\int_{\S_3}\frac{G_3}{2\pi}=Q'_5$.  So the coupling turns out to be
    \be\label{zetterform}-\i Q'_5 \int_{W\times \S^1} \frac{\tr F\wedge F}{4\pi}.\ee
    The integral is $2\pi \I$, where $\I$ is the instanton number of the gauge field on $W\times \S^1$.   Thus the action is
    \be\label{letterform} -2\pi\i \I Q'_5.\ee
    As $\I$ is in general an arbitrary integer, this formula explains the interpretation of $Q'_5$ as the sigma-model ``level.''   In other words,
    the formula shows that the effective $B$-field is $Q'_5$ times a minimal $B$-field that would satisfy the appropriate Dirac quantization.
    
    This characterizes the sigma-model $B$-field topologically, but the reader may wonder if it is possible to write an explicit formula for the gauge-invariant field strength
    $H=\d B$ as a three-form on $\M$.  Indeed, to understand the supersymmetry of the sigma-model, we will need to understand explicit formulas.   We explain
    some necessary tools in section \ref{systematic} and eventually arrive at the final formula in section \ref{hkt}.
    
    \subsubsection{Theta-Angles And The Second Betti Number}\label{secondbetti}
    
    Another important question is whether the $\sigma$-model admits a theta-angle.   In general, a (continuous)  theta-angle is associated to a term in the Euclidean action of the 
    form $-\i\theta \Gamma$, where $\Gamma$ is an integer-valued topological invariant and therefore $\theta$ is an angular variable, 
    $\theta\cong \theta+2\pi$.  In the present context, there is exactly one  suitable invariant, namely the
    integral of the third Chern class,\footnote{For $Q_5=2$, the third Chern class vanishes and the sigma-model with target the $\SU(2)$ moduli space
    does not have a continuous theta-angle. (It does have a discrete one because $\pi_5(\SU(2))=\Z_2$.)   For $Q_5=2$, possibly
    the parity-odd supergravity modulus found in \cite{GMMS} decouples from the low energy physics. We note that the gauge group of the brane system is really not $\SU(Q_5)$ but $\U(Q_5)$,
    or $\U(2)$ for $Q_5=2$.  The third Chern class is nonzero for $\U(2)$, but at least for most purposes, the center $\U(1)\subset \U(2)$ decouples at low energies because of a  Chern-Simons coupling, as reviewed in the introduction. For $Q_5=2$, perhaps some subtle effects depend on the coupling to the third Chern class.}
    \be\label{thirdch} \Gamma=\frac{1}{24\pi^3}\int_{\R^2\times \S^1\times \S^3} \tr\, F\wedge F\wedge F. \ee
    Thus a single theta-angle will appear as a modulus of the sigma-model with target $\M$.  
   Precisely the same coupling appears in the gauge theory description of Type IIB superstrings on $\T^4$ or $\KK$.   In that context, this coupling
   is interpreted as the dual to a Ramond-Ramond parameter that appears in the supergravity solution \cite{Dijkgraaf}.
   We expect that this coupling has the same interpretation in the case of Type IIB on $\AdS_3\times \S^3\times \S^3\times \S^1$.   Indeed, in that case
   the supergravity solution has precisely one Ramond-Ramond modulus \cite{GMMS}.

    It is straightforward to compare the moduli space of sigma-model parameters to supergravity.
     Since $\Gamma$ is odd under a charge conjugation symmetry of $\SU(Q_5)$ that exchanges the
    fundamental and anti-fundamental representations,
     the angle $\theta$  contributes a factor $\S^1/\Z_2$ to the moduli space.  
     In this article, we will always assume that the metric on $\S^3\times \S^1$ is a product of standard round metrics on the two factors, in general with arbitrary radii $r,r'$.
    Since the instanton equation is conformally invariant, only the ratio $r'/r$ is relevant.  It is a positive number, taking values in the positive half-line $\R_+$.   Including also the theta-angle,
    the sigma-model moduli space is $\R_+\times \S^1/\Z_2$.   This is precisely the moduli space of Type IIB supergravity on $\AdS_3\times \S^3\times \S^3\times \S^1$,
    as determined in \cite{GMMS}.    That gives some preliminary support to the duality conjecture relating the  the sigma-model with target $\M$ to string theory on
    $\AdS_3\times \S^3\times \S^3\times \S^1$.   We have assumed that no unknown dualities are present; this was argued in \cite{GMMS} based on 
    an examination of the ends of the moduli space. 
    
       The parameter $\theta$  has a qualitative effect on the sigma-model spectrum.   
      For any oriented  four-manifold $M$, the moduli space $\M$  of instanton solutions on $M$ is compact and smooth except for the small instanton singularity as well
      as singularities associated with ``un-Higgsing'' -- that is, singularities associated with reducible instanton solutions that do not completely break the gauge symmetry. These singularities
      are ``universal'' -- the small instanton singularity does not depend on the four-manifold in which the small instanton is embedded, and the un-Higgsing singularity depends only
      on the unbroken gauge group and the spectrum of massless charged hypermultiplets, not on details of the four-manifold.\footnote{These statements fail in some exceptional cases described in section \ref{conformal}.}  In the case of $\T^4$ or $\KK$, 
      there are four parameters, related to each other
       by supersymmetry, that control the singularities.   Three of these are parameters associated to 
       Neveu-Schwarz (NS) $B$-field modes,
        which resolve the small instanton singularity via a noncommutative deformation of the instanton
   equation \cite{NS}.  The same parameters also resolve the un-Higgsing singularities as long as $Q_1$ and $Q_5$ are relatively prime.   (If $Q_1$ and $Q_5$ are not relatively
   prime, the model has  unavoidable un-Higgsing singularities.)
    The fourth is a Ramond-Ramond mode, which in the sigma-model becomes the theta-angle associated to the third Chern class \cite{Dijkgraaf}.  It resolves
    the same singularities that the NS parameters resolve in the abstract sense of giving the sigma-model a discrete spectrum (not in the sense of classically resolving the 
    singularities of $\M$).   
   In the case of $\S^3\times\S^1$, the $B$-field modes that could resolve the singularity in a classical sense are absent globally, since $H^2(\S^3\times \S^1;\R)=0$.   
   However, the theta-angle is still present.  One expects the small instanton singularity to be resolved as long as $\theta\not=0$, and (though this point deserves a more careful
   study) one expects the un-Higgsing singularities
   to be resolved as long as the triple $Q_1,Q_5,Q'_5$ has no common divisor.   
      
The reader might notice the following gap in our reasoning.   In the sigma-model with target $\M$, we are not interested in arbitrary gauge fields on $\R^2\times\S^3\times \S^1$
    but only in those that can be interpreted in terms of maps of $\R^2$ to $\M$.   Concretely, these are gauge fields on $\R^2\times \S^1\times \S^3$
    that when restricted to $p\times \S^1\times \S^3$, for any point $p\in\R^2$, satisfy the instanton equation on $\S^1\times \S^3$.   With this restriction, is it still true that $\Gamma$
    can take arbitrary integer values?   General results that will be described in section \ref{systematic} imply  that, for any $Q_5$, this is true for sufficiently large $Q_1$.
    This allows the possibility that, for example, when we approximate the six-dimensional gauge theory by a two-dimensional sigma-model, $\Gamma$ might vanish
    identically for some small values of $Q_1$.   If this happens (and the duality conjecture that we are discussing is correct), 
    it would mean that the modulus that is seen in supergravity and is related to the theta-angle of the sigma-model
    decouples from the low energy physics for those
   particular values of $Q_1$.
   
   A similar question concerns the coupling (\ref{neffcouple}) that we interpret in terms of a two-form field $\h B$ on $\M$.   If $A$ is a completely general gauge field on
   $W\times \S^1\times \S^3$, then the expression in eqn. (\ref{betterform}) can equal an arbitrary integer multiple of $2\pi Q'_5$, and therefore the coupling (\ref{neffcouple}) is well-defined
   precisely mod $2\pi Q_5'$.      
     That is the basis for interpreting $Q_5'$ as the sigma-model ``level.''   However, if we constrain 
   $A$ so that its restriction to $p\times \S^1\times \S^3$ satisfies the instanton equation for every $p$, then is it still true that  eqn. (\ref{betterform}) can equal an arbitrary integer
   multiple of $2\pi Q_5'$?   Again, for sufficiently large $Q_1$, this is true by the general results that will be explained in section \ref{systematic}.
   
   One can also consider the possibility of discrete theta angles associated to torsion in $H^2(\M;\U(1))$.   There is no evidence in supergravity that any such discrete parameters should exist.
   It seems quite likely that for gauge group $\U(Q_5)$, there is no torsion in $H^2(\M;\U(1))$.  This would follow from results in \cite{AB} together with the Atiyah-Jones conjecture
   described in section \ref{systematic}.

    \subsubsection{First Betti Number}\label{firstbetti}
    
    We will also need to understand $H^1(\M;\R)$ and its dimension, which is the first Betti number $b_1(\M)$.   In general, a generator of $H^1(\M;\R)$ is a closed 1-form $\lambda$
    that is not exact.   A closed 1-form $\lambda$ can always be written locally as $\lambda=\d \varphi$ with some function $\varphi$, but $\varphi$ may not be single-valued; it may be
    well-defined only modulo a constant.  Conversely, if $\varphi$ is a function on $\M$ that is well-defined modulo a constant, then $\lambda=\d\varphi$ is a nonzero
    element of $H^1(\M;\R)$.
    
    For the case that $\M$ is the instanton moduli space on $\S^3\times \S^1$, a real-valued function that is only well-defined modulo a constant is
    \be\label{zofox}\varphi=\frac{1}{2\pi}\int_{\S^3\times \S^1}\CS(A) \wedge \frac{\d\phi}{2\pi}   ,\ee
    with $\phi$ an angular variable on $\S^1$.   Here we have chosen to integrate over $\phi$, but for the purpose of finding a generator of $H^1(\M;\R)$, it would not matter if
    we instead set $\phi$ to a specific value.   The function $\varphi$ defined in eqn. (\ref{zofox})  is multi-valued because of the usual multi-valuedness of the Chern-Simons form.
    
    Therefore, the exterior derivative of $\varphi$  is a generator of $H^1(\M,\R)$. 
    Let $\A$ be the space    of all gauge connections on
    $\S^3\times \S^1$ (not just the ones that satisfy the instanton equation), and let $\G$ be the group of gauge equivalences (locally, this    is simply the group of maps of $M$ to the gauge group $\SU(Q_5)$). 
    We can view $\varphi$ as a multi-valued function defined on the space $\A/\G$ of gauge fields
  modulo gauge equivalences, and likewise we can define the exterior derivative of $\varphi$ as a 1-form
    on $\A/\G$.   We will reserve the symbol $\d$ for the exterior derivative on a finite-dimensional manifold such $\S^3\times \S^1$ or $\M$, and write $\delta$ for the exterior derivative
    on the infinite-dimensional manifold $\A$ or its quotient $\A/\G$.   We also define
    \be\label{fdef} \psi(y) = \delta A(y),~~~~~y\in \S^3\times \S^1.\ee
    We  can compute an explicit formula for $\lambda=\delta\varphi$:
    \be\label{unef} \lambda=\frac{1}{(2\pi)^2}\int _{\S^3\times \S^1}\tr\, F\wedge \delta A \wedge  \frac{\d\phi}{2\pi}  =\frac{1}{(2\pi)^2}\int_{\S^3\times \S^1}\tr\, F\wedge \psi\wedge  \frac{\d\phi}{2\pi}  \ee
    
     One might wonder how to explicitly  describe a loop in $\M$ on which $\lambda$ has a nonzero integral.   This can be done as follows.  Let $u$ denote a point in  $\S^3$ and let $A(u,\phi)$ be any instanton solution on $\S^3\times \S^1$ of instanton number $Q_1$, representing a point in $\M$.   Introduce a second circle
     $\t \S^1$ parametrized by another angular variable $\t\phi$.   Then $A(u,\phi+\t\phi)$ is, for each fixed $\t\phi$, an instanton solution describing a point in $\M$; as $\t\phi$ varies over $\t\S^1$, $A(u,\phi
     +\t\phi)$ varies over a loop $\gamma\in \M$.     It is also true that for fixed $\phi$, $A(u,\phi+\t\phi)$ can be viewed as an instanton solution, again of instanton number $Q_1$,
     on $\S^3\times \t\S^1$.   Using this, we can evaluate the integral that defines $\oint_\gamma \lambda$:
          \be\label{zimbox} \oint_\gamma \lambda = Q_1. \ee
          Here, since $\int_{S^1}\frac{\d \phi}{2\pi}=1$, we have to evaluate
          \be\label{pimbox} \frac{1}{(2\p)^2}\int_{S^3\times \t S^1}\tr \, F\wedge \psi.\ee
          Here we can view $\psi$ as a two-form with one index along $S^3$ and one along $\t S^1$ (the reader might want to 
          return to this point after reading section (\ref{systematic})), so we can replace $\tr \,F\wedge \psi$ by $\frac{1}{2}\tr\,F\wedge F$.  (There is  a factor of $\frac{1}{2}$ here
          because either factor of $F$ in $\tr\,F\wedge F$ might have an index along $\t S^1$.)   So we arrive at the integral that computes the instanton number on $S^3\times\t S^1$.
     
    \subsubsection{Zeroth Betti Number}\label{zero}
    
    Though this will play less of a role in the present article,
    one  may also ask whether $\M$ is connected, that is, whether its zeroth Betti number vanishes.   It appears that the fact that $S^3\times S^1$ is elliptically fibered can be used to prove
    that $\M$ is connected.   To see that $S^3\times S^1$ is elliptically fibered, one can use the Hopf fibration $S^3\to S^2$ with fibers $S^1$, implying a fibration $S^3\times S^1\to S^2$
    with fibers $S^1\times S^1$.  Since  $S^2\cong \CP^1$ and since $S^1\times S^1$ can be  regarded as an elliptic curve, the fiber and base of this fibration are both complex manifolds. The description
    of $S^3\times S^1$ in section \ref{spherex} shows that the total space of the fibration is also a complex manifold.  So $S^3\times S^1$ is a complex (but not Kahler) elliptic fibration.
     The use of the elliptic fibration to prove that $\M$ is connected would follow
    ideas explained in \cite{BH}, generalized to higher instanton number and gauge groups of higher rank.
    
    The generalization to $S^3\times S^1$ of the Atiyah-Jones conjecture, described in section \ref{systematic}, would also imply that $\M$ is connected.
    
    These two remarks may be related as it may be possible\footnote{E. Gasparim, private communication.}
     to use the elliptic fibration to prove the Atiyah-Jones conjecture for $S^3\times S^1$.

    \subsection{Symmetries Of $\M$}\label{freeaction}
    
Here we will discuss properties of the instanton moduli space  $\M$ that are associated to   symmetries of $\S^3\times \S^1$.

First of all, $\S^3\times \S^1$ has discrete symmetries that act by a reflection of $\S^3$ and/or $\S^1$.   But separate reflections of the two factors reverse the
orientation of $\S^3\times \S^1$, so they are not symmetries of the instanton equation and do not lead to symmetries of $\M$.   Only a combined reflection
of $\S^3$ and $\S^1$ leads to a symmetry of $\M$.  We will denote such a joint reflection of the two factors as $\rho$.   This only characterizes $\rho$ up to a rotation of the
two factors, but the precise choice of $\rho$ will never be important.

The  restriction to a joint reflection of the two factors agrees with what one would expect based on a presumed duality with Type IIB superstrings on $\AdS_3\times \S^3\times \S^3\times \S^1$. 
  As analyzed originally in \cite{EFGT}, the superstring
 solution on $\AdS_3\times \S^3\times \S^3\times \S^1$ has a nonzero three-form field on each of the first three factors, $\AdS_3$ and the two $\S^3$'s.    This three-form is invariant under all continuous symmetries of $\AdS_3\times \S^3\times \S^3\times \S^1$ and is  of the same type (Neveu-Schwarz, Ramond-Ramond, or a combination) on all three factors.  $S$-duality implies that the
type of three-form does not affect the following remarks, and for definiteness we will assume a Neveu-Schwarz three-form $H$.
 A reflection of one of the $\S^3$'s will
reverse the sign of $H$.   To get a symmetry, we can compensate for this by reversing the worldsheet orientation, which also reverses the sign of $H$.
But as there is $H$ flux on each of the first three factors of $\AdS_3\times \S^3\times \S^3\times \S^1$, to get a symmetry that involves reversing the world-sheet orientation, we must make a simultaneous reflection of all three of those factors.
Finally, as Type IIB superstring theory does not have a symmetry that reverses the spacetime orientation, to get a symmetry we must also simultaneously reflect the $\S^1$.
This explains that in a dual description involving gauge fields on the product $\S^3\times \S^1$ of the last two factors, we should only expect to see a joint reflection of the two factors as a  symmetry, not a reflection of just one factor.
But we also see that this discrete symmetry of $\S^3\times \S^1$ is accompanied in the full string theory by a reflection of $\AdS_3$, which will reverse the boundary orientation.
So therefore the joint reflection $\rho$ of the two factors of $\S^3\times \S^1$ will be a parity symmetry of the dual CFT -- a symmetry reverses the orientation and  exchanges left- and right-moving modes.

Now we move on to discuss continuous symmetries of $\S^3\times \S^1$ and their action on $\M$.
For a convenient model of $\S^3\times \S^1$, describe $\S^3$ with real variables $y_\lambda,~\lambda=0,\cdots, 3$ satisfying $\sum_{\lambda=0}^3 y_\lambda^2=1$, and parametrize
$\S^1$ by a periodic variable $\tau$ with $\tau\cong \tau+T$ for some $T$.    Choose a metric $g$ on  $\S^3\times \S^1$ that is described by the line element
\be\label{nucc} \d s^2=\sum_{\lambda=0}^3 \d y_\lambda^2+\d \tau^2.\ee
In what follows, indices $i,j,k$ are tangent to  $\S^3\times \S^1$, while 
indices $a,b,c=1,\cdots,3$ are tangent to $\S^3$, and an index $\tau$ is tangent to $\S^1$. 
We denote the orientation of  $\S^3$ via the antisymmetric Levi-Civita tensor $\epsilon_{abc}$ and orient $\S^3\times \S^1$ so that the instanton equation reads
\be\label{instanteq} F_{ab}+\epsilon_{abc}g^{cc'}F_{c'\tau} =0.\ee

Let us look at the vector field $V=\frac{\partial}{\partial\tau}$ that generates a rotation of $\S^1$, and the corresponding vector field $\h V$ on $\M$.   In general, the action of a vector field $V$ on a gauge field $A$, with field strength $F=\d A+A\wedge A$, is only uniquely determined
up to an infinitesimal gauge transformation and takes the form
\be\label{ohappy}\delta A_i=V^j F_{ji}-D_i\sigma, \ee
where $\sigma$ is the generator of a gauge transformation.   Let us prove that the particular vector field  $V=\frac{\partial}{\partial\tau}$ acts without fixed points on the instanton
moduli space $\M$ (assuming that the instanton number is nonzero).   A zero or fixed point of the vector field $\h V$ on $\M$ that corresponds to the vector field $V$ on $\S^3\times \S^1$ would be an instanton solution such that, for some $\sigma$, eqn. (\ref{ohappy}) reduces to
\be\label{nohappy}0= F_{\tau i} -D_i\sigma. \ee
Squaring, taking a trace, and integrating, we get
\be\label{zohappy}0=-\int_{\S^3\times \S^1}\d^4 x\sqrt g \sum_i \tr\,(F_{\tau i}-D_i\sigma)^2 .\ee
The term linear in $\sigma$ vanishes after integrating by parts, since an instanton connection satisfies the second order Yang-Mills equation $D^i F_{ji}=0$.   Hence
\be\label{lohappy} 0=-\int_{\S^3\times \S^1}\d^4 x\sqrt{g}\sum_i \tr\, \left( F_{\tau i}^2+(D_i\sigma)^2\right). \ee
In particular, a fixed point satisfies $F_{\tau i}=0$.   Since the instanton equation (\ref{instanteq}) then implies that $F_{ij}=0$ for all $i,j$, it follows that a fixed point is actually a flat connection
and can only exist if the instanton number is zero.

Note that $V$ has constant length as a vector field on $\S^3\times \S^1$.
In section \ref{large}, after defining the metric of $\M$, we will show that the vector field  $\h V$ on $\M$ that is associated to $V$ also has a  constant length, which
will determine one of the central charges in the large $\N=4$ algebra.  Of course, the assertion that $\h V$ has (nonzero) constant length is  much more precise  than the statement that we just proved
showing that $\h V$ has no zeroes or fixed points.    Similarly,  writing  $\SU(2)_\ell$ and $\SU(2)_r$ for the left and right actions of $\SU(2)$ on $\S^3=\SU(2)$, a generator $T_\ell$ or $T_r$ of $\SU(2)_\ell$ or $\SU(2)_r$ is a vector field on $\S^3\times \S^1$ of constant length.
In section \ref{large}, we will learn that the corresponding vector fields $\h T_\ell$ and $\h T_r$ on $\M$ likewise have (nonzero) constant  length, and hence act without fixed points.
   More generally, if $T_\ell$ and $T_r$ have unequal
lengths (as vector fields on $\S^3$), then $\h T_\ell+\h T_r$ acts  on $\M$ without fixed points.
It  will turn out as well that any linear combination $u \h T_\ell+v \h T_r +w \h V $ with $w\not=0$ has no fixed point.
 The fact that $\M$ has many symmetry generators that act without fixed points is likely to mean that
some attempts at supersymmetric localization will give a trivial result.

However, if $T_\ell$ and $T_r$ have equal length, then $\h T_\ell+\h T_r$ does have fixed points in $\M$, as analyzed in 
 \cite{BH}.   We discuss  some consequences  in section \ref{notso}.   For now, we just note
that an example of a vector field $T=T_\ell+T_r$ on $\S^3$ such that  $T_\ell$ and $T_r$ have equal length is
\be\label{itto} T= y_2\frac{\partial }{\partial y_3}- y_3\frac{\partial}{\partial y_2}. \ee
The corresponding vector field $\h T=\h T_\ell+\h T_r$ on $\M$ does have fixed points.

    \subsection{Is The Moduli Space A Symmetric Product?}\label{notso}
    
    As explained in the introduction, in the literature there are primarily two proposals for the dual of Type IIB superstring theory on $\AdS_3\times \S^3\times \S^3\times \S^1$:
  the dual might be a symmetric product of $n$  copies of $\S^3\times \S^1$ for some $n$, denoted $\Sym^n(\S^3\times \S^1)$, or it might be  a sigma-model with target the instanton moduli space $\M$ on $\S^3\times \S^1$.
    
    One can also ask -- and this question has been raised as well -- whether the two conjectures are different.   
    Or is $\M$, which according to an index theorem has real dimension $4 Q_1 Q_5$,  the same as, or possibly deformation equivalent to,
  $\Sym^{Q_1Q_5}(\S^3\times \S^1)$?   A motivation for this question is that the instanton moduli spaces on $\T^4$ or $\KK$ are indeed
such  symmetric products.   Is the same true for  $\S^3\times \S^1$, at least for some values of the charges?
    
    The answer to this question is that in fact (for $Q_5\not=1$)
    the instanton moduli spaces on $\S^3\times \S^1$ are not symmetric products of copies of $\S^3\times\S^1$ or deformation equivalent to such  products.
   One can show this  by considering the fixed points of the $\U(1)$ symmetry of $\S^3\times \S^1$ that is generated by
    the vector field $T$ of eqn. (\ref{itto}).   It is shown in \cite{BH} that  instantons on $\S^3\times \S^1$ that are invariant under this $\U(1)$ 
    are equivalent to BPS monopoles on a hyperbolic
    three-manifold $\HH_3/\Z$, where $\HH_3$ is hyperbolic three-space (the Euclidean version of $\AdS_3$), and a generator of $\Z$ acts by a hyperbolic element of the symmetry group of $\HH_3$; thus  $\HH_3/\Z$ is simply the Euclidean version of a BTZ black hole.  
    For $\GG=\SU(2)$, the magnetic charge is a single integer $q$, which as shown in \cite{BH} must be a divisor of the instanton number $Q_1$.   For every divisor $q$, the monopole moduli space is non-empty
    according to Corollary 5.3 in \cite{Braam}.  For any\footnote{For $\SU(2)$ gauge theory with instanton number 1, the instanton modulli space
    is described explicitly in \cite{BH} and does not appear to be a symmetric product.} $Q_1\geq 2$, there are at least two divisors, namely 1 and $Q_1$.    Thus the fixed point set always has at least
    two topological components.  For $Q_5>2$, the magnetic charge of a monopole is classified by $Q_5-1$ integers, and the number of topological components of the fixed point set 
    grows rapidly.   
    
    By contrast, the zeroes of the vector field $T$ on $\S^3$ comprise the circle $y_2=y_3=0$, so its zeroes on $\S^3\times \S^1$ make up the connected manifold $\S^1\times \S^1$,
    a two-torus.
    The zeroes of $T$ on $\Sym^{Q_1 Q_5}(\S^3\times \S^1)$ make up $\Sym^{Q_1 Q_5}(\S^1\times \S^1)$.
    This is still a connected manifold, and will remain connected after any deformation or resolution.  So the instanton moduli space on $\S^3\times \S^1$ is not deformation
    equivalent to $\Sym^{Q_1Q_5}(\S^3\times \S^1)$.
    
    We should remark that for $Q_5=1$, there is reasonable evidence that the dual of string theory on $\AdS_3\times \S^3\times \S^3\times \S^1$ is a sigma-model
    with target $\Sym^{Q_1}(\S^3\times \S^1)$, with each factor taken at level $Q_5'$   \cite{EGGL}.  Of course, for $Q_5=1$, the gauge group is $\U(1)$ and instantons only exist after a noncommutative deformation \cite{NS}.   In the case of $\T^4$
    or $\KK$, there are four parameters associated to such a deformation.  Three of these parameters arise from  Neveu-Schwarz $B$-field modes and are studied in \cite{NS} and the fourth,
    related to these by supersymmetry, is the expectation value of a certain Ramond-Ramond  field.   The NS $B$-field modes have no analog for $\S^3\times \S^1$, because
    $H^2(\S^3\times \S^1;\R)=0$.   But the RR mode does have an analog \cite{GMMS}, and it is plausible that with this mode turned on, the instanton moduli space on
    $\S^3\times \S^1$ for $Q_5=1$ is a symmetric product,  potentially providing a framework for the results of \cite{EGGL}.

    \subsection{Cohomology: Systematic Approach}\label{systematic}
    
   At the end of section \ref{secondbetti}, we  asked whether certain topological statements about gauge fields on $\R^2\times \S^1\times \S^3$ (with prescribed behavior at infinity
   along $\R^2$) are modified if one requires
   that the restriction of the gauge field to $p\times \S^1\times \S^3$, for any $p\in \R^2$, is an instanton solution.   These are questions about the relationship between
   the topology of the space $\A/\G$ of gauge fields modulo gauge transformations on a given $\GG$-bundle $E\to \S^3\times \S^1$,
    and  the topology of the moduli
    space $\M$ of instanton connections on $E$. Here $\A$ is the space of all connections on $E$ and $\G$ is the group of gauge equivalences, so 
    $\M$ is a subspace of $\A/\G$, namely the subspace that parametrizes gauge equivalence classes of instanton solutions.

The original mathematical statement about questions of this nature was     
    the Atiyah-Jones conjecture \cite{AJ} about $\SU(2)$ gauge fields on $\S^4$.  It asserts, roughly speaking, that in the limit that the instanton number is
    large, $\M$ and $\A/\G$ are topologically equivalent. 
   This equivalence means, for instance, that for any $d\geq 0$, the restriction map  $j^*:H^d(\A/\G;\Z)\to H^d(\M;\Z)$ associated to the embedding $j:\M\to \A/\G$ is an isomorphism.
    The original Atiyah-Jones conjecture was proved in \cite{BHMM} and its generalization to $\SU(n)$ bundles on $\S^4$ was proved in \cite{T}.   For a short survey of related results and questions, see \cite{G}.
    
   In this article, we are primarily interested in instantons on $\S^3\times \S^1$, rather than $\S^4$.   Hence a generalization of
   the Atiyah-Jones conjecture is relevant.  
  A plausible generalization  would say that for any compact gauge group $\GG$ and any oriented compact  four-manifold $M$, $\M$ and $\A/\G$  are topologically equivalent in the limit of large
  instanton number.   What has actually been proved for general $\GG$ and $M$ is a somewhat weaker statement (Theorem 2$^*$ in \cite{Taubes}), which says in particular
  that for any $M$ and $\GG$, the map on homology $j_*:H_d(\M;\Z)\to H_d(\A/\G;\Z)$ is a surjection if the instanton number is large enough compared to $d$.   This implies that for any manifold $N$,
  constraining a gauge field on $N\times \S^1\times \S^3$
 to satisfy the instanton equation when restricted to $p\times \S^1\times \S^3$ for any $p\in N$ places no topological restriction if the instanton number is sufficiently large.  
 So it answers the questions raised at the end of
 section \ref{secondbetti}, which involved the case that $N$ is of dimension 2 or 3.
 
  However, several issues that arise in the present article require further information concerning the relation between the topology of $\M$ and of $\A/\G$.   In particular, it would
  be nice to know that the restriction maps  $j^*:H^i(\A/\G;\Z)\to H^i(\M;\Z)$ are isomorphisms, at least for $i=1,2$ and suitable $Q_1,Q_5$.
 The relevance to arguments in this paper is as follows.   In general, the number of continuous theta-angles in a sigma-model with target
$\M$ is the dimension of $H^2(\M;\R)$.   The invariant defined in eqn. (\ref{thirdch}) can be viewed as a generator of $H^2(\A/\G;\Z)$ and therefore (forgetting its integrality) of $H^2(\A/\G;\R)$.
As we explain shortly,
 this invariant generates $H^2(\A/\G;\R)$. It can be restricted to $\M$ and this restriction is nonzero according to \cite{Taubes}.  In comparing the sigma-model moduli space to the supergravity moduli space, we assumed that this restriction generates $H^2(\M;\R)$, leading to a unique theta-angle.   This is true if the map $j^*$ is an isomorphism on the degree two cohomology. 
   Somewhat similarly, in eqn. (\ref{unef}), we defined a closed one-form $\lambda$ that is a generator of $H^1(\A/\G;\R)$.
As we explain shortly,   $\lambda$ generates $H^1(\A/G;\R)$. 
We proved that the restriction of $\lambda$ to $\M$ is nonzero by exhibiting a curve in $\M$ on which $\lambda$ has a nonzero integral.
  In section \ref{conformal}, in one approach to proving conformal invariance
  of the sigma-model with target $\M$, we will want to know that $\lambda$  actually generates $H^1(\M;\R)$.   This is true 
  if the map $j^*$ is an isomorphism on the degree one cohomology.   In the present article, we will assume that 
some sort of analog of the Atiyah-Jones
  conjecture is true and justifies these statements,  at least for suitable $Q_1,Q_5$.     Hopefully, such a statement holds at least in the regime in which
  supergravity is valid, namely $Q_1\gg Q_5\gg 1$.    

  The context for some statements in the last paragraph is the following.     
      In \cite{AB}, Atiyah and Bott explicitly described a set of generators
      of the real cohomology ring  of $\A/\G$, along lines that we will explain shortly.\footnote{Atiyah and Bott were studying
     the case that $M$ is a two-manifold.   However, as long as one considers only the real cohomology of $\A/\G$, not the integer cohomology, their considerations carry over to
     the case that $M$ is of any dimension.   For the group $U(n)$, they described the classifying space. The upshot was to show that on any manifold $M$, the cohomology of
     $\A/\G$ is generated by certain classes associated to Chern classes.   For the real cohomology, one can use differential forms and this leads to the generators that they
     described and that are introduced presently.   Since $\U(n)=(\SU(n)\times \U(1))/\Z_n$, the same result applies to the real cohomology of $\A/\G$ for gauge group $\SU(n)$.}   
      For the case of $M=\S^3\times\S^1$,
     in  degree 1 and degree 2,
     the generators that they describe are precisely the ones that we considered in eqns. (\ref{thirdch}) and (\ref{unef}).

    We will need some knowledge of the explicit generators of the real cohomology of $\A/\G$ constructed by Atiyah and Bott, mainly because we will in section \ref{hkt} need
    to recognize a certain explicit three-form on $\M$ that represents the three-dimensional cohomology class that was described somewhat implicitly in section \ref{moreb}.
    The following discussion is in the spirit of \cite{AB}, as later refined and extended for applications to four-manifolds \cite{Donaldson} and  interpreted more physically \cite{EW2}.
    Recall that we denote the exterior derivative on $\A$ as $\delta$ and we also define $\psi=\delta A$.  
The components $A_\mu^a(x)$ ($x$ is a point in $M$ and $\mu$ and $a$
    are tangent space and Lie algebra indices, respectively) are understood as functions on $\A$, while $\psi_\mu^a$(x) is a one-form on $\A$.   A general function $F(A,\psi)$ that is homogeneous
    of degree $k$ in $\psi$ represents a $k$-form on $\A$.      We will extend the definition of $\delta$ with other fields included so that $\delta^2$ will generate a gauge transformation and therefore   will vanish on gauge-invariant functions.
        
    If we were interested in differential forms on $\A$,  we would simply define $\delta\psi=0$
 and then $\delta$ would represent the exterior derivative acting on functions of $A$ and $\psi$.   
    But  in order to construct differential forms on $\A/\G$, it is necessary to take into account
    the gauge group.   For this purpose, we introduce a Lie algebra valued scalar field $\sigma$ on $M$.  The action of $\delta$ on the three fields $A,\psi,\sigma$ is defined by
    \begin{align}\label{multo} \delta A & = \psi \cr
     \delta \psi &=-\d_A\sigma \cr
    \delta\sigma&=0. \end{align}
    Here $\d_A=\d+[A,\cdot]$ is the gauge-covariant extension of the exterior derivative.   These formulas imply that $\delta^2$ is equivalent to a gauge transformation
    generated by $\sigma$; for example, $\delta^2 A=\delta \psi=-\d_A\sigma$ is the infinitesimal transformation of $A$ under a gauge transformation generated by $\sigma$, and similarly
    $\delta^2\psi=[\sigma,\psi]$, $\delta^2\sigma=[\sigma,\sigma]=0$.    Since we want $\delta$ to increase the degree of a differential form by 1, and we have assigned degree 0 to $A$ and
    degree 1 to $\psi$, we have to assign degree 2 to $\sigma$.   We will call this degree the $\delta$-degree; it has also been called ghost number.   In keeping with the fact that
    $\sigma$ has $\delta$-degree 2, we will eventually learn that it can be converted for many purposes  to a two-form on $\A/\G$.
    
    Since in general $\delta^2$ is the generator of a gauge transformation, it follows that acting on gauge-invariant
    functions $F(A,\psi,\sigma)$, $\delta^2=0$.   We will explain explicitly how to convert a gauge-invariant function $F(A,\psi,\sigma)$ that is homogeneous of degree $k$ to a $k$-form on $\A/\G$.
A function $F$ that is $\delta$-closed, that is, one that  satisfies $\delta F=0$, will map  to a closed  form on $\A/\G$.   What we have described is in fact the Cartan model of the equivariant cohomology
    of $\G$ acting on $\A$.   Because $\G$ acts freely\footnote{Except for the center of $\GG$ and except for subtleties involving gauge connections on $M$ that are reducible, that is, those that do not 
   completely break the gauge symmetry down to the center of $\GG$.   For our purposes, these subtleties are unimportant, because reducible instantons on $\S^3\times \S^1$
arise only in very high codimension, and the center of $\GG$ acts trivially on gauge fields.  However, a more careful treatment of the cohomology
    might be important for more delicate questions beyond the scope of the present article.} 
     on $\A$, this equivariant cohomology is the same as  the cohomology of $\A/\G$. 
    
    Before explaining\footnote{The reader might prefer to return to the following  discussion
after reading section \ref{hkt}.}  how to map functions $F(A,\psi,\sigma)$ to differential forms on $\A/\G$, we will describe the important gauge-invariant and $\delta$-invariant functions on $\A$.
    The most obvious possibility is to pick a point $x\in M$ and a homogeneous polynomial $P$ of degree $s$ on the Lie algebra $\g$ of $\GG$, and consider the function 
    $P(\sigma(x))$.  This is gauge-invariant and $\delta$-invariant and is of $\delta$-degree $2s$, so it will map to a closed $2s$-form on $\A/\G$.   This form is not $\delta$-exact, since 
    $P(\sigma(x))$ is not $\delta \O$ for any $\O$, so it will potentially lead to a nontrivial cohomology class on $\A/\G$.   
    
    To minimize the notation that is required in the following discussion, let us assume that $\GG=\SU(Q_5)$, in which case the ring of invariant polynomials on $\g$ has a simple
    set of generators that we will denote $P_s^{(0)}(\sigma)=\tr \,\sigma(x)^s$, $s=2,\cdots, Q_5$.  
     Now with $\d$ as the exterior derivative on $M$, we compute $\d P_s^{(0)}= s\,\tr\,\sigma^{s-1}\d_A \sigma$.   This does not vanish, but we find $s\,\tr\,\sigma^{s-1}\d_A\sigma=-
     \delta P_s^{(1)}$ with $P_s^{(1)}=s \tr\,\sigma^{s-1}\psi$:
          \be\label{icco} \delta P_s^{(1)} =-\d P_s^{(0)}. \ee
     In this formula, $P_s^{(0)}$ is a scalar function on $M$, which can be defined at any point $x\in M$, and has $\delta$-degree $2s$; on the other hand $P_s^{(1)}$ is a one-form on $M$
     of $\delta$-degree $2s-1$. 
     
      The formula (\ref{icco}) can be
     read in two ways.   First of all,  reading the formula from right to left, 
     it says that although $P_s^{(0)}$ is not $\delta$-exact, its derivative along $M$ is $\delta$-exact.   Hence, once we learn how to interpret
     $\tr\,\sigma(x)^s$ as a $2s$-form on $\A/\G$, this $2s$-form will be independent of $x$, up to an exact form.   
     
    On the other hand, reading the formula from left to right, it says that although the 1-form $P_s^{(1)}$ on $M$ is not $\delta$-closed, its variation under $\delta$ is $\d$-exact.
    Hence if $\gamma\subset M$ is a closed loop, then $\oint_\gamma P_s^{(1)} $ is $\delta$-closed, since
    \be\label{zimbo}\delta \oint_\gamma P_s^{(1)}=-\oint_\gamma\d P_s^{(0)}=0. \ee
    Hence, once we learn to convert functions $F(A,\psi,\sigma)$ that are gauge-invariant and $\delta$-invariant to closed differential forms on $\A/\G$, the function  $\oint_\gamma P_S^{(1)}$ 
    of $\delta$-degree $2s-1$ will
    correspond to a closed differential form of degree $2s-1$ on $\A/\G$.
    
    What we have described so far is the first step in a ``descent'' procedure.   For $m\geq 0$, one finds inductively gauge-invariant polynomials $P_s^{(m)}(A,\psi,\sigma)$
    such that $P_s^{(0)}=\tr\,\sigma^s(x)$ and $\d P_s^{(m)}=-\delta P_s^{(m+1)}$.   These relations imply that if $\Sigma\subset M$ is any $m$-cycle, then
    $\int_\Sigma P_s^{(m)}$ is $\delta$-closed (indeed, $\delta \int_\Sigma P_s^{(m)}=-\int_\Sigma\d P_s^{(m-1)} =0$).    So   $\int_\Sigma P_s^{(m)}$, which has $\delta$-degree
    $2s-m$, will correspond to a closed form on $\A/\G$ of degree $2s-m$.    The cohomology class of this form on $\A/\G$ only depends on the homology class of $\Sigma$,
    since if $\Sigma=\partial B$ is a boundary, then $\int_\Sigma P_s^{(m)} =\int_B \d P_s^{(m)}=-\delta \int_B P_s^{(m+1)}$ is $\delta$-exact.
    
        The construction just sketched
     has been important in the theory of smooth four-manifolds \cite{Donaldson}.   As follows from the arguments in  \cite{AB}, the cohomology classes corresponding to 
      $\int_\Sigma P_s^{(m)}$ for various $\Sigma$ and $m$ generate the (real)  cohomology ring of $\A/\G$.  All these classes can be restricted from $\A/\G$ to $\M$.   An analog
      of the Atiyah-Jones conjecture would potentially say that these restrictions generate the real cohomology of $\M$, but at any rate, the cohomology classes of $\M$ that
      will be important in this article can be obtained as such restrictions.

      For the particular case $M=\S^3\times \S^1$, we can only make one generator of $H^1(\M;\R)$ this way, namely $\int_{\S^3\times q} P_2^{(3)}$ for an arbitrary 
      point  $q\in \S^1$, and one generator
      of $H^2(M;\R)$, namely $\int_{\S^1\times \S^3}P_3^{(4)}$.    These  correspond to the generators described in more direct terms in sections \ref{secondbetti} and \ref{firstbetti}.
      In degree 3, there are two possible generators, namely $\int_{p\times \S^1}P_2^{(1)}$, with $p\in \S^3$, and $\int_{\S^3\times q} P_3^{(3)}$, $q\in \S^1$.  The first
      of these corresponds as we will see to the field strength  of the $B$-field on $\M$, and the second will not play an important role in the present
      article.   
      
      Before explaining how to concretely interpret $\int_\Sigma P_s^{(m)}$ as a closed differential form on $\M$, we will make the descent procedure more explicit in the
      case that is actually important for understanding the $B$-field on $\M$.   Setting $s=2$ and changing the normalization, we set $P^{(0)}=\frac{1}{8\pi^2}\tr\,\sigma^2$.
      Inductively solving $\d P^{(m)}=- \delta P^{(m+1)}$, we find
      \begin{align}\label{descent}  P^{(0)}    & =\frac{1}{8\pi^2}\tr\,\sigma^2   \cr
       P^{(1)}&=  \frac{1}{4\pi^2}\tr\,\sigma\psi \cr
       P^{(2)}&= \frac{1}{4\pi^2}\tr\,\left(\frac{1}{2}\psi\wedge \psi - \sigma F\right) \cr
        P^{(3)}& = -\frac{1}{4\pi^2} \tr\,\psi\wedge F \cr
        P^{(4)}& = \frac{1}{8\pi^2}\tr\,F\wedge F. \end{align}
  So $P^{(4)}$ is the four-form whose integral over $M$ is the instanton number  $\int_\Sigma c_2(E)$, where $c_2$ is the second Chern class of the gauge bundle $E\to M$.
    This statement can be generalized as follows.   Consider a family of gauge connections on  $E\to M$, parametrized by some parameter space $S$.   These will fit together\footnote{For a more complete description, see section \ref{hkt}. We disregard a potential
  obstruction involving the center of the gauge group,
which can be circumvented by considering the bundle ${\rm ad}(E)$ associated to $E$ in the adjoint representation of $\GG$ and is inessential for our purposes.  Note that rather
than a single $\GG$-bundle $E\to M$, we really should begin this construction with a family of $\GG$ bundles $E\to M$, parametrized by $S$. }  
     into
    a gauge connection in the $M$ direction on a bundle  $\h E\to M\times S$ (for a point $s\in S$, $\h E$ restricts on $s\times M$ to the original $E\to M$, up to isomorphism).  In section \ref{moreb}, in discussing the $B$-field on $\M$, with $M=\S^3\times \S^1$, we made this construction for $S=\R^2$, and
    subsequently for $S$ a three-manifold $W$.  The universal choice of $S$ is simply $\A/\G$, which parametrizes all possible gauge fields on $M$, up to gauge transformation.
   Any other $S$ is obtained by mapping some other space into this one.
   
   The bundle $\h E\to M\times \A/G$ has a second Chern class of degree 4.   We will  consider this second Chern class as a 
   real cohomology class, valued in $H^4(M\times \A/\G;\R)$.   This cohomology group has a decomposition
   $H^4(M\times \A/\G;\R)=\oplus_{m=0}^4 H^m(M;\R)\otimes H^{4-m}(\A/\G;\R)$.   The geometrical interpretation of the descent procedure \cite{AJe,BaS} is
   that $P^{(m)}$ corresponds to the part of $c_2(\h E)$ that is valued in $H^m(M;\R)\otimes H^{4-m}(\A/G;\R)$.    Then if we are given an $m$-cycle $\Sigma\subset M$,
   we can integrate\footnote{This procedure, which was introduced in \cite{AB}, was used to great effect by Donaldson \cite{Donaldson} in studying four-manifolds.}  $P^{(m)}$ over $\Sigma$ to get a $\delta$-closed function of $\delta$-degree $4-m$, namely $\int_\Sigma P^{(m)}$, representing an element of $H^{4-m}(\A/\G;\R)$.
   This element can be restricted from $\A/\G$ to $\M\subset \A/\G$ to get a $\delta$-closed function on $\M$, which will correspond to a closed form of degree $4-m$.
   
   To get a three-dimensional cohomology class on $\A/\G$, we have to implement this procedure with $m=1$.   So we pick a point $p\in \S^3$ and integrate $P^{(1)}$ over
   $p\times \S^1$ to get $\frac{1}{4\pi^2} \oint_{p\times \S^1}\tr \, \sigma\psi$.   Once we learn how to interpret $\sigma$ as a two-form on $\A/\G$, this will indeed become a closed
   three-form on $\A/\G$, normalized so that its periods (its integrals over closed three-cycles in $\A/\G$) are integers.  That integrality will follow from  the integrality of the second Chern class.
   The cohomology class of $\frac{1}{4\pi^2} \oint_{p\times \S^1}\tr \, \sigma\psi$ does not depend on $p$, as explained in the discussion of eqn. (\ref{zimbo}).   Any choice of $p$
   will give a representative of the same cohomology class, but we can get a particularly nice representative by averaging over $p$.  If $\d^3\Omega$ is a ``round''
   volume form on $\S^3$ normalized so that the total volume is 1, then the average is just
  $\int_{\S^3\times \S^1}\d^3\Omega\,P^{(1)}$.   After converting $\sigma$ to a two-form, this will become a three-form on $\A/\G$ -- and therefore, by restriction, on $\M$ -- with integer periods.
  Since the field strength of the $B$-field is supposed to have periods that are $2\pi$ times integers, a minimal three-form that satisfies Dirac quantization will be $2\pi$ times this or
  \be\label{iddo}H_0=2\pi\int_{\S^3\times \S^1}\d^3\Omega \, P^{(1)}. \ee

  In a moment, we will explain how to turn the expression (\ref{iddo})
   into a concrete three-form on $\A/\G$, by eliminating $\sigma$ in favor of $A$ and $\psi$.   
  But for now, let us compare the expression (\ref{iddo})  to eqn. (\ref{etterform}).   Since eqn. (\ref{etterform}) is a formula for $\int_W H$ (rather than a description of a three-form $H$), to make this comparison
  we will integrate eqn. (\ref{iddo}) over a three-cycle $W\subset \A/\G$, corresponding to a family of gauge connections on $\S^3\times \S^1$.   We 
  get
  \be\label{liddo}\int_W H_0=2\pi \int_{W\times \S^3\times \S^1} \d^3\Omega P^{(1)}.\ee
  Since $\int_{\S^3}\d^3\Omega=1$, we can do the integral over $\S^3$, giving $\int_W H_0 =2\pi \int_{W\times \S^1} P^{(1)}.$
  But since $P^{(1)}$ just represents the part of $\frac{1}{8\pi^2}\tr F\wedge F\sim c_2(\h E)$ that is of degree 1 on $\S^3\times \S^1$ and degree 3 on $\A/\G$, we have
  $\int_W H_0=2\pi \int_{W\times \S^1}\frac{\tr F\wedge F}{8\pi^2}$.    Comparing this to eqn. (\ref{zetterform}), we see that the relation between $H_0$ and the field strength $H=\d B$ of
  the sigma-model is
  \be\label{kiddo}H = Q'_5 H_0+\d(\cdots).\ee
The derivation of this statement involved integrating over an arbitrary cycle $W$, so the statement only holds modulo an unknown exact form, denoted here as $\d(\cdots)$. 
  
  The reader who has gotten this far may well feel that we have not learned much, since we already asserted at the end of section \ref{moreb} that the $B$-field of the sigma-model
  is $Q'_5$ times a minimal $B$-field that obeys Dirac quantization.   However, by turning $\sigma$ into an explicit differential form, we can now do what was not available previously
  and describe  explicit three-form representatives of $H_0$.
  
  Actually, though the formula for $H_0$ that we will get makes sense and is correct as a three-form on $\A/\G$, it is possibly easier to get some intuition if we restrict to $\M$, which is in any event
  what we want to do for our application.   In this analysis, we make use of the way this construction fits into an actual quantum field theory, as explained in \cite{EW2}.
  
  When we restrict to $\M$, the gauge field $A$ on $\S^3\times \S^1$ satisfies the instanton equation\footnote{We denote the selfdual and anti-selfdual parts of a two-form
  $b$ as $b^+$ and $b^-$.}  $F^+=0$.  (In the construction in \cite{EW2}, the restriction to $\M$ is made  in the sense of supersymmetric localization.)
   Therefore $\psi=\delta A$ satisfies the linearization of the
  instanton equation, namely $(\d_A \psi)^+=0$.   However, being a one-form in four-dimensions, $\psi$ has four components, and (since the bundle of selfdual two-forms has
  rank three) the equation $(\d_A\psi)^+=0$ is only three equations.  Therefore in any physical model in which this construction is embedded, $\psi$ will obey a fourth equation.
  That equation is model-dependent, and different choices will actually lead to different three-forms all representing the same cohomology class on $\M$.
  
  In \cite{EW2}, the theory considered was a twisted version of $\N=2$ super Yang-Mills theory.   In that theory, in addition to  $A,\psi,\sigma$, the important fields for our present
  purposes are a pair $\bar\sigma,\eta$. These  are  spin zero fields in the adjoint representation; $\bar\sigma$ is a boson of $\delta$-degree $-2$ and $\eta$ is a fermion of $\delta$-degree
  $-1$.  They transform under $\delta$ as
  \begin{align} \delta \bar\sigma & = \eta\cr
                                    \delta \eta & = [\sigma,\bar\sigma]. \end{align}
   We note that these formulas are consistent with $\delta$ having $\delta$-degree 1, and with $\delta^2$ being a gauge transformation generated by $\sigma$ (thus, $\delta^2\bar\sigma
   =[\sigma,\bar\sigma]$, and similarly for $\eta$).
   
   The action of the theory is $\delta$-invariant,  and the part of the action that is relevant for our purposes is $\delta$-exact.   There is much arbitrariness in the $\delta$-exact part of
   the action.   The minimal choice made in \cite{EW2} amounted to
   \be\label{chock}-\delta\int_M\d^4x\sqrt g \,\tr\, \bar\sigma D_i\psi^i=\int_M \d^4x \sqrt g\, \tr\left( \bar\sigma D_i D^i\sigma  -\bar\sigma [\psi_i,\psi^i] -\eta D_i\psi^i \right). \ee
   Assuming that these are the only terms by which $\bar\sigma$ and $\eta$ appear in the action (or the only terms that are relevant in lowest order, which is good enough
   in the localization argument), we can read off the equations of motion:
   \begin{align}  \label{domo} D_i\psi^i & = 0 \cr
                             D_i D^i \sigma      & = [\psi_i,\psi^i].\end{align}
    We have simultaneously accomplished two things.   First, we have found the desired extra equation of motion for $\psi$, namely $D_i\psi^i=0$.   And second,
    we have learned how to express $\sigma$ in terms of $\psi$, namely by solving the equation
    $D_i D^i\sigma =[\psi_i,\psi^i]$, implying that
    \be\label{delfox} \sigma(x) =\int_{\S^3\times \S^1}\,B(x,y)[\psi_i,\psi^i](y)\d^4y \sqrt{g_y} , \ee
    where $B(x,y) $ is the Green's function of the Laplace operator:
    \be\label{nelfox}\frac{D}{D x^i}\frac{D}{D x_i} B(x,y)=\delta^4(x,y) \mathrm{Id}. \ee
    Here $\sqrt{g_y}$ is just $\sqrt g$ regarded as a function of $y$, and ${\mathrm{Id}}$ is the identity operator acting on the fiber  of the adjoint bundle $\ad(E)$.   
     
    Using in eqn  (\ref{iddo}) the explicit formula for $P^{(1)}$ from eqn. (\ref{descent}) and using eqn. (\ref{delfox}) to eliminate $\sigma$, 
    we get an explicit formula for $H_0$ as a three-form on $\M$:
    \be\label{iminy}H_0=\frac{1}{2\pi}\int_{\S^3\times \S^1\times \S^3\times \S^1}\d^3\Omega_x \d\tau_x \d^4 y\sqrt{g_y} \tr\otimes \tr \, \psi_\tau(x) B(x,y) [\psi_\mu,\psi^\mu](y). \ee
    Here the two copies of  $\S^3\times \S^1$ are  parametrized respectively  by $x$ and $y$.
   We have used the fact that the only part of $\psi(x)$ that contributes in the evaluation of eqn. (\ref{iddo}) is $\psi_\tau(x)\d\tau$.

The preceding derivation can be generalized by replacing  $D_i\psi^i$ on the left hand side of 
 eqn. (\ref{chock}) with another  expression.  Then the auxiliary condition $D_i\psi^i=0$ in  eqn. (\ref{domo})
will be replaced by another condition.  As we will learn in section \ref{hkt},
to understand the moduli space $\M$ of instantons on $\S^3\times \S^1$, a different auxiliary condition is more useful, namely \be\label{namely}D_i\psi^i+\star(H\wedge \psi)=0\ee
 (where
$\star$ is the Hodge star).
To modify the preceding derivation to incorporate this condition, we merely add another $\delta$-trivial term to the action, taking it to be
   \begin{align}\label{nock} -\delta\int_M\d^4x\sqrt g\, \tr\, \bar\sigma\bigl( D_i\psi^i+\star(H\wedge\psi)\bigr)  =\int_M&\d^4x\sqrt g\, \tr\bigl(\bar\sigma (D_i D^i \sigma
   +\star(H\wedge\d_A\sigma) ) \bigr.\cr &\bigl.-\eta( D_i\psi^i+\star(H\wedge \psi))-\bar\sigma[\psi_i,\psi^i]\bigr)\end{align}    
 The equation of motion for $\eta$ gives the desired condition (\ref{namely}).   But the equation of motion for $\bar\sigma$  also changes and is
 now
 \be\label{defo} W\sigma=[\psi_i,\psi^i], \ee
 where the operator $W$ is defined by
 \be\label{efo} W\sigma =D_i D^i \sigma +\star(H\wedge \d_A\sigma). \ee
 The formula (\ref{iminy}) is still valid, except that $B(x,y)$ is now the Green's function of\footnote{We show in section \ref{hypercomplex} that $W$ has no kernel or cokernel, so
 this Green's function exists.}   $W$, obeying
 \be\label{mimo} W_x B(x,y) =\delta^4(x,y){\mathrm{Id}},\ee
 where $W_x$ is the operator $W$ acting on the variable $x$.  An important detail, though, is that as $W$ is not self-adjoint, its Green's function $B(x,y)$ is not symmetric
 in $x$ and $y$, and the equation that it obeys in the $y$ variable is actually
 \be\label{zimox} W_y^\dagger B(x,y)=\delta^4(x,y){\mathrm{Id}},\ee
 where $W^\dagger$ is the adjoint operator
 \be\label{pimox}  W^\dagger\sigma =D_\mu D^\mu \sigma -\star(H\wedge \d_A\sigma). \ee
 One way to prove eqn. (\ref{zimox})  is to observe that eqn. (\ref{mimo}) implies that
 \be\label{helpme}\int_{M}\d^4 y \sqrt {g_y}\, B(x,y) W_y B(y,z) = B(x,z). \ee
 On the other hand, we can write the left hand side of eqn. (\ref{helpme}) as 
 \be\label{zelpme}\int_M\d^4 y \sqrt{ g_y} \, W_y^\dagger B(x,y) \cdot B(y,z),\ee 
 and eqn. (\ref{zimox}) follows by comparing this with eqn. ({\ref{helpme}).
 
 Regardless of whether we take $B(x,y)$ to be the Green's function of the Laplace operator, the operator $W$, or some other operator related to another valid condition for $\psi$,
 eqn. (\ref{iminy}) is an explicit formula for a three-form on $\M$ in the appropriate cohomology class.   But as will eventually become clear, the torsion of the generalized hyper-Kahler metric of $\M$ is the representative of this cohomology class that  we get if $B(x,y)$ is the Green's function of the operator $W$.

  \section{Some Background}\label{review}
 
 In this section, we will review relevant background material on three topics:   supersymmetric sigma-models with a $B$-field; the relevant geometry of $\S^3\times \S^1$;
 and the large $\N=4$ superconformal algebra in two dimensions.   The first two topics were reviewed recently in \cite{PW};  more information can be found there on some aspects,
 along with further references.
 
 \subsection{Supersymmetric Sigma-Models With a $B$-Field}\label{sigmab}
 
 The most basic type of  two-dimensional supersymmetric sigma-model with a $B$-field is a model with $(0,1)$ supersymmetry.\footnote{\label{noworry}We will treat these models
 classically, not worrying about sigma-model 
 anomalies \cite{MooreNelson}, as they will be absent in the $\N=4$ models that we consider eventually.   Likewise we will not worry about a possible 
 anomaly in the $\U(1)$ chiral symmetry in the $(0,2)$ models that we discuss presently; this is again absent when the model is extended to $\N=4$ and the global
 symmetry is extended from $\U(1)$ to $\SU(2)$.}
 We  will briefly review such models following
 \cite{HullWitten}.   We consider a model with a general target space $\eX$, which we describe by local coordinates $X^P$, $P=1,\cdots,\dim\,\eX$, and endowed with a metric
 $G_{PQ}$ and a $B$-field $B_{PQ}$, with field strength $H_{PQR}=\partial_P B_{QR}+\partial_Q B_{RP}+\partial_R B_{PQ}$.   On a superspace $\Sigma$ of dimension $2|1$ with
 even coordinates $x^-, x^+$ and one odd coordinate $\theta$, we introduce a  supersymmetry generator $Q_+=\i\left(\frac{\partial}{\partial  \theta}-\i \theta\frac{\partial}{\partial x^+}\right)$, satisfying $Q_+^2=\i\partial_+$ and commuting with $D=\frac{\partial}{\partial\theta}+\i\theta\frac{\partial}{\partial x^+}$, which is used in writing Lagrangians.
 The sigma-model map $\Phi:\Sigma\to \eX$ can be described concretely in terms of superfields $\X^P=X^P+\i\theta\psi^P$.  The supersymmetry variation of these fields is
 is $\delta \X^P=Q_+\X^P$ or 
 \be\label{supervar} \delta X^P=\i\psi^P, ~\delta \psi^P=\partial_+ X^P.\ee
 A natural supersymmetric action for these fields  is
 \be\label{zelfog}S=-\frac{\i  }{2} \int\d^2 x\,\d\theta \,\left(G_{PQ}+B_{PQ}\right) D \X^P \frac{\partial}{\partial x^-}\X^Q.    \ee
 After integrating over $\theta$, the bosonic part of the action is the standard 
 \be\label{hopeful} S_b=\frac{1}{2}\int\d^2x (G_{PQ}+B_{PQ})   \partial_+X^P \partial_- X^Q.\ee  The fermion action turns out to be
  \be\label{fermac}S_f =\frac{\i}{2}\int \d^2x\, G_{PQ}\psi^P\left(\delta^Q_S\frac{\partial}{\partial x^-} + \frac{\partial X^R}{\partial x^-} \t \Gamma^Q_{RS}\right)\psi^S,\ee
 where  $\t\Gamma^Q_{RS}$ is not the usual Riemannian affine connection $\Gamma^Q_{RS}$ but has also a torsion term proportional to $H$:
 \be\label{newconn} \t\Gamma^Q_{RS} =\Gamma^Q_{RS} +\frac{1}{2}G^{QT}H_{TRS}. \ee
 Using this new affine connection, we
  define a connection $\nabla $ with torsion that differs from the Riemannian connection $D$. The covariant derivative of a vector field $V$ with respect to
 $\nabla$ is 
 \be\label{ewcon}\nabla_P V^R=\partial_P V^R+\t\Gamma^R_{PQ}V^Q=D_P V^R+\frac{1}{2} G^{RS}H_{SPQ}V^Q. \ee
 This connection has completely antisymmetric torsion $H$.  Accordingly it is metric compatible, meaning that the Riemannian metric $G_{PQ}$ is covariantly constant.
 
 As an example of the usefulness of this connection, let us ask a question whose importance in relation to the large $\N=4$ algebra was explained in \cite{BPS}.
 What condition should a vector field $V$ on $\eX$ satisfy so that the sigma-model field $\Lambda=\sum_K V_K\psi^K$ obeys $\partial_-\Lambda=0$ and thus is a chiral free fermion?  
 Since the equation of motion for $\psi$ is $\partial_- \psi^P=-\partial_- X^R \tilde \Gamma^P_{RS}\psi^S$, we have
 $\partial_-\Lambda=\partial_-  X^R\partial_R V_S\, \psi^S  - V_P\partial_- X^R \tilde \Gamma^P_{RS}\psi^S$.
 The condition for this to vanish is $\partial _R V_S-\tilde\Gamma^P_{RS} V_P=0$, or in other words \be\label{constkilling}\nabla_R V_S=0,\ee that is, $V$ is covariantly constant for the connection $\nabla$.  
 If this condition is satisfied, then in addition to the chiral free fermion $\Lambda$ of spin 1/2, we get a free chiral current of spin 1:
 \be\label{freej} J=\{Q,\Lambda\}= V_K\partial_+ X^K+\i \partial_L V_K \psi^L\psi^K. \ee

 Up to this point, $G_{PQ}$ and $B_{PQ}$ were not subject to any particular constraint.   That changes if we ask for the sigma-model to have $(0,2)$ supersymmetry \cite{AF,GHR}.
 The $\N=2$ superconformal algebra has a $\U(1)$ $R$-symmetry under which the two supercharges have charge $\pm 1$.  We do not want to restrict the target
 space  $\eX$ to have a  $\U(1)$ symmetry,
 so we assume the bosons to have $R$-charge 0 and the fermions to have $R$-charges $\pm 1$.
Thus we  assume a $\U(1)$ symmetry that acts on fermions only, generated by 
 \be\label{delmo}\delta \psi^P=\I^P{}_Q \psi^Q ,\ee
 with some linear operator $\I$.   In a generic local coordinate system on $\eX$, the matrix elements of $\I$ are position-dependent, but as the components of $\psi$  are supposed
 to have $R$-charges $\pm 1$, we can normalize $\I$ so that $\I^2=-1$, making  $\I$ an almost complex structure on $\eX$. This implies in particular that $\eX$ has
 even dimension $2p$.  We denote the $+\i$ and $-\i$ eigenmodes
 of $\I$ as $\psi^\alpha$ and $\psi^{\bar \beta}$, $\alpha,\beta=1,\cdots ,p$.   The condition that the fermion action $S_f$ of eqn. (\ref{fermac})
 actually does have the symmetry of eqn. (\ref{delmo}) gives two conditions.   First, the metric tensor $G$ must be of type $(1,1)$ with respect to $\I$, meaning that
 its nonzero components are $G_{\alpha\bar\beta}=G_{\bar\beta\alpha}$.  (The metric $G$ is therefore said to be hermitian with respect to $\I$.) 
  Second, since $S_f$ is constructed in terms of the covariant derivative of $\psi$ with
 respect to the connection $\nabla,$ invariance of $S_f$ requires that the tensor $\I^P{}_Q$ should be covariantly constant with respect to this connection:
 \be\label{covcon}\nabla \I = 0. \ee
 Thus both $G$ and $\I$ are covariantly constant with respect to $\nabla$, implying that ithe structure group of $\nabla$ reduces to $\U(p)$.  
The $R$-symmetry conjugates the original supersymmetry (\ref{supervar}) to a second one that acts by 
\be\label{secondvar}\delta X^P=\I^P{}_Q \psi^Q,~~~\delta \psi^P= -\I^P{}_Q\partial_+ X^Q.
\ee
 Requiring that this squares to a translation generator $\i\partial_+$ gives a condition that is quadratic in $\I$ and its first derivative.   This condition is precisely the
 vanishing of the Nijenhuis tensor.  Thus $\I$ is actually an integrable complex structure on $\eX$, and one can introduce on $\eX$ local holomorphic coordinates $x^\alpha,\, \bar x^{\bar \beta}$,
 $\alpha,\beta =1,\cdots, p$.   In such a coordinate system, the nonzero components of $\I$ are simply $\I^\alpha{}_\beta=\i \delta^\alpha{}_\beta$, 
 $\I^{\bar\alpha}{}_{\bar\beta}=-\i\delta^{\bar\alpha}{}_{\bar\beta}$, and the  condition (\ref{covcon}) simplifies to $\tilde\Gamma^\alpha_{\beta\bar\gamma}=\tilde\Gamma^\alpha_{\bar\beta\bar\gamma}=0$ (and the complex conjugate of this).   From this we learn that $H_{\alpha\beta\gamma}=H_{\bar\alpha\bar\beta\bar\gamma}=0$, saying that  $H$ is of type $(2,1)\oplus (1,2)$ with respect to $\I$, and that 
 \be\label{hform} H_{\alpha\beta\bar\gamma}=-\partial_\alpha G_{\beta\bar\gamma}+\partial_\beta G_{\alpha\bar\gamma}. \ee
 For $H$ to be the field strength of some two-form field  $B$,
 one requires\footnote{Here the remark in footnote  \ref{noworry} is relevant. As sigma-model anomalies will eventually be canceled
 by adding left-moving fermions, we are not interested in canceling them by a Green-Schwarz mechanism with $\d H\not=0$, as is appropriate in the context of the heterotic string.} $\d H=0$, which here tells us that
 \be\label{murky} \frac{\partial^2}{\partial x^\alpha \partial \bar x^{\bar \beta}}G_{\gamma\bar\delta}\d x^\alpha\d x^\gamma \d {\bar x}^{\bar\beta}\d {\bar x}^{\bar\delta}=0.\ee
 So locally 
 \be\label{knod} G_{\alpha\bar\beta}=\partial_\alpha K_{\bar \beta}+\partial_{\bar\beta} K_\alpha,\ee
  where the one-form $K$ is the closest analog in this context of a Kahler potential.
 This observation is natural in a certain superspace construction of this class of models \cite{Hull}.   
 
 Associated to this data we can define a hermitian form $\omega$ by $\omega_{PR}=G_{PQ}\I^Q{}_R$, or in local holomorphic coordinates $\omega_{\alpha\bar\beta}=-\i G_{\alpha\bar\beta}
 =-\omega_{\bar\beta\alpha}$, with other components of $\omega$ vanishing. In particular $\omega$ is of type $(1,1)$; moreover, $\omega$  is covariantly constant with respect to $\nabla$,
 since $G$ and $\I$ are.     Note that the two-form $\omega$ associated to an antisymmetric tensor $\omega_{PR}$ is defined as\footnote{The factor of $\frac{1}{2}$ ensures that if $A$ is a one-form and $F=\d A$, then $F=\frac{1}{2}\sum_{P,R}\d X^P \d X^R F_{PR}$ where the definition of $F_{PR}$ is the
 standard $F_{PR}=\partial_P A_R-\partial_R A_P$. Note that as $\d=\sum_P \d X^P\partial_P$ and $A=\sum_R \d X^R A_R$, we have $\d A=\sum_{P,R}\d X^P \d X^R\partial_P A_R
 =\frac{1}{2}\sum_{P,R}\d X^P\d X^R F_{PR}$.} 
 \be\label{noteme}\omega=\frac{1}{2} \sum_{P,R}\d X^P \d X^R \omega_{PR}.\ee
   If $\d\omega=0$, then $\eX$ is called a Kahler manifold and eqn. (\ref{hform}) implies that $H=0$.
 Indeed, if as usual we expand the exterior derivative in parts of type $(1,0)$ and $(0,1)$ by writing $\d=\partial+\bar\partial$, then eqn. (\ref{hform})   tells us that
 \be\label{urky}H=-\i(\partial-\bar\partial)\omega. \ee
 An alternative way to write this formula is the following.   One defines an action of $\I$ on differential forms by $\I(f(X) \d X^P)=f(X) \I^P{}_Q \d X^Q$ and more generally
 \be\label{bozo}\I(f(X) \d X^{P_1}\cdots \d X^{P_s})=f(X) \I^{P_1}{}_{Q_1}\d X^{Q_1}\cdots \I^{P_s}{}_{Q_s}\d X^{Q_s}. \ee
 Then, since $\omega$ is of type $(1,1)$ and $H$ is of type $(2,1)\oplus (1,2)$,   the formula (\ref{urky}) for $H$ is equivalent to
 \be\label{rky} H=-\I \d\omega =-\I \d\I\omega. \ee
 The formulas (\ref{urky}) and (\ref{rky}) are useful because they express $H$ just in terms of the gauge-invariant data $\omega$, $\I$, as opposed to the usual formula
 $H=\d B$, where $B$ is gauge-dependent.     For completeness, we also include a formula for the Lee form (which appears in the one-loop beta function of a model
 of this type \cite{Buscher,Hull2})
 \be\label{leeform}   \theta_K=-\frac{1}{2}\I^L{}_K H_{LPQ} \I^{PQ}.\ee
The Lee form will be relevant in section \ref{conformal}, though we will not make any use of the formula (\ref{leeform}).
 
 Having gotten this far, the further extension to $(0,4)$ supersymmetry is straightforward.   In this case, one wants an $\SU(2)$ $R$-symmetry that acts only on the fermions.
 This group has three symmetry generators $\I,\J,{\mathcal K}$, and as the fermions should all transform as spin $1/2$ under $\SU(2)$, the generators can be normalized so that
 they obey the quaternion relations $\I^2=\J^2={\mathcal K}^2=-1$, $\I\J=-\J\I={\mathcal K}$.  This implies in particular that $\eX$ must have dimension $D=4q$ for some integer $q$.
 Using $\J$ or $\K$ instead of $\I$ in eqn. (\ref{secondvar}) will give two more supersymmetries, making four supersymmetries in all.
   The preceding considerations apply to each of $\I,\J,{\mathcal K}$ separately, so in particular each of $\I,\J,{\mathcal K}$
 is an integrable complex structure.   Being endowed with three complex structures that satisfy the quaternion relations, $\eX$ is called a hypercomplex manifold.\footnote{A hypercomplex manifold actually has a family of complex structures parametrized by $\S^2$, since if $a,b,c$ are real numbers
satisfying $a^2+b^2+c^2=1$, then the quaternion relations imply that $a\I+b\J+c\K$ is an almost complex structure, and it is integrable if $\I$, $\J$, and $\K$ are integrable.} Moreover, the metric $G$ is of type $(1,1)$ for each of $\I,\J,{\mathcal K}$, and is therefore said to be hyper-hermitian.    Each of $\I,\J,$ and ${\mathcal K}$
must be covariantly constant for the connection $\nabla$ constructed as in eqn. (\ref{ewcon}) in terms of $G$ and $H$.  This implies that the structure group of $\nabla$
reduces to $\Sp(q)$. For each complex structure, we define a corresponding hermitian form $\omega^{(\I)}_{ PR}=G_{PQ}\I^Q{}_R$, $\omega^{(\J)}_{\,PR}=G_{PQ}\J^Q{}_R$,
$\omega^{(\K)}_{PR}=G_{PQ}\K^Q{}_R$.  The derivation of eqn. (\ref{rky}) applies equally to any of $\I,\J,\K$,  so we must have
\be\label{forgo}H=-\I\d\omega^{(\I)}=-\J\d\omega^{(\J)}=-\K\d\omega^{(\K)}. \ee  
For the same three-form $H$ to satisfy these three different formulas is a very strong constraint. 

So far we have discussed $(0,n)$ worldsheet supersymmetry with $n=1,2,4$.   Now we discuss the extension to $(n,n)$ supersymmetry.   Roughly speaking, one just
adds additional fermions of opposite chirality and one requires two copies of the structure that has just been described, one copy for positive chirality and one for negative
chirality.   Since both chiralities of fermions will be present, we write henceforth $\psi^P_+$ for the fermions that we have been considering so far and $\psi^P_-$ for
the fermions of opposite chirality.  For $n=1$, the main point to consider is that exchanging the two chiralities involves reversing the worldsheet orientation, and under
this operation $H$ is odd.   Therefore, fermions of the two chiralities  will see  connections that differ in  the sign
of the torsion, which is proportional to $H$ and therefore is odd under exchanging the two chiralities.  Since two different connections with torsion will play a role,
we will be more precise in the notation.  
Previously we defined a connection $\nabla$ 
that appears in the kinetic energy of the positive chirality fermions and is described by the affine connection (\ref{newconn}).   Reversing the sign of $H$, we get the affine connection 
\be\label{toldox} \t\Gamma'{}^Q_{RS} =\Gamma^Q_{RS} -\frac{1}{2}G^{QT}H_{TRS} \ee
associated to the connection $\nabla'$ that appears in the kinetic energy of negative chirality fermions.
 The respective definitions of covariant derivatives are
  \be\label{zewcon}\nabla_P V^R=D_P V^R+\frac{1}{2} G^{RS}H_{SPQ}V^Q, ~~\nabla'_P V^R=D_P V^R-\frac{1}{2} G^{RS}H_{SPQ}V^Q. \ee
  It will be important that these definitions depend only on the metric $G$ and the torsion $H$, not on the complex structure or hermitian form.
 
 To get $(2,2)$ supersymmetry, we want separate $\U(1)$ $R$-symmetries for both positive chirality and negative chirality fermions.   We denote the
 corresponding symmetry generators as $\I$ and $\I'$.   An important insight of \cite{GHR} is that in general no relation between $\I$ and $\I'$
 must be assumed.   In particular, in general they need not commute, though models in which they do commute are simpler in some respects
 (the models considered in the present article do not have that property).  The derivations that we have explained up to this point apply separately
 for positive chirality fermions and $\I$ and for negative chirality fermions and $\I'$.  In particular, both $\I$ and $\I'$ are integrable complex
 structures; $\I$ is covariantly constant for the connection $\nabla$ and $\I'$ is covariantly constant for the connection $ \nabla'$. So both
 connections have holonomy in $\U(p)$.   Defining the
 two hermitian forms $\omega_{PR}=G_{PQ}\I^{Q}{}_R, ~~\omega'_{PR}=G_{PQ}\I'^Q{}_R$, the derivation of eqn. (\ref{forgo}) applies for
 each, with a sign change:
 \be\label{norgo}H=-\I \d\omega=+\I' \d\omega'. \ee
 The structure just described has been interpreted as generalized Kahler geometry \cite{Gualtieri}.   It has also been called strong bi-KT geometry (KT refers to
 Kahler geometry generalized to allow torsion; the prefix ``bi'' refers to the presence of separate KT structures for the two fermion chiralities; and ``strong'' means that
 $\d H=0$).  
 
 To get $(4,4)$ supersymmetry, we need separate hypercomplex structures $\I,\J, \K$ for positive chirality fermions, with closed torsion $H$,  and $\I',\J',\K'$
 for negative chirality fermions, with torsion $-H$.   Here $\I,\J,\K$ must be covariantly constant for $\nabla$, and $\I',\J',\K'$ must be covariantly
 constant for $ \nabla'$.   So both connections have holonomy in $\Sp(q)$.   Defining the various hermitian forms by $\omega^{(\I)}_{PR}=G_{PQ}\I^Q{}_R$,
 etc., the derivation of the relation between $H$ and the hermitian forms goes through as before to give formulas for $H$ in terms of any one
 of $\I,\,\J,\,\K$ or $\I',\,\J',\K'$:
  \be\label{norgox}H=-\I \d\omega^{(\I)}=-\J\d\omega^{(\\J)}=-\K\d\omega^{( \K)} =+\I' \d\omega^{( \I')}=+\J'\d\omega^{(\J')}
  =+\K'\d\omega^{(\K')}. \ee
 Such geometry has been called generalized hyper-Kahler geometry, or strong bi-HKT geometry (where HKT refers to hyper-Kahler geometry generalized to allow torsion).

\subsection{Large And Small Algebras}\label{largesmall}

A sigma-model with any (complete) generalized hyper-Kahler target $M$ of dimension $D=4q$  has superconformal symmetry with what is known as the small $\N=4$ algebra.   Obvious
generators are the stress tensor $T$ and the four supercurrents $G_a$ associated to the four supersymmetries that were constructed in section \ref{sigmab}.  The other generators of the chiral algebra, for generic $M$, are the currents
of an $\SU(2)$ current algebra at level $q$ that acts only on the fermions.
These $\SU(2)$ current algebras exist, for both left- and right-moving modes, because the fermion kinetic energy is constructed using connections $\nabla$ or $\nabla'$
whose $\Sp(q)$ holonomy groups each commute with a corresponding $\SU(2)$ action on the tangent bundle of $M$.   For both left-movers and 
right-movers of the sigma-model, it is possible to define
fermion bilinears that generate the corresponding $\SU(2)$.  Along with $T$ and the $G_a$, these bilinears generate the small $\N=4$ superconformal algebra.  

In the small $\N=4 $ algebra,  the $G_a$ are primaries of the $\SU(2)$ current algebra; they transform in the spin $1/2$ representation of $\SU(2)$, viewed as a
four-dimensional real representation.  There is a natural $\SO(4)$ action on the four real fields $G_a$.   Here $\SO(4)=(\SU(2)\times SU(2))/\Z_2$, where the $\SU(2)$ that
is part of the small $\N=4$ algebra is one of the two factors.    In particular, this  $\SU(2)$ action on the $G_a$ commutes with a second $\SU(2)$.  
This second $\SU(2)$ is a group of outer automorphisms of the
small $\N=4$ algebra; it commutes with all the algebra generators except the $G_a$.     In many familiar models with small $\N=4$ superconformal symmetry, the outer automorphism group is not realized as a symmetry.  
For example, a sigma-model with target $\T^4$ or $\mathrm{K3}$ has small $\N=4$ superconformal symmetry, and the outer automorphism group is not realized as a symmetry.

A simple example of a theory with small $\N=4$ superconformal symmetry in which the outer automorphism group is realized is the theory of a single free hypermultiplet,
or in other words a sigma-model with target the flat hyper-Kahler manifold $M=\R^4$.   There are four massless free bosons $\phi_{AX}$, $A,X=1,2$ with a reality condition
$\bar\phi^{AX}=\epsilon^{AB}\epsilon^{XY}\phi_{BY}$ (here $\epsilon$ is the antisymmetric tensor with $\epsilon^{12}=1$); similarly there are four massless free chiral fermions $\psi_{RX}$, $R, X=1,2$, with -- in Lorentz signature -- a similar reality condition.
(For simplicity we consider fermions of one chirality only.)   The four supercurrents are $G_{AR}=\partial\phi_{AX} \psi_{RY}\epsilon^{XY}$ and the generators of the
$\SU(2)$ current algebra are $J_{RS}=\epsilon^{XY}\psi_{RX}\psi_{SY}$.   Clearly, those currents generate a current algebra that acts on the fermions only, by the natural $\SU(2)$
action on the first index of $\psi_{RX}$,  and just as clearly, the $G_{AR}$
transform as spin $1/2$ under this $\SU(2)$.  This current algebra is at level 1; the analogous current algebra in a model with $q$ hypermultiplets 
 is at
level $q$.      The outer automorphism group of the small $\N=4$ algebra, in this presentation, is an $\SU(2)$ symmetry that acts on the first
index of $G_{AR}$.   It is realized in this model by a global $\SU(2)$ symmetry that acts on the first index of the scalar field $\phi_{AX}$, with trivial action on $\psi_{RX}$.   
However, it is not possible to define a holomorphic current that generates this symmetry, and therefore the currents that generate it are not part of the superconformal algebra
of the theory.   

To extend the small $\N=4$ algebra associated to a hypercomplex structure $\I,\J,\K$ on some generalized hyper-Kahler manifold to
a large $\N=4$ algebra, this manifold   should have an $\SU(2)$ symmetry that rotates $\I$, $\J$, and $\K$ and that is generated by holomorphic currents.  The simplest example
is $M=\S^3\times \S^1$, to which we turn next.    One of the main goals of the present article is to show that the moduli space $\M$ of instantons on $M$ is another example.
    
 \subsection{Geometry of $\S^3\times \S^1$}\label{spherex}
 
 To understand the unusual geometry of $\S^3\times \S^1$, it is convenient to start with $\R^4$ minus the origin with the scale invariant metric
 \be\label{undef}\d s^2=\frac{\d \vec Y^2}{\vec Y^2}, \ee
 where $\vec Y=(Y_0,Y_1,Y_2,Y_3),$ $\vec Y^2=\sum_{\lambda=0}^3  Y_\lambda^2$.   Because the metric is scale-invariant, we can divide by the equivalence relation $\vec Y\cong e^T\vec Y$ for any
 fixed $T>0$.   Setting $\tau=\log |\vec Y|$, the  quotient is $\S^3\times \S^1$ with metric
 \be\label{tuggo}\d s^2=\d\Omega^2+\d\tau^2, ~~\tau\cong \tau+T,\ee
 where $\d\Omega^2$ is the metric of a round three-sphere of radius 1.  If $Y_\lambda= y_\lambda  e^\tau$, $\sum_\lambda y_\lambda^2=1$, then $\d\Omega^2=\sum_\lambda \d y_\lambda^2$.
 
 To see a complex structure on $\S^3\times \S^1$, let $Z_1=Y_0+\i Y_1$, $Z_2=Y_2+\i Y_3$.   Then we can parametrize $\S^3\times \S^1$ by complex variables $(Z_1,Z_2)$
 with $(Z_1,Z_2)\cong e^T(Z_1,Z_2)$, exhibiting a complex structure that we will call\footnote{In discussing $\S^3\times \S^1$, to
 take advantage of the symmetries and write some formulas more economically, we denote a triple of complex structures that obey the quaternion relations
 as $\I_1,$ $\I_2$, $\I_3$ rather than $\I,\J,\K$.}  $\I_1$.   In this complex structure, the metric of $\S^3\times \S^1$ is hermitian (that is, of type $(1,1)$)
 but not Kahler.   Indeed $\S^3\times \S^1$ cannot be Kahler as its second Betti number vanishes.   $\S^3\times \S^1$ with this complex structure is known as a Hopf surface, a prototype
 of a compact manifold that is complex but not Kahler.   
 
 Complex structure $\I_1$ is not invariant under all of the symmetries of $\S^3\times \S^1$, so those symmetries can be used to generate many more complex structures.
 Parametrize $\S^3=\SU(2)$ as
 \be\label{dolme}g=\begin{pmatrix} z_1 & -\bar z_2 \cr z_2 & \bar z_1\end{pmatrix},~~|z_1|^2+|z_2|^2=1,~~Z_i = z_i e^\tau. \ee
We denote  $\SU(2)$ acting on itself on the left or right 
as $\SU(2)_\ell$ or $\SU(2)_r$.   Clearly, $\I$ is invariant under $\SU(2)_\ell$, which maps $z_1$ and $z_2$
to complex linear combinations of themselves, but not under $\SU(2)_r$.   Rather, $\SU(2)_r$ maps $\I_1$ to a family of complex structures on $\S^3\times \S^1$
that make up a hypercomplex
structure (explicit formulas are given presently).  Moreover the discrete symmetry $\rho$ that acts by a joint reflection of the two factors of $\S^3\times \S^1$ exchanges $\SU(2)_\ell$ and $\SU(2)_r$,
and so maps  this hypercomplex structure to a second one that is invariant under $\SU(2)_r$ and rotated by $\SU(2)_\ell$. 

To write explicit formulas,   view $\S^3\times \S^1$ as the group manifold $K=\SU(2)\times \U(1)$.  The metric  (\ref{tuggo})
  is invariant under the left and right action of $K$ on itself, so it is possible to choose orthonormal bases consisting of left-invariant or right-invariant one-forms.
  As $\U(1)$ is abelian, $L^0=R^0=\d\tau$ is both left-invariant and right-invariant.   To find left-invariant forms on $\SU(2)$, we simply note that $g^{-1}\d g$ is
  left-invariant.  
Expanding this in components, a basis of left-invariant one-forms is given by
  \be\label{defone}L^1=y_0\d y_1-y_1\d y_0+y_2 \d y_3-y_3\d y_2,\ee
  and two more forms  $L^2$ and $L^3$ that differ from this by cyclic permutation of indices $1,2,3$.   The $L^a$ are an orthonormal basis of the cotangent bundle of $\S^3$,
  since  $\sum_{a=1}^3 L^a\otimes L^a=\d\Omega^2$.
  Because of the $\SU(2)\times \SU(2)$ symmetry of $\S^3$, to verify this and similar statements later, it suffices to verify that the statement is true at the
  point $p$ with $(y_0,y_1,y_2,y_3)=(1,0,0,0)$; this is immediate.  The $L^a$ obey
  \be\label{leftob}\d L^1=2 L^2\wedge L^3\ee
  and cyclic permutations of this statement.  Of course, we also have 
  \be\label{oneb}\d L^0=0. \ee Similarly, $ \d g g^{-1}$ is right-invariant and can be expanded in terms of the right-invariant one-form
   \be\label{defonez}R^1=y_0\d y_1-y_1\d y_0-y_2 \d y_3+y_3\d y_2,\ee
   and two more right-invariant forms  $R^2$ and $R^3$ differing from this by cyclic permutation of indices $1,2,3$.  
   Another way to see that the  $L^a$ and $R^b$ are respectively left-invariant and right-invariant is to observe that
   they are constructed from $4\times 4$ matrices that are respectively selfdual or anti-selfdual.  The $R^b$ satisfy
   \be\label{rightob}\d R^1=-2R^2\wedge R^3,\ee
   and cyclic permutations, again with
   \be\label{zeldob}\d R^0=0.\ee   As the $L^a$ and $R^b$ for $a,b=1,2,3$ are  orthonormal bases of the cotangent bundle of $\S^3$, $L^1\wedge L^2\wedge L^3$
   and $R^1\wedge R^2\wedge R^3$ are equal up to sign.   A short check at the point $p$ shows that they are equal.
   
   Including $L^0$ and $R^0$, we get orthonormal bases $L^i$ and $R^j$, $i,j=0,\cdots, 3$, of the cotangent bundle of $M=\S^3\times \S^1$:
   \be\label{ingot}\d s^2=\sum_{i=0}^3 L^i\otimes L^i =\sum_{j=0}^3 R^j\otimes R^j. \ee
   We can define a connection
   $\nabla$ on the tangent bundle of $M$ by saying that the $L^i$ are covariantly constant, and another connection $\nabla'$
   by saying that the $R^j$ are covariantly constant.\footnote{In \cite{PW}, these connections are called $\hat\nabla$ and $\breve\nabla$.  A similar remark applies for the complex structures and
   hermitian forms introduced below. In the present article, the notation is different, as we reserve the ``hat'' (as in $\hat\nabla$) for a structure on the moduli space $\M$ as opposed
   to the four-manifold $M$.}     Both of these connections are metric compatible, since the $L^i$ and $R^j$ are orthonormal bases.  
   
     To place $\nabla$  in the sigma-model context  of
 section \ref{sigmab}, we need to find a three-form $H$ such that for any one-form $V$,    $\nabla_P V_Q = D_P V_Q-\frac{1}{2}H_{PQR}V^R$, where $D$ is the
 Riemannian connection.   It suffices to verify this condition if $V$ is one of the $L$'s.
 In this case $\nabla V=0$, so we need $D_P V_Q=\frac{1}{2}H_{PQR}V^R$.  If this equation is symmetrized in $P$ and $Q$, the right hand side  vanishes
 because $H$ is totally antisymmetric, 
 and the left hand side also vanishes since the vector fields dual to the $L^i$ are Killing vector fields, implying that  $D_P V_Q+D_Q V_P=0$.    The antisymmetric
 part of the equation is $\partial_PV_Q-\partial_Q V_P=H_{PQR}V^R$, and this is satisfied for $H=2 L^1\wedge L^2\wedge L^3$ by virtue of eqns. (\ref{leftob}) and (\ref{oneb}).
 
 To place $\nabla'$ in the sigma-model context, we need the same equation to hold with $H$ replaced by $-H$ and $\nabla $ replaced by $\nabla'$.
 In other words, for every one-form $V$, we need $\nabla' _P V_Q =D_P V_Q +\frac{1}{2} H_{PQR} V^R$.  
 It suffices to verify this if $V$ is one of the $R$'s; this verification, again for $H=2 L^1\wedge L^2\wedge L^3=2 R^1\wedge R^2\wedge R^3$,  can be carried out 
using eqns. (\ref{rightob}) and
 (\ref{zeldob}).  
 
 Since a three-sphere of radius 1 has volume $2\pi^2$, the fact that $H=2L_1L_2L_3$ is twice the Riemannian volume form of the sphere implies
 that $\int_{\S^3}H=4\pi^2$.   This means that to construct a model in which  $\int_{\S^3}H=2\pi k$, we  have to multiply the metric of $\S^3$ by\footnote{The formula
 (\ref{hform}) shows that rescaling $G$ will rescale $H$ by the same factor. Since the radius of $\S^1$ does not affect $\int_{\S^3}H$, it does not matter whether
 we rescale it or not. Only the rescaling of $\S^3$ is relevant.}  $k/2\pi$.   The line
 element of $\S^3$ will then be
 \be\label{inoc} \d s^2=\frac{k}{2\pi} \d\Omega^2.\ee 
 This normalization will be important in section \ref{large}, but  for now we continue with the case of unit radius.
 
The complex structure $\I_1$ that we defined at the outset acts on the $L_i$ by
 \be\label{infox} \I_1(L^0)=L^1,~~\I_1(L^1)=-L^0, ~ \I_1(L^2)=L^3,~~\I_1(L^3)=-L^2. \ee
 Because the $L_i$ are left-invariant, it suffices to verify this formula at the point $p$.
 The group $\SU(2)_r$ can act, in particular, by a cyclic permutation of $L_1,L_2,L_3$, thus conjugating $\I_1$ to two more complex structures 
 $\I_2$, $\I_3$ that differ  by cyclic permutation of indices $1,2,3$.      A short calculation at the point $p$
 shows that the quaternion relations are obeyed.  And since the $\I_a$ act by constant linear transformations on the one-forms $L^i$ which are covariantly constant
 with respect to $\nabla$, they are also covariantly constant.  
 
 The corresponding hermitian forms are
 \be\label{hermforms}\omega^{(1)} =L^0\wedge L^1+L^2\wedge L^3,\ee
 with $\omega^{(2)}$, $\omega^{(3)}$ obtained by cyclic permutation of indices $1,2,3$.   To establish the $\N=4$ supersymmetric structure for positive
 chirality modes, it remains only to verify that eqn. (\ref{norgox}) is satisfied.
 We have $\d\omega^{(1)}=2L^0 \wedge L^2\wedge L^3$, and indeed $\I_1(2L^0\wedge L^2\wedge L^3)=2L^1\wedge L^2\wedge L^3=H$.

 To define a similar structure for negative chirality modes, we need a hypercomplex structure that is defined by  constant linear transformations of the $R^j$ and hence  
 is covariantly constant for $\nabla'$.   Here we have a sign choice to make as we can define either
 \be\label{winfox}\I'_1(R^0)=R^1,~\I'_1(R^1)=-R^0,~\I'_1(R^2)=R^3,~\I'_1(R^3)=-R^2,\ee
 or 
  \be\label{zinfox}\I''_1(R^0)=R^1,~\I''_1(R^1)=-R^0,~\I''_1(R^2)=-R^3,~\I''_1(R^3)=R^2,\ee
 with  $\I_2',\I_3'$ and $\I''_2$, $\I'_3$ obtained by cyclic permutations.  The hermitian forms are then
 \be\label{winherm}\omega'{}^{(1)}=R^0\wedge R^1+R^2\wedge R^3,\ee
or
 \be\label{tinherm}\omega''^{(1)}=R^0\wedge R^2-R^2\wedge R^3,\ee
 along with cyclic permutations.
 
 The origin of this sign choice is as follows.   Starting with one hypercomplex structure defined by the $\I_a$, to get a second hypercomplex structure with equal and opposite
 torsion, one way is to apply a reflection of $\S^3$, doing nothing to $\S^1$, and a second way is to apply a joint reflection to both $\S^3$ and $\S^1$.  Under either operation, the
 three-form $H=2 L^1\wedge L^2\wedge L^3=2R^1\wedge R^2\wedge R^3$  changes sign, so either operation gives a second hypercomplex structure with opposite torsion.
 The joint reflection (which is unique only up to a rotation of $\S^3\times \S^1$)
 can be chosen to lead to eqn. (\ref{winfox}) and the reflection of only $\S^3$  can be chosen to lead to eqn. (\ref{zinfox}).   We have seen in section \ref{freeaction} that only the joint
 reflection of $\S^3\times \S^1$ is natural in our problem, so we will use the hypercomplex structure defined by the $\I'_a$, not the one defined by the $\I''_a$.
 
 In section \ref{hypercomplex}, when we explain how a hypercomplex structure on a four-manifold leads to a hypercomplex structure on the instanton moduli space, we will
 see that (assuming instantons are defined to have anti-selfdual curvature) the construction only works if the hermitian forms associated to the hypercomplex structure are selfdual.
 A look back at the previous formulas shows that, if we orient $\S^3\times \S^1$ using the four-form $L^0\wedge L^1\wedge L^2\wedge L^3$,
  the $\omega^{(k)}$ and $\omega'^{(k)}$ are selfdual but the $\omega''^{(k)}$ are anti-selfdual.   Therefore, the $\I_k$ and $\I'_k$ are the choices for which the construction will work.
   
    A consequence of asking for the $\omega^{(k)}$ and $\omega'^{(k)}$ to be all selfdual is that the $\I_k$ and $\I'_k$ do not commute.   Supersymmetric sigma-models with
 torsion are simpler in some respects when the left and right complex structures do commute \cite{GHR}, but we will not be in that situation.
 
   While invariant under $\SU(2)_\ell$, the $\I_k$ and $\omega^{(k)}$ transform in the adjoint representation of $\SU(2)_r$.  Conversely,
   while invariant under $\SU(2)_r$, the $\I'_k$ and $\omega'^{(k)}$ transform in the adjoint representation of $\SU(2)_\ell$.

 \subsection{The Large $\N=4$ Algebra}\label{largefour}
 
 The model we have arrived at -- a sigma-model with a ``round'' metric on the target space $\S^3\times \S^1$, and with $H$ a multiple of the volume 
 form of $\S^3$ -- has a very special structure.   First of all, it is an exactly soluble conformal field theory, since the $\S^3$ model with $H$ a multiple of the volume
 form is simply an $\SU(2)$ WZW model, and the $\S^1$ factor leads to a free superconformal field theory with abelian symmetry.   Beyond that, this
 particular model is actually a prototype of a model that exhibits the ``large'' $\N=4$ superconformal algebra \cite{STP,Sch}.   
We will follow the presentation of  \cite{GMMS}.   (We also follow their notation, which differs slightly from notation used in the rest of the present article, and their conventions. Note
that in those conventions, currents are antiholomorphic, which accounts for some minus signs in the following formulas.)
  
 The supersymmetric WZW with target $\SU(2)$ is actually equivalent to a purely bosonic $\SU(2)$ WZW model with three decoupled free fermions (of each chirality)
 in the adjoint representation of $\SU(2)$ \cite{New,SG}.\footnote{To be more precise, the  $\SU(2)$ WZW model has $\SU(2)_\ell\times \SU(2)_r$ symmetry,
 with the two factors corresponding to the left and right action of $\SU(2)$ on itself.  With an appropriate choice of orientation, the  $\SU(2)_r$ currents are chiral
 (holomorphic) and are accompanied by three positive chirality fermions in the adjoint representation of $\SU(2)_r$, and similarly the $\SU(2)_\ell$ currents
 are antichiral (antiholomorphic) and are accompanied by three negative chirality chiral fermions in the adjoint representation of $\SU(2)_\ell$.}  
  Likewise the supersymmetric model with target $\S^1$ is equivalent to a purely bosonic model with target $\S^1$
 with a decoupled free fermion.  For generic radius of the $\S^1$, the chiral algebra of the $\S^1$ model is just generated by a single abelian current $\partial U$.
 (For special radii, that is for particular values of the circumference of the $\S^1$, the $\S^1$ theory has additional chiral fields -- exponentials of $U$.  We will
 not consider that case.)   Thus overall the supersymmetric model with target $\S^3\times \S^1$ has a chiral algebra that includes  $\SU(2)$ currents $J^i$, $i=1,\cdots,3$
 of some integer level\footnote{The case $\kappa=0$ is possible \cite{DS}.   In that case, the bosonic WZW model with target $\SU(2)$ becomes trivial and drops out;
 the supersymmetric WZW model with target $\SU(2)$ is then just equivalent to three free fermions in the adjoint representation of $\SU(2)$.}
  $\kappa\geq 0$ associated with the bosonic $\SU(2)$ WZW model, four free fermions $\psi^a, ~a=0,\cdots 3$ (one can consider $\psi^0$ to come from the
  $\S^1$ model and the others from the $\SU(2)$ WZW model), and an abelian current $J^0$ that we can normalize to satisfy $J^0(z)J^0(w)\sim-\frac{1}{2(z-w)^2}$. 
 The chiral algebra generated by these fields is actually the large $\N=4$ algebra.
 
  To describe the realization of this algebra, it is  convenient to introduce selfdual and anti-selfdual $4\times 4$ matrices
 \be\label{zioo}\alpha_{ab}^{\pm,i}=\frac{1}{2}\left(\pm \delta_{ia}\delta_{b0}\mp \delta_{ib}\delta_{a0}+\epsilon_{iab}\right),~~i=1,\cdots,3,\ee
 where $\epsilon_{iab}$ is antisymmetric with $\epsilon_{123}=1$. These matrices obey
 \be\label{igloo}[\alpha^{\pm,i},\alpha^{\pm,j}]=-\epsilon^{ijk} \alpha^{\pm k},~~[\alpha^{\pm,i},\alpha^{\mp,j}]=0, ~~\{\alpha^{\pm i},\alpha^{\pm j}\}=-\frac{\delta^{ij}}{2}.\ee
 The generators of the large $\N=4$ algebra in this model are then\footnote{The discussion in section \ref{largesmall} would make one anticipate the presence of generators
 $T$, $G_a$, and $A_{\pm,i}$, but does not make clear why $U$ or $Q^a$ are needed.   In fact \cite{SG}, it is possible to construct a ``smaller'' version of the large $\N=4$ algebra
 in which $U$ and $Q^a$ are omitted.  This algebra is a generalized $W$-algebra (the short distance singularity
 in a product of generators is in general a  nonlinear function of the generators,
 in contrast to simple chiral algebras like the Virasoro algebra -- or the version of the large $\N=4$ algebra with $U$ and $Q^a$ included --
 that can be presented in such a way that OPE singularities are linear in the generators).
 The symmetry of string  theory on $\AdS_3\times \S^3\times \S^3\times \S^1$ is, however \cite{EFGT,BPS}, the ``large'' version of the large $\N=4$ algebra, with extra generators $U$ and $Q^a$ and a simpler structure of the OPE's.      So that is the relevant version for our application.  The existence of the extra generators is a consequence of the symmetries of
$\AdS_3\times \S^3\times \S^3\times \S^1$.}
 \begin{align}\label{thosegen}   T& =-J^0J^0-\frac{\sum_{i=1}^3J^i J^i }{\kappa+2}-\sum_{a=0}^3 \partial\psi^a\psi^a \cr
         G_a& = 2 J^0\psi_a+\frac{4\alpha_{ab}^{+,i} J^i\psi^b}{\sqrt{\kappa+2}}  -\frac{2\epsilon_{abcd}\psi^b\psi^c\psi^d}{3\sqrt{\kappa+2}} \cr
           A^{-,i}& = \alpha_{ab}^{-,i} \psi^a\psi^b \cr
            A^{+,i} &=J^i+ \alpha^{+,i}_{ab}\psi^a\psi^b \cr
                  U & = -\sqrt{\kappa+2} J^0\cr
                   Q^a& =\sqrt{\kappa+2} \psi^a.\end{align}
The operators $A^{-,i}$ and $A^{+,i}$ generate $\SU(2)$ current algebras respectively at level 1 and level $\kappa+1$.
$A^{-,i}$ is constructed  from fermions only and its existence depends only on the fact that the relevant connection on
the tangent bundle of the sigma-model target space $M=\S^3\times \S^1$ has holonomy
 $\SU(2)\subset \SO(4)$, so in particular $A^{-,i}$ would have an analog, also at level 1, in a sigma-model with target a hyper-Kahler four-manifold such
 as $\T^4$ or $\mathrm{K3}$.   By contrast, $A^{+,i}$ contains the currents that generate bosonic symmetries of $\S^3\times \S^1$ (namely the
right action of $\SU(2)=\S^3$ on itself) and $U\sim J^0$ generates a rotation of the second factor of $\S^3\times \S^1$.  Those generators, and similarly
the $Q^a$, which are multiples of the free fermions $\psi^a$, have no analog if the target space is $\KK$. (A $\T^4$ target has  translation symmetries
that lead to the existence of holomorphic currents and free fermions in the chiral algebra of the sigma-model.   Relative to $\S^3\times \S^1$, the important difference is that
the  continuous symmetries of $\T^4$ are commutative and do not enhance the small $\N=4$ algebra to a large one.)

In general, a representation of the large $\N=4$ algebra is characterized by two integers $k^+$ and $k^-$, which are the levels of the $A^{+,i}$ and $A^{-,i}$ current 
algebras.  The Virasoro central charge is
\be\label{foddo}c=\frac{6 k^+k^-}{k^++k^-}.\ee
What we see in eqn. (\ref{thosegen}) is a realization of the algebra with $(k^+,k^-)=(\kappa+1,1)$.   By taking the tensor product of several such realizations, possibly
with different values of $\kappa$, one can get realizations of the large $\N=4$ algebra with any $k^+\geq k^-$.   The tensor product procedure is not straightforward
and is described in section 4.7 of \cite{GMMS}. Since the $\N=4$ algebra has an outer automorphism that exchanges $A^{+,i}$ and $A^{-,i}$, and therefore
exchanges $k^+$ and $k^-$, there is no essential loss of generality in assuming $k^+\geq k^-$.   
For the record, though we will not use these formulas in any detail, the singular 
OPE's of the large $\N=4$ algebra, apart from standard OPE's involving the stress tensor $T$,
are as follows, with $\gamma=\frac{k^-}{k^++k^-}$:
\begin{align} G^a(z) G^b(0) & = \frac{2c\delta^{ab}}{3z^3}-\frac{8\gamma \alpha^{+,i}_{ab}A^{+,i}(0)+8(1-\gamma)\alpha^{-,i}_{ab} A^{-,i}(0)}{z^2}\cr
                                             &~~~-\frac{4\gamma \alpha_{ab}^{+,i} \partial A^{+,i}(0) +4(1-\gamma)\alpha^{-,i}_{ab}\partial A^{-.i}(0)}{z}+\frac{2\delta^{ab} T(0)}{z}+\cdots \cr
                                         A^{\pm,i}(z)A^{\pm,j}(0)& = -\frac{k^{\pm}\delta^{ij}}{2z^2}+\frac{\epsilon^{ijk} A^{\pm, k}(0)}{z}+\cdots \cr   
                                              Q^a(z) Q^b(0)& =-\frac{(k^++k^-)\delta^{ab}}{2z}+\cdots\cr
                                                  U(z) U(0)& = -\frac{k^++k^-}{z^2}+\cdots \cr
                                                    A^{\pm,i}(z) G^a(0) & = \mp \frac{2 k^\pm \alpha^{\pm, i}_{ab} Q^b(w)}{(k^++k^-)z^2}+\frac{\alpha^{\pm,i}_{ab} G^b(0)}{z}+\cdots\cr
                                                         A^{\pm, i} Q^a(0)& =\frac{\alpha^{\pm, i}_{ab} Q^b(0)}{z}+\dots\cr
                                                          Q^a(z) G^b(0)& = \frac{2\alpha^{+,i}_{ab} A^{+,i}(0) -2\alpha^{-,i}_{ab}(0)+\delta^{ab} U(0)}{z}\cr
                                                            U(z) G^a(0) &= \frac{Q^a}{z^2}+\cdots. \end{align}

  In the case of the moduli space $\M$ of instantons on $\S^3\times \S^1$, 
 we will explain enough in sections \ref{hypercomplex} and \ref{hkt} 
to show that the sigma-model with target $\M$ has $\N=4$ supersymmetry with the small $\N=4$ algebra.  To extend this result to get a large
 $\N=4$ algebra, we need to ``find'' the extra generators $U,\,Q^a,\, A^{+,i}$ for both positive and negative chirality.  The strategy for finding them  is simple.
 $\S^3\times \S^1$ has many Killing vector fields, associated with the rotation symmetries of the two factors, 
 and associated to these are Killing vector fields on the instanton
 moduli space $\M$.   We will show that these Killing vector fields on $\M$ are covariantly constant for the 
 appropriate connections with torsion, and this, as in the analysis
 of eqn. (\ref{freej}), leads to the existence of additional chiral operators in the sigma-model with target $\M$.   In that way we will find the extra generators that
 are needed to fill out the large $\N=4$ algebra.  This analysis will be the topic of section \ref{large}, but first we need to understand why a sigma-model with
 target $\M$ has at least small $\N=4$ supersymmetry.
 
 \section{Hypercomplex Structure Of The Instanton Moduli Space}\label{hypercomplex}
 
In this section, largely following \cite{book,Hitchin,MV}, 
we will show that if the oriented four-manifold $M$ has a hypercomplex structure, with selfdual hermitian forms and closed torsion,
then the moduli space $\M$ of instantons on $M$ also carries a hypercomplex structure (which we will show in section \ref{hkt} to have closed torsion).   This is true 
for every simple compact  gauge group $\GG$,  and every component of instanton moduli space.   To be more precise, we define these structures on the smooth part of $\M$.
That means in particular that we always consider instanton solutions that are irreducible in the sense that the equation
\be\label{kolo}\d_A\sigma=0,\ee
where $\sigma$ is a section of $\ad(E)$, has no non-zero solutions.   Such solutions, which generate unbroken gauge symmetries (continuous automorphisms of $E$), arise only at singularities of $\M$.

If  $M$ has two hypercomplex structures compatible with the same metric, both with selfdual hermitian forms, and
with opposite torsion, then  we will see that $\M$ likewise has two hypercomplex structures 
with the same properties.   (The proof that the two hypercomplex structures of $\M$  have opposite torsion is given in section \ref{hkt}.)  
In our application, we have $M=\S^3\times \S^1$ and $\GG=\SU(Q_5)$,
but presenting the arguments in greater generality poses no difficulty.\footnote{Actually, a compact hypercomplex four-manifold that is not hyper-Kahler is
locally isomorphic to $\S^3\times \S^1$ \cite{Boyer}.  But the following analysis can likely be extended to other examples that are complete but not compact.}

To begin with, we assume that the oriented four-manifold $M$ has simply a strong KT  structure, namely a complex structure $\I$, a metric $g$ that is of type $(1,1)$
and (therefore) a hermitian form $\omega_{IL} =g_{IK}\I^K{}_L$ that is also of type $(1,1)$, and that we assume to be selfdual,\footnote{\label{always}The definition $\omega_{IL} =g_{IK}\I^K{}_L$ implies that $\omega$ is  selfdual or anti-selfdual, depending on how $M$ is oriented.   Assuming that the instanton equation is going to be $F^+(A)=0$, we require $\omega$ to be selfdual so that
the $(1,1)$ part of the instanton equation can be written as in eqn. (\ref{nubic}) below.}  and with closed torsion $H=\i(\bar\partial-\partial)\omega$.   
Given this, we will define a similar hermitian structure on $\M$ (the closedness of the torsion will be shown in section \ref{hkt}).

First of all, the space $\A$ of all connections on the $\GG$-bundle $E\to M$ itself has a natural metric, with the length squared of a variation $\delta A$ of a connection $A$ being
\be\label{ibo}|\delta A|^2=-\frac{1}{4\pi^2}\int_M\tr \,\delta A\wedge \star \delta A=-\frac{1}{4\pi^2}\int_M\d^4 x \sqrt g   g^{ij}\tr\,\delta A_i\delta A_j. \ee
The factor $1/4\pi^2$ is arbitrary for now; when we compute the torsion on the moduli space, it will be convenient.
The space $\A$ also has a natural complex structure $\h \I$, defined by saying that the $(0,1)$ part of $\A$ is holomorphic and the $(1,0)$ part is antiholomorphic.  
Thus if $\delta A$ is a variation of $A$, and  $\delta A=\delta A^{1,0}+\delta A^{0,1}$ is its expansion  in parts of type 
$(1,0)$ and $(0,1)$, then 
\be\label{terf}\h\I(\delta A)=\i \delta A^{(0,1)}-\i \delta A^{(1,0)}.\ee

 The
reason to define the $(0,1)$ -- rather than $(1,0)$ -- part of $A$ to be holomorphic is as follows.   First of all, if we decompose the Yang-Mills curvature $F(A)$
in pieces of type $(p,q)$ with respect to the complex structure $\I$ of $M$, then the instanton equation reads
\be\label{juju}F^{(0,2)}=F^{(2,0)} = \omega \wedge F^{(1,1)}=0.\ee
In particular, vanishing of $F^{(0,2)}$ means that 
 $\bar\partial_A =\bar\partial+[A^{(0,1)},\,\cdot\,]$
satisfies $\bar\partial_A^2=0$. This makes  the bundle $E$ --  or more precisely its complexification $E_\C$ -- a holomorphic  bundle over the complex manifold $M$.   We would
like to define the complex structure of $\A$ in a way that ensures that, after eventually reducing to the instanton moduli space $\M$,  the holomorphic  bundle
$E_\C$ varies holomorphically over $\M$. For this we have to define the complex structure of $\A$ so that $\bar\partial_A=\bar\partial+A^{(0,1)}$ varies holomorphically with $A$. This is true with the sign choice that we made.
In fact, a theorem of \cite{Buchdahl,LiYau}  identifies $\M$ as the moduli space of stable holomorphic structures on $E$ (where here the notion
of stability is a generalization of the usual stability condition for a holomorphic  bundle over  a Kahler manifold).  We will not need this  difficult theorem. For our purposes
it will suffice to know how to define the complex structure of $\M$, which is  much easier.

Having a metric and a complex structure, $\A$ acquires  a hermitian form:\footnote{Here we include a factor of $1/2$ that was introduced in eqn. (\ref{noteme}).}
\be\label{zobo}\h\omega=-\frac{1}{8\pi^2}\int_M\tr\,\omega \wedge \delta A\wedge \delta A.\ee
Viewing $\delta A$ as a 1-form on $\A$ (valued in 1-forms on $M$), we see that $\h\omega$ is a two-form on $\A$, and it is of type $(1,1)$ because $\omega$ is of type $(1,1)$.

\def\ASD{{\mathrm{ASD}}}
Within the infinite-dimensional space $\A$, there is an infinite-dimensional submanifold $\A^\ASD$ that parametrizes gauge fields that solve the instanton equation.
The instanton moduli space is $\M=\A^\ASD/\G$, where $\G$ is the group of gauge equivalences (if $E$ is trivialized, then $\G$ is the group of
maps from $M$ to $\GG$).  
There is no problem to restrict the metric $|\delta A|^2$ or  the two-form $\h\omega$ from $\A$ to $\A^\ASD$.  However, there is no immediate
way to extract from these objects a corresponding metric or two-form on $\M$. The difficulty is as follows.   Let $A$ be an instanton solution describing a point in $\M$, and let $\zeta$ be a tangent
vector to $\M$ at that point.   {\it Up to gauge transformation}, $\zeta$ corresponds to a solution of the linear equation obtained by linearizing the instanton equation $F^+(A)=0$.
This linear equation is
\be\label{lineq}(\d_A\delta A)^+   =0,\ee
where $\d_A=\d+[A,\,\cdot\,]$ is the gauge-covariant extension of the exterior derivative.
This equation possesses a gauge invariance
\be\label{bineq} \delta A\to \delta A-\d_A\sigma,\ee
where $\sigma$, a section of the adjoint bundle $\ad(E)$,  is an arbitrary generator of a gauge transformation.
The metric and hermitian forms defined in (\ref{ibo}) and eqn. (\ref{zobo}) are not invariant under such a change\footnote{\label{precise} To be more precise, the metric
is not invariant under $\delta A\to \delta A-\d_A\sigma$.   The hermitian form $\h\omega$ is invariant if and only if $\d\omega=0$, that is, if and only if $M$ is Kahler.
To prove this, substitute $\delta A\to \delta A-\d_A\sigma$ in eqn. (\ref{zobo}), integrate by parts, and use $F^+=(\d_A\delta A)^+=0$, giving a result proportional to $\d\omega$.}
 of $\delta A$.   Therefore, if we want to use
eqns. (\ref{ibo}) and (\ref{zobo}) to define a metric and a hermitian form on the tangent space $T_A\M$ to $\M$ at $A$, we need to impose some sort of gauge condition on $\delta A$.
We will only consider gauge conditions that are invariant under gauge transformations of the background gauge field $A$ (but not under the transformation (\ref{bineq}) of $\delta A$).
For example, though it is not the gauge condition that we will ultimately use,
an obvious gauge condition that is invariant under gauge transformations of the background 
is the Landau gauge condition $D_i \delta A^i=0$, where $D_i=\partial_i+[A_i,\cdot\,]$.

Once a gauge condition on $\delta A$ is picked, the tangent space $T_A\M$ to $A$ in $\M$
 becomes a subspace of the tangent space  $T_A\A$ to $A$ in $\A$.   
If and only if the conditions we impose on $\delta A$ -- the equation (\ref{bineq}) plus the gauge condition that we have not yet chosen -- are invariant under $\delta A\to \I\delta A$,
it will make sense to restrict the complex structure  $\h\I$ defined in (\ref{terf}) from $T_A\A$ to $T_A\M$.   This will define an almost complex structure on $\M$, which we will eventually show to 
be integrable. 

With this aim, let us discuss the  Hodge decomposition
of the instanton equation $F^+(A)=0$, and its linearization $(\d_A\delta A)^+=0$.  
The equation $F^+(A)=0$ has parts of type $(2,0)$ and $(0,2)$, which are complex conjugates of each other, and a part of type $(1,1)$.
The $(0,2)$ part of the instanton equation is explicitly
\be\label{makelin}\bar\partial A^{(0,1)}+A^{(0,1)}\wedge A^{(0,1)}=0 \ee
and the $(2,0)$ part is just the complex conjugate of this. 
Similarly the $(0,2)$ part of the linearized equation is
\be\label{zakelin}\bar\partial\delta A^{(0,1)}+[A^{(0,1)},\delta A^{(0,1)}]=0. \ee
Since the complex structure on $\A$ is defined by saying that $A^{(0,1)}$ is holomorphic, the eqn. (\ref{makelin}) is holomorphic.   Similarly, eqn. (\ref{zakelin}) is linear in 
$\delta A^{(0,1)}$ with no dependence on $\delta A^{(1,0)}$, so under $\delta A\to \I\delta A$, it is just multiplied by $\i=\sqrt{-1}$.  Hence the equation $(\d_A\delta A)^{(0,2)}=0$ is invariant
under $\h\I$.   Its complex conjugate, namely $(\d_A\delta A)^{(2,0)}=0$, is of course also invariant.

However, the $(1,1)$ part of the instanton equation is not holomorphic or antiholomorphic and its linearization is not invariant.   That $(1,1)$ part is
\be\label{nubic}\omega \wedge F =0. \ee
As $\omega$ is of type $(1,1)$, it is equivalent to write $\omega\wedge F^{(1,1)}=0$, and because $\omega$ is selfdual, it is also equivalent to write $\omega\wedge F^+(A)=0$.
 It is to ensure that the $(1,1)$ part of
the instanton equation can be written as in (\ref{nubic}) that we require $\omega$ to be selfdual.
Eqn. (\ref{nubic}) is neither holomorphic nor antiholomorphic, and similarly its linearization, namely
\be\label{ubic}\omega\wedge \d_A\delta A=0,\ee
depends on both $\delta A^{(1,0)}$ and $\delta A^{(0,1)}$ and is not invariant under $\delta A\to \I\delta A$.

The only hope of getting a set of conditions on $\delta A$ that is invariant under $\I$ is to interpret whatever eqn. (\ref{ubic}) transforms into under $\delta A\to \I\delta A$ as
the gauge condition that we are going to impose.   In other words (as in section 5.3 of \cite{book}), the gauge condition will have to be
\be\label{zubic}\omega\wedge \d_A(\I \delta A)=0.\ee
If this does make sense as a gauge condition, then this gauge condition plus eqn. (\ref{ubic}) is a set of equations that is invariant under $\delta A\to \I\delta A$, since
if we substitute $\delta A\to \I\delta A$ in eqn. (\ref{zubic}), we get back eqn. (\ref{ubic}).   Imposing on $\delta A$ the linearized instanton equation together with eqn.
(\ref{zubic}), we can then restrict $\h\I$ from $\A$ to $\M$ to provide an almost complex structure on $\M$, 
and eqns. (\ref{ibo}) and (\ref{zobo}) define a metric and hermitian form on $\M$,  both of type $(1,1)$.   

In general, to show that a proposed gauge condition is valid, we have to demonstrate two facts: (1) it should be possible to impose this gauge condition,
in the sense that by a suitable gauge transformation, it can always be satisfied; (2) it fully fixes the gauge, in the sense that the gauge transformation  that
ensures that the condition is satisfied is unique.   We have to decide whether the gauge condition (\ref{zubic}) satisfies those two conditions.

As a preliminary, we will put the candidate gauge condition (\ref{zubic}) in a more convenient form.
 More explicitly, the left hand side of eqn. (\ref{zubic}) is
 \be\label{tubic} \frac{1}{2}\epsilon^{ijkl}\omega_{ij} D_k(\I^m{}_l \delta A_m)
 =\frac{1}{2}\epsilon^{ijkl}\omega_{ij}\I^m{}_l D_k\delta  A_m +\frac{1}{2}\epsilon^{ijkl}\omega_{ij}(D_k \I^m_l)
 \delta A_m.\ee 
 Being selfdual, $\omega$ satisfies $\frac{1}{2}\epsilon^{ijkl}\omega_{ij}=\omega^{kl}$, so the first term on the right in eqn. (\ref{tubic}) is $\omega^{kl}\I^m_l  D_k \delta A_m
 = G^{km}D_k\delta A_m=D_k\delta A^k$.  Using $\omega_{ln}=G_{lp}\I^p{}_n$, the last term on the right can be written  $\frac{1}{2}\epsilon^{ijkl} \omega_{ij} (D_k \omega_{ln})
 \delta A^n$.   To express this in terms of the torsion, we recall that $\omega$ is covariantly constant for the connection $\nabla$ with torsion, $\nabla_k\omega_{ln}=0$.
 Using this and the definition (\ref{ewcon}) of $\nabla$,  we get 
 \be\label{funone}\frac{1}{2}\epsilon^{ijkl} \omega_{ij} (D_k \omega_{ln})
 \delta A^n=\frac{1}{4}\epsilon^{ijkl}\omega_{ij} \left(H_{kl}{}^p \omega_{pn}-H_{kn}{}^p\omega_{pl}\right)\delta A^n. \ee
 The right hand side  as written is homogeneous and quadratic in $\omega$, but rather surprisingly it actually turns out to be entirely independent of $\omega$.
 We will see that this  surprising fact is a necessary input in defining a hypercomplex structure on $\M$.  To show that the right hand side of eqn.  (\ref{funone}) is
 really independent of $\omega$ is a matter of doing some group theory in the tangent space to an arbitrary point $p\in M$ at which we want to prove the statement.
 The expression in question is a bilinear function of  a one-form $\delta A$ and a three-form $H$.  This bilinear function  depends on the metric tensor $g$ of $M$ at $p$ (used to raise
 and lower indices in eqn. (\ref{funone})) and on $\omega$, but nothing else (the antisymmetric tensor $\epsilon^{ijkl}$ is determined by the metric and  orientation of $M$;
 the orientation is determined by $\omega$, which is assumed to be selfdual).   So this expression is invariant under linear transformations of the tangent space to $M$
 at $p$ that preserve $g$ and $\omega$.   The group of such linear transformations is $\U(2)$.   However, the expression that we are trying to analyze is an even function
 of $\omega$, invariant under $\omega\to -\omega$.  So this expression  is also invariant under  linear transformations of the tangent space that map $\omega$ to $-\omega$.
Including such linear transformations extends $\U(2)$ to a double cover 
that we will call $\U^*(2)$ (this group is the group of $2\times 2$ unitary complex matrices extended by the operation of complex conjugation,
 which reverses the sign of $\omega$).   The one-form $\delta A$ transforms in a four-dimensional representation of $\U^*(2)$ that is irreducible even over the complex
 numbers (this statement would not be true if we consider $\U(2)$ instead of  $\U^*(2)$).   Because the Hodge duality map $\star $ from three-forms to one-forms commutes with $\U^*(2)$, the three-form $H$ transforms
 in the same irreducible four-dimensional representation.   Therefore, the right hand side of eqn. (\ref{funone}) is an invariant bilinear form on an irreducible representation
 of $\U^*(2)$.    Such an invariant bilinear form is unique up to a constant multiple, so the expression in question must be a multiple of any conveniently chosen $\U^*(2)$ invariant that is
 bilinear in $H$ and $\delta A$.   A convenient invariant is $\star( H\wedge \delta A)=\frac{1}{6}\epsilon^{ijkl}H_{ijk}\delta A_l$.   So the right hand side of eqn. (\ref{funone})
 must be a multiple of this.   By checking an example,\footnote{In local coordinates $x^1,\cdots, x^4$, one can take the metric of $M$ at the point $p$ to be $\delta_{ij}$, the hermitian form to
 have non-zero coefficients $\omega_{12}=-\omega_{21}=\omega_{34}=-\omega_{43}=1$, and $H$ and $\delta A$ to have nonzero elements $H_{123}=\delta A_4=1$.}
  one can verify that the coefficient is 1.  Putting these facts together, the candidate gauge condition is\footnote{This assertion is Lemma 8 in \cite{Hitchin}, where a rather
  different proof is given.}
  \be\label{omely}D_m\delta A^m+\star( H\wedge \delta A)=0.\ee 
  
  It is now relatively simple to determine whether this is a satisfactory gauge condition.    Under $\delta A\to \delta A-\d_A\sigma$, the gauge condition
  transforms by
  \be\label{zomely} D_m\delta A^m+\star (H\wedge \delta A)\to D_m\delta A^m+(\star H\wedge \delta A) -D_i D^i\sigma -\star( H\wedge\d_A\sigma).\ee
  Assuming that the right hand side of eqn. (\ref{zomely}) vanishes  for some $\sigma$, the condition that this choice of $\sigma$ is unique is that there is
  no non-zero solution of the equation 
  \be\label{gimel}-D_i D^i\sigma -\star (H\wedge \d_A\sigma)=0. \ee   
  This equation can be written
  \be\label{himel} W\sigma=0,\ee
  where the linear operator $W$ was introduced in eqn. (\ref{efo}).
  Given any solution of this equation, we can multiply by $\sigma$, take a trace,
  and integrate to get
  \be\label{domely}0=-\int_M\d^4x\sqrt g \,\tr\,\sigma(-D_i D^i\sigma) -\int_M \tr \, \sigma H\wedge \d_A\sigma.\ee
  Here the second term is 
  \be\label{helpmex} -\int_M H\wedge \tr \,\sigma \d_A\sigma =-\frac{1}{2}\int_M H\wedge \d\,\tr\,\sigma^2=0,\ee  where in the last step we integrate by parts and use $\d H=0$.
  Integrating by parts in the first term we then learn that
  \be\label{pomely}  0=-\int_M\d^4x\sqrt g \tr\, D_i\sigma D_j\sigma g^{ij},\ee
  which implies that $D_i\sigma=0$.  As remarked earlier (see eqn. (\ref{kolo})), at a smooth point of $\M$, this implies that $\sigma=0$.  That establishes the desired uniqueness.
  
  It remains to verify that the right hand side of eqn. (\ref{zomely}) always does vanish for some $\sigma$, which is true if the equation 
   \be\label{timely} -D_i D^i\sigma -\star( H\wedge\d_A\sigma)= w \ee
   has a solution for every section $w$ of $\ad(E)$.   If this is not the case, then  sections of $\ad(E)$ 
   of the form  $-D_i D^i\sigma -\star( H\wedge\d_A\sigma)$ generate a proper 
   subspace  of the Hilbert space $\H$ of all $L^2$  sections of $\ad(E)$, and there is some section $w$ that is orthogonal to this subspace.
   The orthogonality condition is
   \be\label{imely}0=-\int_M\d^4x \sqrt g\tr \, w \left(-D_i D^i\sigma -\star( H\wedge\d_A\sigma)\right),\ee
   and the condition that this is true for all $\sigma$ is
   \be\label{pimely}-D_i D^i w +\star (H\wedge \d_A w)=0.\ee
   This is the same equation (\ref{gimel}) that we have already analyzed, but with $H\to -H$.  The same argument as before shows that there are no nonzero solutions.
      Actually, the operator that appears in eqn. (\ref{pimely}) is simply the adjoint  $W^\dagger$ of the operator $W$ introduced earlier, so to prove that the gauge condition is satisfactory,
      what we have had to prove is that $W$ and $W^\dagger$ both have trivial kernel.  Since $W$ and $W^\dagger$ 
      are related by $H\leftrightarrow -H$, this pair of statements is invariant under reversing the sign of $H$.  Instead of saying that $W$ and $W^\dagger $ have trivial kernel,
      an equivalent statement is that either one of them has trivial kernel and cokernel.

At this point, having verified that the gauge condition is a good one, we have defined an almost complex structure $\h \I$ on $\M$.  It remains to verify
   that this structure is integrable.  A simple argument is available.   First of all, the equation $F^{0,2}(A)=0$ is  holomorphic in $A$, so if we impose
   only this equation, we get an (infinite-dimensional) complex submanifold $\B$ of $\A$.   However, the conditions $\omega\wedge \d_A\delta A=\omega\wedge \d_A(\I\delta A)=0$
   appear to spoil holomorphy.  To prove that $\h\I$ is integrable, we will show that imposing these nonholomorphic equations is equivalent to a certain holomorphic operation.
   
   The idea is to exploit the fact that, given the complex structure of $M$, the action on $\A$ of the group $\G$ of gauge transformations can be analytically continued
   to a holomorphic action of the complexification $\G_\C$ of this group.  A generator $\sigma$ of $\G_\C$ is a section of $\ad(E)\otimes_\R \C$, the complexification of $\ad(E)$.
   More explicitly, such a generator is $\sigma=\sigma_1+\i\sigma_2$ where $\sigma_1$, $\sigma_2$ are sections of $\ad(E)$.  We define an action of $\sigma$ on $A$ by
   $A\to A-\d_A\sigma_1-\I \d_A\sigma_2$.   In particular, the holomorphic variable $A^{(0,1)}$ tranforms by $A^{(0,1)}\to A^{(0,1)}-\bar\partial_A \sigma_1-\I \bar\partial_A\sigma_2
   =A^{(0,1)}-\bar\partial_A \sigma$, where we use the fact that $\I$ acts as $\i$ on the $(0,1)$-form $\bar\partial_A \sigma_2$.   Since $A^{(0,1)} -\bar\partial_A\sigma$ is
   holomorphic in $A$ and $\sigma$, this does define a holomorphic action of $\G_\C$ on $\A$.   
   
   The claim now is that the conditions    $\omega\wedge \d_A\delta A=\omega\wedge \d_A(\I\delta A)=0$ can be viewed as gauge-fixing conditions that fix the action of
   $\G_\C$ on $\A$, so that $\M$ can be interpreted as $\B/\G_\C$, making obvious the complex structure of $\M$ and thus the integrability of $\I$.   We already know
   that the condition $\omega\wedge \d_A(\I\delta A)=0$ can be viewed as a gauge-fixing condition for the ordinary gauge symmetry $A\to A-\d_A\sigma_1$; it remains to
   verify that the other condition $\omega \wedge \d_A\delta A=0$ can be viewed as a gauge-fixing condition for the imaginary gauge symmetry $A\to A-\I\d_A\sigma_2$.  
   Following the same logic as before, we have to show that the operator $\sigma\to -\omega \d_A(\I \d_A\sigma)$ has trivial kernel and cokernel.  But this is actually up to sign
   the same operator that we have already studied so indeed the kernel and cokernel are trivial.
   
   At this point, then, we know that $\M$ is a complex manifold with metric and hermitian forms given by equations (\ref{ibo}) and (\ref{zobo}).   Now as in \cite{MV},  let us assume
   that $M$ has not just the single complex structure $\I$ that we have assumed so far, but a hypercomplex structure, with three complex structures $\I,\J,\K$ that satisfy
   the quaternion relations and the conditions described in section \ref{sigmab}.  In particular, this means that the metric $g$ of $M$ is of type $(1,1)$ for each of $\I$, $\J$, and $\K$,
   and that $\I,\J,$ and $\K$ as well as $g$ are all covariantly constant for the same connection $\nabla$  whose torsion is a closed three-form $H$.  The three hermitian
   forms $\omega_\I=g\I$, $\omega_\J=g\J$, and $\omega_\K= g \K$ are, of course, different.    However, because of the crucial fact that the gauge condition (\ref{omely}) depends only
   on $H$ but not on $\omega$, the construction we have described, whether carried out for $\I$, for $\J$, or for $\K$, leads to the same metric on $\M$ and the same
   description of the tangent space $T_p\M$ for any point $p\in \M$ in terms of the common gauge condition (\ref{omely}).
     Therefore,  for any $q\in\M$, restricting $\h\I$, $\h\J$, and $\h\K$ from $T\A$ to $T_q\M$, we get almost 
   complex structures $\h\I$, $\h\J$, $\h\K$ on $T_q\M$. They are all integrable by the argument that was just given, and  they obey the quaternion relations because those were obeyed before restricting to $T_q\M$.  Eqn. (\ref{ibo})
   gives a metric that is of type $(1,1)$ for each of $\h\I$, $\h\J$, $\h\K$, and eqn. (\ref{zobo}), with $\omega$ replaced by $\omega_\I$, $\omega_\J$, or $\omega_\K$,
   gives three hermitian forms $\omega_{\h\I}$, $\omega_{\h\J}$, $\omega_{\h\K}$, each of which is of type $(1,1)$ for the corresponding complex structure.   In section \ref{hkt},
   we will show that each of $\h\I$, $\h\J$, $\h\K$ is associated to the same torsion, since $-\i(\partial_{\h\I}-\bar\partial_{\h\I})\omega_{\h\I}=-\i(\partial_{\h\J}-\bar\partial_{\h\J})\omega_{\h\J}
   =-\i(\partial_{\h\K}-\bar\partial_{\h\K})\omega_{\h\K}$.
   
 Now we specialize to  the case that\footnote{It follows from the classification of compact hypercomplex four-manifolds \cite{Boyer} that $\S^3\times \S^1$ is the unique compact four-manifold with 
 the properties specified in this sentence.}   $M=\S^3\times \S^1$, which admits a second hypercomplex structure $\I',\J',\K'$ such that the same metric $g$ is
 of type $(1,1)$ for each of $\I',\J',\K'$, and the three hermitian forms $\omega'_I$, $\omega'_J$, and $\omega'_K$ are all selfdual, but the torsion is  $-H$, as opposed to the
 previously assumed $+H$.  
 Of course, we can carry out the same construction as before, endowing $\M$ with a new hypercomplex structure $\h\I',\h\J',\h\K'$, and a compatible metric and hermitian 
 forms.   The only problem is that since the torsion is now $-H$,  we have to reverse the sign of $H$ in the gauge condition, which is now
   \be\label{omely2}D_m\delta A^m-\star( H\wedge \delta A)=0.\ee 
Therefore, it seems that the hyperhermitian metric  that we will define on $\M$ will be different.  If so, then  no one metric
on $\M$ will be consistent with all of the structures predicted by the duality conjecture with strings on $\AdS_3\times \S^3\times \S^3\times \S^1$. One metric on $\M$ will
lead to a sigma-model with $(0,4)$ supersymmetry and another metric on $\M$ will lead to a sigma-model with $(4,0)$ supersymmetry.   This would disprove
the duality conjecture.

What saves the day is another striking fact (Lemma 9 in \cite{Hitchin}, where the following proof can be found): reversing the sign of $H$ in the gauge condition actually does
not   change the resulting metric on $\M$.    Consider an instanton connection $A$ representing a point $q\in M$, and consider a tangent vector to $q$ in $\M$ that
can be represented  by a deformation $\delta A$ of $A$ that satisfies 
  \be\label{zomely2}D_m\delta A^m+\star( H\wedge \delta A)=0,\ee 
and also by another deformation $\delta A'$ that satisfies a similar condition with $H$ replaced by $-H$:
  \be\label{domely2}D_m\delta A'{}^m{}-\star( H\wedge \delta A')=0,\ee 
The statement that $\delta A$ and $\delta A'$ represent the same tangent vector to $q$ in $\M$ means that they are gauge equivalent,
\be\label{vomel}\delta A=\delta A'-\d_A\sigma\ee
for some generator $\sigma$ of a gauge transformation.  
Based on this information, we want to show that
\be\label{omigo}|\delta A|^2=|\delta A'|^2,\ee
where these expressions are defined via eqn. (\ref{ibo}).    This will show that the length of a tangent vector at any point $q\in\M$ is the same
regardless of which of the two gauge conditions is used to compute it, or in other words that the two gauge conditions lead to the same metric on $\M$.
The difference between the left and right of eqn. (\ref{omigo}), in view of the relation (\ref{vomel}) between $\delta A$ and $\delta A'$, is
\be\label{tomigo}-\frac{1}{4\pi^2}\int \d^4 x\sqrt g\,\tr \left(D_m\sigma D^m\sigma +2 D_m\sigma \delta A^m\right),\ee
and we will show that this vanishes.
Using eqn. (\ref{vomel}) to solve for $\delta A'$ in eqn. (\ref{domely2}),
we get
\be\label{tomely}D_m\delta A^m+ D_m D^m \sigma-\star(H\wedge \delta A) - \star(H\wedge \d_A\sigma)=0. \ee 
Adding  eqn. (\ref{zomely2}) to this, we find 
\be\label{impelt} 2 D_m\delta A^m +D_m D^m\sigma -\star(H\wedge \d_A\sigma)=0.\ee
Multiplying by $\sigma$, taking a trace, and integrating, we have
\be\label{zimbel} 0=\int_M\d^4x \sqrt g\tr\,\left(2 \sigma D_m\delta A^m  +\sigma D_m D^m\sigma\right) +\int_M H\wedge \tr \,\sigma \d_A \sigma.\ee 
The last term on the right hand side vanishes, as we have already seen in eqn. (\ref{helpmex}).   Dropping this term and integrating by parts in the 
remaining terms,  eqn. (\ref{zimbel}) becomes equivalent to the desired result
(\ref{tomigo}),  completing the proof.

The theory of generalized Kahler reduction gives a possibly more conceptual explanation of the success of this calculation \cite{BCG}, at least in the untwisted case (the case that
$H$ is exact). Similarly,
a more conceptual explanation of why the relevant gauge condition turned out to depend only on $H$ and not $\omega$ might conceivably be found in the theory of
generalized hyper-Kahler reduction.  

 \section{$B$-Field On The Instanton Moduli Space}\label{hkt}

 For a complex structure $\I$ on $M$ satisfying certain conditions, 
 we defined in section \ref{hypercomplex}
 a corresponding complex structure $\h\I$ on $\M$, along with a metric and hermitian form defined in eqns. (\ref{ibo}) and (\ref{zobo}).   In particular,
 the hermitian form is
 \be\label{tellmot}\omega_{\h\I}=-\frac{1}{8\pi^2}\int_M \tr\,\omega_{\I}\wedge \delta A\wedge \delta A,\ee
 where $\delta A$ satisfies the gauge condition $W\delta A=0$, as in eqn.  (\ref{omely}).   

In the present section, 
 we will compute the corresponding torsion $\h H=-\h\I \d\omega_{\h\I}$.    A formula for $\h H$ was computed in \cite{book}, section 5.3; another useful reference is \cite{Hitchin}.
 The important properties of $\h H$ for the strong HKT  geometry of $\M$ were pointed out in \cite{MV}.
 
 It turns out that $\h H$ has  several key properties: (i) it depends only on the metric $g$ and torsion  $H$ of $M$ 
 and not on $\I$ or  $\omega_\I$; (ii) it satisfies  $\d\h H=0$, assuming $\d H=0$;  (iii) it is homogeneous and linear in $H$, and therefore changes sign under $H\to -H$. 
 All three conditions are needed to ensure that for $M=\S^3\times \S^1$, $\M$ has the properties that are predicted by the conjectured duality with Type IIB superstrings.
 Condition (i) means that the same torsion form $\h H$ is compatible with the three complex structures $\h\I$, $\h\J,$ $\h\K$ of $\M$, defined in section \ref{hypercomplex},  that generate the hypercomplex
 structure  of $\M$.  Hence one can define on $\M$ a metric compatible connection $\h\nabla$, with torsion $\h H$,  for which $\h\I, \h\J$, and $\h\K$ are all covariantly constant.
 We say that $\h\nabla$ corresponds to the connection $\nabla $ on $M$.   This gives $\M$ an HKT structure, which according to condition (ii) is strong if the HKT structure of $M$ is strong.
 To extend this structure to a (twisted) generalized hyper-Kahler structure on $\M$, we need to define on $\M$ a second HKT structure with equal and opposite torsion.
 The ability to do this is what we get from condition (iii), which enables us to define a metric compatible connection $\h\nabla'$ on $\M$, with torsion $-\h H$, such
 that $\h \I',\h\J'$, and $\h\K'$ are all covariantly constant.
 
 Furthermore, we will show that with the normalization that was chosen in  eqn. (\ref{ibo}), the periods of $\h H$
 are valued in $2\pi \Z$, so  $\h H$ is the curvature of a $B$-field $\h B$ on $\M$.  $\h B$ is uniquely determined up to the possibility of adding to it a flat $B$-field.   Shifting
 $\h B$ by a flat $B$-field that is not pure gauge
  is a modulus of the sigma-model with target $\M$; it amounts to shifting the theta-angle that was introduced in section \ref{secondbetti}.

Before getting into too many details, let us note that
 it may seem surprising that $\d\omega_{\h\I}\not=0$.   Viewed as a differential form on the infinite-dimensional manifold $\A$, $\omega_{\h\I}$ is a differential
form with constant coefficients on a linear space and it certainly satisfies $\d\omega_{\h\I}=0$.   We can in a completely natural way restrict $\omega_{\h\I}$ from $\A$ to
$A^\ASD$, the subspace consisting of all connections that satisfy the instanton equation $F^+(A)=0$, and it remains closed.   However, there is no natural way to push $\omega_{\h\I}$ down
to a two-form on the moduli space $\M=\A^\ASD/\G$.   The key point is that, assuming\footnote{As explained in footnote \ref{precise}, if $\d\omega_\I=0$, meaning that $M$ is
Kahler, then $\omega_{\h\I}$ is a pullback from $\M$ after all.   This can be used to prove that if $M$ is Kahler, then $\d\omega_{\h\I}=0$ and $\M$ is Kahler.  The
fact that $\d\omega_{\h\I}=0$ if $\d\omega_\I=0$ will be clear from the formula obtained below.}
 $M$ is not Kahler,  $\omega_{\h\I}$ does not vanish if we substitute $\delta A=-\d_A\sigma$.
This substitution amounts to contracting $\omega_{\h\I}$ with a vector field tangent to the fibers of $\A^\ASD\to \M$, and the fact that it does not vanish means that 
$\omega_{\h\I}$ is not the pullback of a form on $\M$.   Because of this, in section \ref{hypercomplex}, to define $\omega_{\h\I}$ as a form on $\M$, we had to stipulate that the
formula (\ref{tellmot}) should be evaluated only for variations $\delta A$ that satisfy the gauge condition.
In what follows, we will work  through  in more detail 
 what is involved in interpreting $\omega_{\h\I}$ as a form on $\M$ with the help of the gauge condition; this will hopefully make it clearer why  $\d\omega_{\h\I}\not=0$
 and 
how to compute $\d\omega_{\h\I}$.   

The following considerations are local along $\M$ so we can work in a small
 neighborhood of a general  point $r\in\M$.  We pick local coordinates $m^\alpha$, $\alpha=1,\cdots,\dim\,\M$ on $\M$, with $m^\alpha(r)=0$.    Let us first formulate
 a condition that, if it could be satisfied, would suffice for proving that $\d\omega_{\h\I}=0$.  Then we will describe what happens instead.
 For any choice of the $m$'s,  there is a gauge field $A(x;m)$ that, in its dependence on $x\in M$, satisfies the instanton equation.  Up to gauge transformation,
 $A(x;m)$ is the solution of the instanton equation determined by  $m$.   Of course, $A(x;m)$ is only uniquely determined up to a gauge transformation,
 and this gauge transformation can depend on  $m$.   Suppose that we could make a gauge choice such that, at any point in $\M$ and for any choice
 of $\beta$, the quantity  $\frac{\partial A(x;m)}{\partial m^\beta}$ obeys the gauge condition that is used in defining the metric of $\M$:
 \be\label{gilby}W \frac{\partial A(x;m)}{\partial m^\beta}=0. \ee
 If so, then in the definition (\ref{tellmot}) of $\omega_{\h\I}$, we could interpret $\delta A$ as an explicit one-form on $\M$ (valued in one-forms on $M$), namely
 $\delta A=\sum_\beta \d m^\beta \frac{\partial A(x;m)}{\partial m^\beta}$.   Substituting this in eqn. (\ref{tellmot}), we would get an explicit formula for $\omega_{\h\I}$ as
 a two-form on $\M$, namely
 \be\label{wellmot}\omega_{\h\I}\overset{?}{=}-\frac{1}{8\pi^2}\sum_{\alpha,\beta}\d m^\alpha \,\d m^\beta \int_M\omega_{\I} \wedge
 \,\tr\,\frac{\partial A(x;m)}{\partial m^\alpha}\wedge \frac{\partial A(x;m)}{\partial m^\beta}.   \ee
 Then using $\d=\sum_\gamma \d m^\gamma\partial_{m^\gamma}$, we would get
  \be\label{wzellmot}\d\omega_{\h\I}\overset{?}{=}-\frac{1}{4\pi^2}\sum_{\alpha,\beta}\d m^\alpha \,\d m^\beta \d m^\gamma
  \int_M\omega_{\I} \wedge\tr\,\frac{\partial^2 A(x;m)}{\partial m^\alpha\partial m^\gamma}\wedge \frac{\partial A(x;m)}{\partial m^\beta}=0, \ee
  where the vanishing holds because $\frac{\partial^2 A(x;m)}{\partial m^\alpha\partial m^\gamma}$ is symmetric in $\alpha$ and $\gamma$.  
  
  However, eqn. (\ref{gilby}) is unrealistic.  It represents $\dim\,\M$ gauge conditions, one for each choice of $\beta$, while in actuality at each point in $M\times \M$
  we are only entitled to impose one gauge condition.   An example of a gauge condition that is not too restrictive is
  \be\label{gurv}\sum_i\left[\frac{\partial}{\partial x^i}+A_i(x;0), A^i(x;m)\right]+\star(H\wedge (A(x;m) -A(x;0)))=0  , \ee
  where $A(x;0)$ is simply $A(x;m)$ evaluated at $m=0$.   The condition (\ref{gurv}) is trivial at $m=0$, so it places
  no gauge condition on $A(x;0)$.   To first order in $m$, eqn. (\ref{gurv}) says that 
 $\left.\frac{\partial A(x;m)}{\partial m^\alpha}\right|_{m=0}$ obeys the gauge condition (\ref{zomely}) that was used in  
defining $\omega_{\h\I}$ as a differential form on $\M$.   Therefore, assuming that $A(x;m)$ satisfies eqn. (\ref{gurv}), eqn. (\ref{wellmot}) is actually
correct at $m=0$ (but only at $m=0$, as we will see).    The analysis in section \ref{hypercomplex} implies that the gauge condition (\ref{gurv}) is a valid one to first order in $m$, and this argument
can be slightly extended to show that this gauge condition is a good one to all orders in $m$ (we will only need the result in second order).  

Since it is not possible to satisfy eqn. (\ref{gilby}) beyond first order in $m$, 
for the purpose of evaluating $\omega_{\h\I}$, the vector field $\frac{\partial}  {\partial m^\beta}$ cannot be taken to act on $A$ in the obvious way
$\delta A = \frac{\partial}{\partial m^\beta}A$.   We have to accompany this with a gauge transformation chosen to ensure that $\delta A$ obeys the desired gauge condition. 
Thus we define
\be\label{defme}\frac{D A}{D m^\beta}=\frac{\partial A}{\partial m^\beta}-\d_A\varepsilon_\beta, \ee
where $\varepsilon_\beta$ is the infinitesimal generator of a gauge transformation chosen to satisfy the gauge condition $W \frac{D A}{D m^\beta}=0$, that is
\be\label{homme} \left[\frac{\partial}{\partial x^i}+A_i(x;m),\frac{\partial A^i}{\partial m^\beta}-D^i\varepsilon_\beta\right] +\star\left(H\wedge \left(\frac{\partial A}{\partial m^\beta}-\d_A\varepsilon_\beta\right)\right)=0.\ee  $\frac{D A}{D m}$ is defined intrinsically as the variation of
           $A$ under a change in $m$, accompanied if necessary by a gauge transformation to ensure that the gauge condition is satisfied.
           
We can put eqn. (\ref{homme})  in a more convenient form by subtracting the  condition (\ref{gurv}) that we  assume is satisfied for all $m$.  The result can be conveniently written:
\be\label{zomme} \left[A_i(x;m)-A_i(x;0),\frac{\partial A^i}{\partial m^\beta}\right] -W \varepsilon_\beta=0. \ee 
Near $m=0$, we have
\be\label{hopefun} A(x;m)=A(x;0) + \sum_\alpha m^\alpha \left.\frac{\partial A(x;m)}{\partial m^\alpha}\right|_{m=0}+\O(m^2). \ee
Eqn. (\ref{zomme}) then tells us that
\be\label{pomme} W\varepsilon_\beta = \sum_\alpha m^\alpha \left[\frac{\partial A_i}{\partial m^\alpha},\frac{\partial A^i}{\partial m^\beta}\right]_{m=0}+\O(m^2). \ee
We can solve this for $\varepsilon_\beta$:
\be\label{lomme} \varepsilon_\beta(x)=-\sum_\alpha m^\alpha \int_{M_y} \tr_y\,  B(x,y) \left[\frac{\partial A_i(y;m)}{\partial m^\alpha},\frac{\partial A^i(y;m)}
{\partial m^\beta}\right]\d^4y\sqrt{g_y}+
\O(m^2), \ee
where $B(x,y)$, introduced in eqn. (\ref{mimo}), is the Green's function of the operator $W$, $M_y$ is a oopy of $M$ parametrized by $y$,
and $\tr_y$ represents a trace in the $y$ variable.  Here and in subsequent formulas, $\frac{\partial A}{\partial m}$ is evaluated at $m=0$.  We will
sometimes save space by writing just $A_i$ or $A_i(y)$ rather than
$A_i(y;m)$, and similarly for other quantities.

It is now straightforward to compute $\d\omega_{\h\I}$.  In eqn. (\ref{wellmot}), we just have to replace $\frac{\partial A}{\partial m}$, which in general does not satisfy the
gauge condition,  with $\frac{D A}{D m}$, which does.  Using eqns. (\ref{defme})
and (\ref{lomme}), we can make this quite explicit up to $\O(m^2)$:
\begin{align}\label{ugone} \omega&_{\h\I}=  -\frac{1}{8\pi^2}\sum_{\alpha,\beta}\d m^\alpha \,\d m^\beta \left(\int_M\omega_{\I}\wedge  \,\tr\,\frac{\partial A(x)}{\partial m^\alpha}\wedge \frac{\partial A(x)}{\partial m^\beta}\right.
\cr              &  \left. -2\sum_\gamma m^\gamma   \int_{M_x\times M_y} \omega_{\I}(x) \wedge\,\tr\otimes \tr \,\frac{\partial A(x)}{\partial m^\alpha}\wedge \d_A  B(x,y)  \left[\frac{\partial A^i(y)}{\partial m^\gamma},
              \frac{\partial A_i(y)}{\partial m^\beta}\right] \d^4y\sqrt{g_y}   \right) \cr&    +\O(m^2).                                                                                                                                                                      \end{align} 
Now we can compute $\left.\d\omega_{\h\I}\right|_{m=0},$ just by acting with $\d=\sum_\gamma \d m^\gamma\partial_{m^\gamma}$ and then setting $m=0$:
\begin{align}\label{unuc}\d&\omega_{\h\I}\left.\right|_{m=0}=-\frac{1}{4\pi^2} \sum_{\alpha,\beta,\gamma} \d m^\alpha\d m^\beta \d m^\gamma\times \cr &\int_{M_x\times M_y}\omega_{\I}(x) \wedge\,\tr\otimes \tr \,\frac{\partial A(x)}{\partial m^\alpha}\wedge \d_A  B(x,y) \left[\frac{\partial A^i(y)} {\partial m^\gamma},
              \frac {\partial A_i(y)}   {\partial m^\beta}\right] \d^4y\sqrt{g_y}     .\end{align}
 This can be simplified significantly.   First, integrate by parts in the $x$ variable, using the linearized instanton equation
 $\omega_\I\wedge \d_A\frac{\partial A(x;m)}{\partial m^\alpha}  =0$.  This gives  
   \begin{align}\label{zunuc}\hskip-.5cm&\d\omega_{\h\I}\left.\right|_{m=0}=\sum_{\alpha,\beta,\gamma} \d m^\alpha\d m^\beta \d m^\gamma\times \cr &\frac{1}{4\pi^2} \int_{M_x\times M_y}\d\omega_{\I}(x) \wedge\,\tr\otimes \tr \,\frac{\partial A(x)}{\partial m^\alpha}  B(x,y) \left[\frac{\partial A^i(y)}{\partial m^\gamma},
              \frac{\partial A_i(y)}{\partial m^\beta}\right]  \d^4y\sqrt{g_y}    .\end{align} 
           At $m=0$, $\frac{D A}{D m^\alpha}$ coincides with $\frac{\partial A}{\partial m\alpha}$, so eqn.  (\ref{zunuc}) remains valid at $m=0$ if we replace $\frac{\partial A}{\partial m\alpha}$,
           everywhere with $\frac{D A}{D m^\alpha}$.   But once we make that replacement, the formula is valid for all $m$, not just at $m=0$.  To prove the formula at some given
           value of $m$, we just do the same computation as before, expanding about that value of $m$ rather than about $m=0$.  
        Finally, the formula can be written more economically  in terms of 
             \be\label{ponim}\psi=\sum_\alpha \d m^\alpha  \frac{D A}{Dm^\alpha}.\ee
 The final version of the formula is 
    \be\label{wunuc} \d\omega_{\h\I} =\frac{1}{4\pi^2}\int_{M_x\times M_y}\d\omega_{\I}\wedge \tr\otimes \tr \,\psi(x)  B(x,y) [\psi_i(y),\psi^i(y)]\d^4 y\sqrt{g_y} . \ee

Now it is straightforward to determine the torsion $\h H=-\h\I\d\omega_{\h\I}$, where the action of an almost complex structure on differential forms was defined in eqn. (\ref{bozo}). 
To compute $\h\I\d\omega_{\h\I}$, we  just have to substitute $\psi\to \h\I\psi$  for each of the three occurrences of $\psi$ in eqn. (\ref{wunuc}).   As
$\h\I$ acts by \be\label{doco}\psi^{(0,1)}\to \i \psi^{(0,1)},~~\psi^{(1,0)}\to -\i \psi^{(1,0)},\ee
 and $[\psi_i,\psi^i]$ is of type $(1,1)$, we have $[\h\I\psi_i,\h\I\psi^i]=[\psi_i,\psi^i]$, 
so we only need to consider the action of $\h\I$ on $\d\omega_{\I} \wedge \psi$.    As \be\label{podo}\d\omega_{\I}\wedge\psi=(\d\omega_{\I})^{(2,1)}\wedge \psi^{(0,1)}
+(\d\omega_{\I}^{(1,2)})\wedge \psi^{(1,0)},\ee we have
 \be\label{zopo}\d\omega\wedge \h\I\psi= \i(\d\omega_{\I})^{(2,1)}\wedge \psi^{(0,1)}
-\i(\d\omega_{\I}^{(1,2)})\wedge \psi^{(1,0)}=(\I\d\omega)\wedge\psi=-H\wedge \psi.\ee
So
\be\label{hunuc}\h H=-\h\I\d\omega_{\h\I}=\frac{1}{4\pi^2}\int_{M\times M}H\wedge \tr\otimes \tr\, \psi(x) B(x,y) [\psi_i(y),\psi^i(y)]\d^4y \sqrt{g_y}. \ee

 Now we can verify the three key points that were mentioned at the outset of this section.  Point (i)  is immediate: eqn. (\ref{hunuc}) makes it evident
 that $\h H$ depends only on $H$ and not on $\omega_\I$.   Point (ii) is the assertion that if $\d H=0$, then $\d \h H=0$.   Indeed, in the language of section \ref{systematic},
 and bearing in mind the descent formulas (\ref{descent}) and the formula (\ref{delfox}) for $\sigma$,
 the formula (\ref{hunuc}) can be written
 \be\label{zuhuc} \h H =\int_M H\wedge P^{(1)}. \ee
 So following the logic of section \ref{systematic}, 
 we have
 \be\label{pluhuc}\d \h H=\int_M H \wedge \delta P^{(1)} =-\int_M H\wedge \d P^{(0)}=0,\ee
 where the vanishing in the last step follows by  integrating by parts and using $\d H=0$.   Similarly,  
 the cohomology class of $\h H$ only depends on the
 cohomology class of $H$, since if we change $H$ by $H\to H+\d B$, then the change in $\h H$ is
 \be\label{luhuc}\Delta \h H =\int_M \d B\wedge P^{(1)}=-\int_M B\wedge \d P^{(1)}=\delta \int_M B\wedge P^{(2)},\ee
 and thus is exact.   Finally, concerning point (iii),  the formula (\ref{hunuc}) shows that reversing the sign of $H$ reverses the sign of $\h H$.
 
 Properties (i) and (ii) show that if $M$ has a (strong) HKT structure than so does $\M$.   The addition of property (iii) tells us that if $M$ has a (strong) bi-HKT structure,
 or equivalently a (twisted) generalized hyper-Kahler structure, with all of the hermitian forms being selfdual, then so does $\M$.   The relation of $P^{(1)}$ to the second Chern
 class implies that periods of $\h H$ are integer multiples of periods of $H$, and therefore that $\h H$ obeys Dirac quantization if $H$ does.
 
 Of course, we are mainly interested in the case that $M=\S^3\times \S^1$, with $H$ the pullback from $\S^3$ of a rotationally invariant form with total flux $2\pi Q_5'$.
 In other words, $H=2\pi Q_5'\d^3\Omega$, where $\d^3\Omega$ is a rotationally invariant volume form on $\S^3$ of total volume 1.
 In this case
 \begin{align}\label{punuc} \h H&=
 \frac{Q_5'}{2\pi}\int_{M\times M} \d^3\Omega_x \wedge \tr\otimes \tr \,\delta A(x) B(x,y) [\delta A_i(y),\delta A^i(y)]\d^4y \sqrt{g_y}
 \cr &=\frac{Q_5'}{2\pi}\int_{M\times M} \d^3\Omega_x \wedge \tr\otimes \tr \,\psi(x) B(x,y) [\psi_i(y),\psi^i(y)]\d^4y\sqrt{g_y},\end{align}
 in agreement with eqn. (\ref{iminy}).
  
Finally, we can describe the usual mathematical setting for these formulas and in the process explain claims that were made at the end of section \ref{systematic}.
By definition of the instanton moduli space $\M$, each point $m\in \M$ parametrizes a $\GG$-bundle $E_m\to M$ that carries an instanton connection $A$.
As $m$ varies in $\M$, the $E_m$ vary as fibers of a $\GG$-bundle $\h E\to M\times \M$ (sometimes called the universal instanton bundle).  $\h E$ has
a natural connection in the $M$ direction, namely the instanton connection itself, but it does not have a natural connection in the $\M$ direction.  To define a connection
on $\h E$ in the $\M$ direction, one has to make some sort of a choice.   We made such a choice by placing the gauge condition $W\delta A=0$ on variations of $A$.
With the aid of this choice, we defined in eqns. (\ref{defme}) and (\ref{lomme}) a connection $\frac{D}{D m^\alpha}=\frac{\partial}{\partial m^\alpha}+[\varepsilon_\alpha,\cdot]$ 
on the bundle $\h E$ in the $\M$ direction.   
Taken together,  the original instanton connection $\frac{D}{Dx^i}=\frac{\partial}{\partial x^i}+[A_i,\cdot ]$ in the $M$ direction and the connection $\frac{D}{D m^\alpha}=\frac{\partial}{\partial m^\alpha}+[\epsilon_\alpha,\cdot]$ 
in the $\M$ direction give a full-fledged connection $\h A$
on $\h E\to M\times \M$.   This connection has a curvature $\h F =\d \h A+\h A\wedge \h A$, and a second Chern class, valued in $H^4(M\times \M;\Z)$, that at the level
of differential forms is represented by the four-form $\frac{1}{8 \pi^2}\tr\,\h F\wedge \h F$.   The curvature $\h F$ has components of types $ij$, $\alpha i,$ and $\alpha\beta$
(where indices $i,j$ are tangent to $M$ and indices $\alpha,\beta$ are tangent to $\M$).  The $ij$ part of $\h F$ is simply the curvature $F_{ij}=[D_i,D_j]$ of
the original instanton bundle.  The $\alpha i $ part of $\h F$  is $[D_\alpha, D_i]= \frac{\partial A_i}{\partial m^\alpha} -D_i \epsilon_\alpha$, a quantity that was introduced in eqn.
(\ref{defme}).   The corresponding part of the curvature two-form $\h F$ is  $\sum_{\alpha i}\d m^\alpha \d x^i[D_\alpha,D_i]$.   This is the $(1,1)$-form on $M\times \M$ (a two-form on $M\times \M$ 
with one index tangent to each factor) that was called $\psi$ in section \ref{systematic}. Indeed, $\psi$ was characterized as a general variation of $A$ that obeys the gauge condition (as well as the linearized instanton equation),
or in other words as $\sum_{\alpha i}\d m^\alpha \delta_\alpha A_i \,\d x^i$, where the variations
$\delta _\alpha A_i$ are constrained to be annihilated by the operator $W$; these variations are precisely what we now call $[D_\alpha, D_i]$.   Finally, the $\alpha\beta$ part of the curvature is $[D_\alpha,D_\beta]$.   In eqn. (\ref{lomme}),
we determined $\epsilon_\alpha$ in an expansion around a base-point at $m=0$ and in a gauge with $\left.\epsilon_\alpha\right|_{m=0}=0$.   In that gauge, 
the $\alpha\beta$ part of the curvature reduces at $m=0$ to $\h F_{\alpha\beta}=\partial_\alpha \epsilon_\beta-\partial_\beta \epsilon_\alpha$, which is a quantity that we evaluated as a step in the computation of
$\d\omega_{\h \I}$.  The result can be stated 
\be\label{doofus}\frac{1}{2}\sum_{\alpha\beta}\d m^\alpha \d m^\beta \h F_{\alpha\beta}(x)=-\sum_{\alpha\beta} \d m^\alpha \d m^\beta\int_{M_y} \tr_y\,  B(x,y) \left[\frac{\partial A_i(y)}{\partial m^\alpha},\frac{\partial A^i(y)}
{\partial m^\beta}\right]\d^4y\sqrt{g_y}.\ee  
We derived this formula at $m=0$, but the formula becomes valid for all $m$ if  we replace $\d m^\alpha \frac{\partial A_i(y)}{\partial m^\alpha}$ with $\psi_i=\d m^\alpha F_{\alpha i}$
(which is a gauge-covariant expression that reduces to $\d m^\alpha \frac{\partial A_i(y)}{\partial m^\alpha}$ at $m=0$). 
After this replacement, eqn. (\ref{doofus}) agrees (up to sign) with the formula for $\sigma$ as a two-form on $\M$ that was deduced in eqn. (\ref{delfox}), confirming the interpretation of 
$\sigma$ as part of the curvature of a natural connection on the bundle $\h E\to M\times \M$.   This also confirms the interpretation that was claimed in section \ref{systematic} of
the observables $P^{(n)}$ defined via the descent procedure: they represent various components of  $\frac{1}{8 \pi^2}\tr\,\h F\wedge \h F$.  
  
 \section{Large $\N=4$ Algebra }\label{large}
 
 \subsection{Overview}\label{viewover}
 
 By now, we know that if $M$ is a generalized hyper-Kahler four-manifold, then similarly the  moduli space $\M$ of instantons on $M$ has a generalized hyper-Kahler
 structure, leading\footnote{If we assume conformal invariance, which is discussed more critically in section \ref{conformal}.} 
  to $\N=4$ superconformal symmetry with the small $\N=4$ algebra.   Our next task is to show that,
in the specific case of the instanton moduli space on $\S^3\times \S^1$,  the small $\N=4$ algebras actually are extended to large ones.
  For this, we need to find new holomorphic and antiholomorphic fields
in the sigma-model with target $\M$.   From the discussion of eqns. (\ref{constkilling}) and (\ref{freej}), we know where to find them: vector
fields on $\M$ that are covariantly constant for the connection $\h\nabla$ or $\h\nabla'$ on $\M$ will lead to holomorphic or antiholomorphic currents and free fermions
in the sigma-model, such as  we need.

Let us discuss what it means for a vector field to be covariantly constant for one of these connections.  In general, on a (strong) HKT manifold with connection $\nabla $ and torsion $H$,
the condition for a vector field $V$ to satisfy $\nabla_R V_S=0$ amounts to the following.  The part of the equation that is symmetric in $R$ and $S$ is
\be\label{zoldo} D_R V_S+D_S V_R=0,\ee
saying that $V$ is a Killing vector field.
The antisymmetric part of the equation is
\be\label{oldo}\partial_R V_S-\partial_S V_R=H_{RST} V^T. \ee
This can equivalently be written
\be\label{noldo}\d \Lambda=\iota_V H\ee
where $\iota_V$ is the operation of contracting a differential form with the vector field $V$, and the left hand side is the exterior derivative of the one-form
$\Lambda=\d X^I G_{IJ} V^J$, which we call the one-form dual to $V$.

Eqn. (\ref{noldo})  implies in particular that the Killing vector field $V$ generates a symmetry of $H$. Indeed, the change in a differential form $H$
under an infinitesimal diffeomorphism generated by $V$ is given by the Lie derivative ${\mathcal L}_V$ acting on $H$.  Concretely the definition of $\L_V $ is $\L_V H=\iota_V \d H +\d\iota_V H$.  In the present case,
$\d H=0$  and eqn. (\ref{noldo}) implies that $\d\iota_V H=0$, so $\L_V H=0$.   (Because $\Lambda$ is required to be dual to $V$,
this is not the full content of eqn.  (\ref{noldo}).) 

So  to extend the small $\N=4$ algebra of the sigma-model with target $\M$ to a large one, we need in particular 
Killing vector fields on $\M$ that are symmetries of the torsion
$\h H$ of $\M$.   This should come as no surprise, since extending the small $\N=4$ algebra to a large one involves finding a holomorphic
 $\SU(2)\times \U(1)$ current algebra,
and the obvious way to find one is to find Killing vector fields on $\M$ generating a suitable $\SU(2)\times \U(1)$ symmetry of the sigma-model.

$\S^3\times \S^1$ itself has Killing vector fields generating the group $(\SU(2)_\ell\times \SU(2)_r)/\Z_2\times \U(1)$ of isometries connected to the identity,
where $\SU(2)_\ell$ and $\SU(2)_r$  act on  $\S^3=\SU(2)$ on the left and right, and $\U(1)$ is the group of rotations of $\S^1$.
Moreover, these symmetries transform the complex structures of $\S^3\times \S^1$ in a particular way, described in section \ref{spherex}.
The $(\SU(2)_\ell\times \SU(2)_r)/\Z_2\times \U(1)$ symmetries of $\S^3\times \S^1$ are all symmetries of the instanton equation on $\S^3\times \S^1$,
so automatically they are symmetries\footnote{The group that acts on $\M$ may in general be an extension of
  $(\SU(2)_\ell\times \SU(2)_r)/\Z_2\times \U(1)$ by the center of the gauge group.}  of $\M$.   Moreover, because the complex structures  of $\M$
are directly inherited from the corresponding complex structures of $\S^3\times \S^1$,  $(\SU(2)_\ell\times \SU(2)_r)/\Z_2\times \U(1)$ transforms
the complex structures of $\M$ the same way that it transforms the complex structures of $\S^3\times \S^1$.

So to show that the sigma-model with target $\M$ has large $\N=4$ symmetry, all we need to show is that the Killing vector fields on $\M$
that generate the $\SU(2)_\ell\times \U(1)$ symmetry are covariantly constant for $\h\nabla$, and those that generate the $\SU(2)_r\times \U(1)$
symmetry are covariantly constant for $\h\nabla'$.    The statements about $\SU(2)_\ell\times \U(1)$ and the statements about $\SU(2)_r\times \U(1)$
are exchanged by the discrete symmetry $\rho$ of $\S^3\times \S^1$ that acts by reflection on each factor, so it suffices to analyze one of the two
cases.

Actually, $\S^3\times\S^1$ has only one $\U(1)$ symmetry, which contributes both 
a holomorphic and an                                                                                                                                                                                                                                                                                                                                                                                                                                                                                                                                                                                                                                                                                                                                                                                                                                                                                                                                                                                                                                                                                                                                                                                                                                                                                                                                                                                                                                                                                                                                                                                                                                                                                                                                                                                                                                                                                                                                         antiholomorphic $\U(1)$ current algebra. This is possible because the Killing vector field $V$ that generates this $\U(1)$ is covariantly constant for both $\nabla $ and $\nabla'$,
leading to both holomorphic and antiholomorphic conserved currents.   As we will see, something similar happens on the moduli space $\M$.

   In section \ref{abelian}, we analyze the Killing 
vector field $\h V$ on $\M$ that corresponds to the Killing vector field $V$ on $\S^3\times \S^1$.  We show that it is covariantly constant for both $\h\nabla$ and $\h\nabla'$,  implying that the sigma-model with target $\M$ has the expected holomorphic and
antiholomorphic $\U(1)$ current algebras and free fermions.    It also follows that this sigma-model has a scalar field that is free, at least locally.
In section \ref{nonabelian},  we show that the vector fields that generate the $\SU(2)_\ell$ symmetry of $\M$
are covariantly constant for $\h\nabla$, leading to the expected holomorphic $\SU(2)$ current algebra that is needed to complete the holomorphic  large $\N=4$ algebra.
The analogous statement for $\SU(2)_r$ symmetry and $\h\nabla'$ is an immediate consequence.
 In section \ref{finale} we compute, at least in the semiclassical limit of large $Q_5'$,
 the levels or central charges of the $\SU(2)$ current algebras that are related to the $\SU(2)_\ell$ and $\SU(2)_r$ symmetries.   
 We get the expected   results  $Q_1 Q_5$ and  $Q_1 Q_5'$, as predicted by the duality conjecture  described in the introduction.
 In section \ref{vectorfields}, we show that certain symmetry generators have no fixed point on $\M$. This is potentially relevant to supersymmetric localization 
 of the sigma-model with target $\M$.   In  section
 \ref{concluding}, we show that if $M$ is such that a sigma-model with target $M$ has a large $\N=4$ algebra only for  one chirality, then
 the same is true of $\M$.
  
 \subsection{Rotations of $\S^1$}\label{abelian}
 
 In general, a vector field $V$ on a four-manifold $M$ acts on gauge fields on $M$ by
 \be\label{simoc}\delta A_i = V^j F_{ji}-D_i\sigma,\ee
 where $\sigma$ is an arbitrary generator of a gauge transformation. If $V$ is a Killing vector field on $M$,
 then  eqn. (\ref{simoc}) describes in a gauge-invariant sense the corresponding Killing vector field $\h V$ on the instanton moduli space $\M$.
 However, to use eqn. (\ref{simoc}) together with the differential geometric formulas of sections \ref{hypercomplex} and \ref{hkt}, we must
 pick $\sigma$ so that $\delta A_i$ satisfies the gauge condition $W\delta A=0$.    Actually, there are two gauge conditions of interest, differing in the sign of $H$.
  
Here we consider  the vector field $V=\frac{\partial}{\partial \tau}$ that acts on $\S^3\times \S^1$ by rotating the second factor.
 It turns out that in this specific case,  the relevant gauge conditions are both satisfied with $\sigma=0$.
With $\sigma$ assumed to vanish, eqn. (\ref{simoc}) becomes
\begin{align}\label{zimoc} \delta A_i&= F_{\tau i}, ~~~i=1,\cdots, 3 \cr
                                                      \delta A_\tau& = 0. \end{align}
The relevant gauge conditions are $D_i \delta A^i \pm\star (H\wedge\delta A)=0$.   In the particular case of the $\U(1)$ generator $V$, we want
this equation to be satisfied for each choice of sign, since we expect to get both a holomorphic and an antiholomorphic $\U(1)$ current algebra.
Indeed, $H\wedge \delta A=0$ because $H$ is a pullback from $\S^3$ and $\delta A_\tau=0$, and $D_i \delta A^i=0$ because the instanton equation $F^+=0$ implies
the second order Yang-Mills equation $D_i F^{ij}=0$, which for $j=\tau$ gives $D_i F^{i\tau}=0$.    

% We note that a more abstract way to write eqn. (\ref{zimoc})$ is
%to view $\delta A_i$ and $\delta A_\tau$ as one-forms on the space of connections, and then to express 

We expect to show that $\h V$ is covariantly constant for both connections $\h\nabla$ and $\h\nabla'$ on $\M$.  This means that eqn. (\ref{oldo}) or (\ref{noldo}) must hold
with either sign of the torsion $\h H$, so we really need to establish two conditions
\be\label{dono}\d \h \Lambda = 0=\iota_{\h V}\h H, \ee
where $\h\Lambda$ is the one-form dual to $\h V$.   

The first of the two desired relations is almost immediate.   From  the definition (\ref{ibo}) of the metric of $\M$ and the formula
(\ref{zimoc}) for $\delta A$ (which we are entitled to use since we have verified that $\delta A$ as defined in this formula  satisfies the gauge condition that
was assumed in defining the metric),  it follows that the one-dual dual to $\h V$ is
\be\label{zono}\h\Lambda=-\frac{1}{8\pi^2} \int_M\d^4 x \sqrt g   \sum_i \tr\, F_{\tau i}\delta A^i =-\frac{1}{8\pi^2}\int_M \tr \,F\wedge \delta A \wedge \d\tau,\ee
where the instanton equation was used in the last step.   Up to a constant multiple, $\h\Lambda$ is the basic example discussed in eqn. (\ref{unef})
of a one-form on $\M$ that is closed but not exact.   
 In particular, $\h\Lambda$ is closed, as desired.

To verify the second relation $\iota_{\h V} \h H=0$ requires a more detailed analysis.   We use the formula (\ref{punuc}) for $\h H$.   We note that in this formula,
$\d^3\Omega\, \psi=\d^3\Omega\,\delta A$ 
can be replaced by $\d^3\Omega \,\delta A_\tau \d\tau$, since the part of $\delta A$ tangent to the first factor of $\S^3\times \S^1$ does not contribute
when multiplied by the volume form $\d^3\Omega$ of $\S^3$.  So as $\iota_{\h V} \delta A_\tau=0$ according to eqn. (\ref{zimoc}), we have
$\iota_{\h V}(\d^3\Omega \,\delta A)=0$.
   Hence  in computing $\iota_{\h V}\h H$, we only have to contract
$\h V$ with $[\delta A_i,\delta A^i]$, giving 
\be\label{westful} \iota_{\h V}\h H= \frac{Q_5'}{\pi}\int_{M\times M} \d^3\Omega_x\, \d\tau \, \tr\otimes \tr \,\delta A_\tau(x) B(x,y) [\delta A^i(y),F_{\tau i}]\,\d^4y\sqrt{g_y}.\ee
To proceed farther, we need to study the second order differential equation obeyed by $\delta A$.   This equation can be efficiently found by first observing
that an instanton solution certainly obeys the Yang-Mills equation $D^i F_{ij}=0$.   The variation of this equation is 
\be\label{ibox} D^i (D_i\delta A_j- D_j \delta A_i)+[\delta A^i,F_{ij}]=0.\ee 
To simplify this, we use
\be\label{inzo} D_i D_j \delta A^i= D_j D_i \delta A^i +[D_i,D_j]\delta A^i=[F_{ij},\delta A^i]+ R_{jk} \delta A^k -D_j\star(H\wedge \delta A),\ee
where $R_{jk}$ is the Ricci tensor of $\S^3\times \S^1$ and we used the gauge condition (\ref{omely}) satisfied by $\delta A$.   Setting $j=\tau $, we have $R_{\tau k}=0$ for $\S^3\times\S^1$,
so the term
involving the Ricci tensor drops out.  Since $\delta A_\tau$ is the only component of $\delta A$ that contributes to $H\wedge \delta A$, and $D_\tau$ commutes with $H$, for $j=\tau$ we 
have $D_j \star (H\wedge \delta A)=\star (H\wedge \d_A\delta A_\tau)$ and (\ref{inzo}) becomes $D_i D_\tau\delta A^i =[F_{i\tau},\delta A^i]-\star [H\wedge \d_A \delta A_\tau]$.   
Substituting this in eqn. (\ref{ibox}), we get $D^i D_i\delta A_\tau+2[\delta A^i,F_{i\tau }]+\star [H\wedge D_\tau\delta A_\tau]=0$ or in other words
  \be\label{whoknows}   [ \delta A^i,F_{\tau i}]= \frac{1} {2}W \delta A_\tau,\ee 
 where $W$ is the operator 
 \be\label{thatone}W\phi=D_i D^i \phi+\star [H\wedge \d_A\phi] \ee
 whose Green's function $B(x,y)$ appears in the formula (\ref{westful}).   
 So eqn. (\ref{westful}) can be written 
 \be\label{zestful} \iota_{\h V}\h H= \frac{Q_5'}{2\pi}\int_{M\times M} \d^3\Omega_x\, \d\tau \, \tr\otimes \tr \,\delta A_\tau(x) B(x,y) W_y\delta A_\tau(y)\d^4y \sqrt{g_y}.\ee   
 Integrating by parts and using eqn. (\ref{zimox}) for $W_y^\dagger B(x,y)$, this simplifies to
  \be\label{estful} \iota_{\h V}\h H= \frac{Q_5'}{2\pi}\int_{M} \d^3\Omega_x\, \d\tau\,  \tr \,\delta A_\tau(x)\delta A_\tau(x)=0.\ee
  The vanishing in the last step holds by fermi statistics, since $\delta A_\tau(x)=\psi_\tau(x)$ is a fermionic object -- a scalar function on $M$ valued in one-forms on $\M$ --
  while $\tr\,\delta A_\tau(x)\delta A_\tau(x)$ is symmetric in the two factors of $\delta A_\tau(x)$.
 
 This completes the proof that the vector field $\h V$ is covariantly constant for both connections $\nabla $ and $\nabla'$. 
Therefore, the holomorphic and antiholomorphic chiral algebras of the sigma-model with target $\M$ will contain, at a minimum,
a $\U(1)$ current algebra and a free fermion, beyond the small $\N=4$ algebra that is guaranteed by the generalized hyper-Kahler structure.

  Since $\h V$ is a Killing vector field, it satisfies $D_i\h V_j+D_j\h V_i=0$, and since the dual one-form is closed, we have also $D_i \h V_j-D_j\h V_i=0$.
  So $D_i\h V_j=0$ and $\h V$ is covariantly constant for the Riemannian metric of $\M$.   In Riemannian geometry, if a compact Riemannian $\M$ has a nonzero vector
  field $\h V$ that is covariantly constant, then locally $\M=\M'\times \S^1$ where the two factors are orthogonal and $\h V$ acts by rotating the $\S^1$.   In general
  this may be true only locally and globally one may have $\M=(\M'\times \S^1)/\Z_n$ for some $n\geq 2$, where $\Z_n$ acts by a $2\pi/n$ rotation of $\S^1$ and by 
  some symmetry of order $n$ of $\M'$.   In the present context, we also have a torsion field $\h H$ on $\M$; since $\iota_{\h V}\h H=\L_{\h V}\h H=0$, $\h H$ is a pullback from $\M'$.
  This means that if $\varphi$ is an angle-valued field that parametrizes the $\S^1$, then $\varphi$ is decoupled from the $B$-field of $\M$ (in some gauge, at least locally,
  the $B$-field is independent of $\varphi$ and has no component in the $\varphi$ direction).  So $\varphi$ is a free field, and the holomorphic and antiholomorphic current algebras
  whose existence we have deduced are generated by the holomorphic and antiholomorphic conserved currents $\partial\varphi$ and $\bar\partial\varphi$.
  
  Since $\h V$ is covariantly constant, its length squared $|\h V|^2$  is constant.   We can now easily determine the constant value.  In this
  calculation,  we  use the definition (\ref{ibo}) of the metric of $\M$.  If we take the metric of $\S^3\times \S^1$ to be the usual 
  $\d\Omega^2+\d\tau^2$, then with $\h V$ as described in eqn. (\ref{zimoc}), we get
    \be\label{mivo} |\h V|_0^2=-\frac{1}{4\pi^2} \int_{\S^3\times \S^1}\d^4x\sqrt g \,\tr\,F_{i\tau} F^{i\tau}=-\frac{1}{8\pi^2}\int_{\S^3\times \S^1}\tr \,F\wedge F
  =Q_1,\ee
  where  we used the instanton equation and $Q_1$ is the instanton number.

  The reason for the subscript in $|\h V|_0^2$ in eqn. (\ref{mivo}) is that we actually want to use a different normalization for this calculation.
In sections \ref{nonabelian} and \ref{finale}, we will 
 use the value of $|\h V|^2$ as an ingredient in determining one of the central charges of the large $\N=4$ algebra.
  For this purpose, bearing in mind the brane construction described in section \ref{motivation} that motivates the duality,
  it is important to normalize the metric of $\S^3\times \S^1$ so that $\int_{\S^3}H=2\pi Q_5'$.   According to eqn. (\ref{inoc}), we can ensure
  this by taking the metric of $\S^3\times \S^1$ to be $\frac{Q_5'}{2\pi}\left(\d\Omega^2+\d\tau^2\right)$.  This rescaling of the metric of $\S^3\times\S^1$  
  does not affect the instanton equation, but according to eqn. (\ref{ibo}), it does multiply the metric of $\M$ by $\frac{Q_5'}{2\pi}$.  
Therefore with this normalization, we have 
 \be\label{nivo} |\h V|^2=\frac{Q_1 Q_5'}{2\pi}.\ee
 
 \subsection{Rotations of $\S^3$}\label{nonabelian}
 
Now we consider the case of a vector field that is a generator of, say, the right action $\SU(2)_r$ of $\S^3=\SU(2)$ on itself.
$\SU(2)_r$ leaves fixed the right-invariant one-forms $R^i$ that were introduced in section \ref{spherex}.  On the other hand, the left-invariant one-forms $L^i$ transform
in the adjoint representation of $\SU(2)_r$.

It is convenient to introduce a basis $T_1,\cdots, T_3$ of the Lie algebra of $\SU(2)_r$, normalized so that the Lie algebra takes the canonical form
\be\label{canonical} [T_a,T_b]=\epsilon_{abc} T_c,~~~~a,b,c=1,\cdots, 3. \ee
In the notation of section \ref{spherex}, we  can take
\be\label{wetake}T_1=-\frac{1}{2}\left( y_0 \frac{\partial}{\partial y_1}-y_1 \frac{\partial}{\partial y_0}+y_2 \frac{\partial}{\partial y_3}-y_3 \frac{\partial}{\partial y_2}\right),\ee
with $T_2$ and $T_3$ obtained by cyclic permutations of indices $1,2,3$.
 The one-form dual to  the vector field $T_a$ is  a multiple of $L_a$:
 \be\label{unux}T_a^i G_{ij} \d x^j=-\frac{1}{2} L_a.\ee
 We recall that the metric $G$ and the one-forms $L_a$ are covariantly constant for the connection $\nabla$ on $M$. 
 From eqn. (\ref{unux}), it follows  that also the $T_a$ are covariantly constant for this connection.
 
 Since $T_a$ generates a symmetry of the instanton equation on $\S^3\times \S^1$, it corresponds to a Killing vector field $\h T_a$ on $\M$. The $\h T_a$ generate an $\SU(2)$
 symmetry of $\M$ since the $T_a$ generate an $\SU(2)$ symmetry of $\S^3\times \S^1$. The complex structures $\h\I_a$ of $\M$ transform in the adjoint representation
 of this $\SU(2)$ since the complex structures $\I_a$ of $\S^3\times \S^1$, from which the $\h\I_a$ are deduced by restriction from $T\A$ to $T\M$, transform in this representation.
 
  In a moment, we will show
  that $\h T_a$ is covariantly constant for the conection $\h\nabla$ on $\M$ that corresponds to the connection $\nabla$ on $M$.
 As explained in eqns. (\ref{constkilling}) and (\ref{freej}), this implies that associated to the $\h T_a$ are holomorphic currents with associated free fermions.
Being holomorphic, the $\h T_a$ will generate not just a global action of $\SU(2)_r$ but   a holomorphic $\SU(2)_r$ current algebra.  
Together with the free fermions and  $\U(1)$ modes discussed in section \ref{abelian},
 these holomorphic fields will extend the holomorphic chiral algebra of the $\sigma$-model with target $\M$  from a small $\N=4$ algebra to a large one.  
 
Actually, there is a trivial way to find vector fields on $\M$ that are covariantly constant for the connection $\h\nabla$ and that, like the $\h T_a$, transform in the adjoint
representation of $\SU(2)_r$.   The complex structures $\h \I_a$, $a=1,2,3$, are covariantly constant with respect to $\h\nabla$, and we showed in \ref{abelian} that the $\U(1)$
generator $\h V$ is covariantly constant with respect to $\h\nabla$ (as well as $\h\nabla'$).   So the vector fields $\h\I_a \h V$ are covariantly constant with respect to $\h\nabla$.
We claim that
\be\label{zippo} \h T_a=-\frac{1}{2}\h\I_a \h V, \ee so $\h T_a$ is likewise covariantly constant. This
 will complete the proof that the currents associated to the $\h T_a$ are holomorphic on the sigma-model worldsheet and generate an $\SU(2)_r$ current algebra.\footnote{Without knowing eqn. (\ref{zippo}), we would
still know that, as the vector fields $\h\I_a\h V$ are covariantly constant with respect to $\h\nabla$, they are associated to holomorphic currents
that transform in the adjoint representation of an $\SU(2)$ symmetry group.  We would not know that they are actually the $\SU(2)$ generators.}

First, let us verify the statement on $\S^3\times \S^1$ that is analogous to (\ref{zippo}), namely
\be\label{boldo}T_a=-\frac{1}{2} \I_a V. \ee
From the definition (\ref{infox}) of the $\I_a$, we have $\I_a L_0=L_a$.  $V$ is the vector field dual to $L_0$ and $T_a$ is $-\frac{1}{2}$ times the vector field dual to $L_a$, so
eqn. (\ref{boldo}) is valid.

The corresponding statement on $\M$, namely $\h T_a=-\frac{1}{2}\h\I_a\h V$, is actually an immediate consequence.   First of all, from the statement $T_a=-\frac{1}{2} \I_a V$  on
$\S^3\times \S^1$, it immediately 
follows
 that the corresponding statement $\h T_a=-\frac{1}{2}\h\I_a\h V$ is valid on the space $\A$ of all connections.   To deduce from that a statement
on the moduli space $\M$, we have to take into account that in general the action of a vector field or a complex structure on a tangent vector to $\M$ cannot be trivially deduced from
the action on $\A$; the natural formula in general 
 must be accompanied by a gauge transformation.   However, the gauge condition that we used in embedding the tangent space to $\M$ in that to $\A$ was
chosen to be invariant under the action of the $\h\I_a$, so the natural action of $\h\I_a$ does not need to be accompanied by a gauge transformation.  And we showed in 
section \ref{abelian} that the same is true for $\h V$.   So the formula $\h T_a=-\frac{1}{2}\h\I_a\h V$ can be just restricted from the tangent space to $\A$ to the tangent space to $\M$.
As a bonus, this argument also shows that the naive action of $\h T_a $ by $\delta A_j=T_a^i F_{ij}$ satisfies the gauge condition, with no accompanying gauge transformation needed.

In section \ref{finale}, to  compute the level of the $\SU(2)_r$ current algebra, we will need to know the length squared $|\h T_a|^2 =\h T_a^i G_{ij} \h T_a^j$ of the vector fields
$\h T_a$.   The relation $\h T_a=-\frac{1}{2} \h\I_a\h V$  gives an easy way to do this.   As a complex structure,
$\h\I_a$ is in particular a length-preserving linear transformation of the tangent bundle of $\M$.  So the relation $\h T_a=-\frac{1}{2}\h\I_a \h V $ implies that $|\h T_a|^2 =\frac{1}{4} |\h V|^2$.
In eqn. (\ref{nivo}), we showed that   $|\h V|^2=\frac{Q_1 Q_5'}{2\pi}$, so
\be\label{muskat} |\h T_a|^2 =\frac{Q_1 Q_5'}{8\pi}. \ee

By now, we have shown that the sigma-model with target $\M$ has a holomorphic $\SU(2)\times \U(1)$ current algebra, extending the small $\N=4$ holomorphic superconformal
algebra to a large one.  A similar analysis shows a similar enhancement of the antiholomorphic chiral algebra of the sigma-model.  Indeed, the two analyses are
exchanged by the discrete symmetry of $\S^3\times \S^1$ that acts as a joint reflection on each factor.

 \subsection{The Central Charges}\label{finale}
 
 Finally, we will analyze the central charges of the $\SU(2)$ current algebras in the large $\N=4$ holomorphic (or antiholomorphic) chiral algebra.
 
 One $\SU(2)$ is contained in the small $\N=4$ algebra.   This is the $\SU(2)$ symmetry of the worldsheet fermions that exists because
 the connection $\h\nabla$ has symplectic holonomy. 
The worldsheet fermions all transform as spin 1/2 under this $\SU(2)$. A single set of four real fermions in the spin
 1/2 representation contributes 1 to the $\SU(2)$ current algebra level.   The number of such multiplets in the sigma-model is  the quaternionic dimension of $\M$, or equivalently
 one fourth of its real dimension.   By an index theorem, the real dimension of $\M$ is $4Q_1Q_5$, so the level of the current algebra in the small $\N=4$ algebra is
 $Q_1Q_5$, in accord with expectations from supergravity and the duality conjecture \cite{GMMS}. This formula for the current algebra level 
 is exact, by the usual nonrenormalization theorem for fermion anomalies.
 
 The second $\SU(2)$ is the one that we analyzed in section \ref{nonabelian}, which acts by isometries of $\M$.    As remarked in section \ref{nonabelian}, the semiclassical limit of the sigma-model with target $\M$ is the limit that $Q_5'$ is large,
 keeping $Q_1$ and $Q_5$ fixed (or at least sufficiently small compared to $Q_5'$).   In general, in a sigma-model with a $B$-field, anomalies in target space symmetries receive classical contributions that are large in the semiclassical regime. The WZW model is an example of this. (The semiclassical regime of the WZW model is the regime that the anomaly coefficient
 $k$ is large, and in this regime, $k$ can be computed classically from the WZW action.)   In a sigma-model with fermions as well as a $B$-field, in addition to a large classical contribution, 
 the anomaly can receive  an $\O(1)$ contribution due to the fermions.  In the present context, since we are studying a supersymmetric sigma-model with fermions, such an $\O(1)$
 contribution to the anomaly is possible, and it would be desirable  to know how to compute it.   But here we will only evaluate the classical contribution.  In an expansion in powers of $1/Q'_5$, there can be no further correction
 beyond the $\O(1)$ contribution, because the anomaly coefficient is always an integer.
 
 For a vector field $\h T$ on $\M$  that generates part of a holomorphic chiral algebra, the semiclassical limit of the anomaly coefficient is easy to evaluate, because
 it depends only on the length squared of the vector field.  In the present context, we will take $\h T$ to be one of the $\h T_a$. Its length squared is
 known from  eqn. (\ref{muskat}), and we know that it corresponds to a holomorphic current in the sigma-model.
 
 For large $Q'_5$, the metric of the sigma-model target space $\M$ is scaled up, and in the vicinity of any assumed classical ground state,
   local coordinates $\phi^1,\cdots, \phi^{\dim\,\M }$ can be chosen such that  the metric
 becomes the flat metric $G_{\alpha\beta}=\delta_{\alpha\beta}$ plus a correction of order $1/Q'_5$.  
  This means that the classical contribution to the anomaly can be computed in a theory of free scalar fields with the action 
  \be\label{zobox}I=\frac{1}{2}\int \d^2 x \sum_{\mu=1,2}\,\sum_{\alpha=1}^{\dim\,\M} \partial_\mu\phi ^\alpha \partial^\mu \phi_\alpha.\ee 
  For large $Q_5'$, the vector field $\h T$ becomes a vector field with constant coefficients, generating a symmetry $\delta \phi^\alpha=\h T^\alpha$.
 In the free field
 theory, this symmetry can be generated by the canonical current\footnote{The canonical 
 current is defined by $J^\mu= 2\pi \sum_\alpha\delta \phi^\alpha\frac{\delta I}{\delta \partial_\mu \phi^\alpha}$, where here $\delta \phi^\alpha=\h T^\alpha$.  
  In two-dimensional conformal field theory, the factor of $2\pi$ in the canonical current, and a similar factor in the definition of the stress tensor,
  is conventionally included  to avoid factors of $2\pi$ in operator product coefficients.}
  $J_\mu=2\pi \h T^\alpha\partial_\mu\phi_\alpha$.
    However, if (as in our present application) 
 we know that in the full sigma-model, the current
 generating the symmetry is holomorphic, then in the free field approximation that we are analyzing, the current will have an extra term ensuring this holomorphy and will be
 \be\label{nobo}J_\mu=2\pi\h T^\alpha(\partial_\mu\phi_\alpha+\i \epsilon_{\mu\nu}\partial^\nu\phi_\alpha).\ee
 This extra term, which concretely will originate from the $B$-field coupling of the sigma-model,\footnote{The reason that we do not write any explicit formula for this
 coupling is that there is no canonical local formula; any local formula involves the local form of $\h H=\d\h B$, which does not actually affect the coefficient that we are evaluating.
 Any choice of $\h B$ and $\h H$ that is consistent with the assertion that the current $J_\mu$ is holomorphic will lead to the result analyzed in the text.}
  does not affect the conservation of $J_\mu$ or the fact that
 it acts on $\phi$ by $\delta\phi^\alpha=\h T^\alpha$.  If $z$ is a local complex coordinate on the worldsheet such that $\d^2x=|\d z|^2$, then eqn. (\ref{nobo}) is equivalent to
 \be\label{lobo} J_z=4\pi \h T^\alpha \partial_z\phi_\alpha,~~~J_{\bar z}=0.\ee
 The two-point function of the free scalar field $\phi_\alpha$  is $\langle\phi_\alpha(z)\phi_\beta (0)\rangle=-\delta_{\alpha\beta}\frac{\log |z|}{2\pi},$
 which leads to
 \be\label{tobo} \langle J_z(z) J_z(0)\rangle =-\frac{4\pi\h T^2}{ z^2}. \ee 

The level $k$ of an $\SU(2)$ current algebra is characterized by the statement that if symmetry generators are canonically normalized to satisfy (\ref{canonical}),
then the two-point functions of the corresponding currents have short distance behavior
\be\label{cobo} \langle J_z(z) J_z(0)\rangle =-\frac{k}{2 z^2}. \ee
In the present context, that means that $k=8\pi|\h T|^2$.   From  eqn. (\ref{muskat}), we see that if $\h T$ is any of the $\h T_a$, then $|\h T|^2=\frac{Q_1 Q_5'}{8\pi}$.
So  the level of the $\SU(2)$ current algebra for large $Q_5'$ is
\be\label{obo}k=Q_1 Q_5',  \ee
as predicted by the duality conjecture. 

  As already explained, this formula is conceivably subject to a correction that is $\O(1)$, that is, independent of $Q_5'$.
To compute this correction actually requires among other things
a more precise definition of $Q_5'$ than we have given, because a fermion anomaly can affect the $B$-field flux by an $\O(1)$ amount. 

\subsection{Vector Fields on $\M$ And Their Fixed Points}\label{vectorfields}

In section \ref{freeaction}, we proved that the $\U(1)$ generator $\h V$ has no fixed points on $\M$ -- something we now understand more deeply, having proved that $\h V$ actually has
constant length.   We can also now justify certain analogous assertions that were made in section \ref{freeaction}.

Consider a linear combination  $T =\sum_a c_a T_a$ of the $\SU(2)_r$ generators.    The
corresponding symmetry generator of $\M$ is $\h T =\sum_{a=1}^3 c_a \h T_a=-\frac{1}{2}\sum_{a=1}^3 c_a \h\I_a \h\V$.    Since $\frac{1}{\sqrt{\sum_b c_b^2}} \sum_a c_a\h\I_a$
is a complex structure and acts on  the tangent bundle of $\M$ as an orthogonal transformation,  preserving lengths, it follows  that $\h T$
has constant length $\frac{1}{2}\sqrt{\sum_a c_a^2}|\h V|$.   In other words, any $\SU(2)_r$ generator has a (nonzero) constant length and hence
acts on $\M$ without fixed points. 

Now consider an $\SU(2)_\ell$ generator $T=\sum_a c'_a T'_a$.   The corresponding vector field on $\M$ is $\h T' =\sum_{a=1}^3 c_a' \h T'_a=-\frac{1}{2}\sum_{a=1}^3 c_a' \h\I_a' \h\V$.
Reasoning as  before, its length is $\frac{1}{2}\sqrt{\sum_a c'_a{}^2}|\h V|$ and  it acts on $\M$ without fixed points.

What about a sum  $\h T+\h T'$ of generators of $\SU(2)_r$ and $\SU(2)_\ell$?    This vector field does not have constant length, as in general the angle between the two vector
fields $\h T$ and $\h T'$ is variable.  However, the triangle inequality gives a lower bound on the length of $\h T+\h T'$ which shows that $\h T+\h T'$ acts without fixed points
as long as $|\h T|\not=|\h T' |$, in other words, as long as the original vector fields $T$ and $T'$ on $\S^3\times \S^1$ have unequal length.
If for example $\sum_a c_a^2>\sum_a c'_a{}^2$, then $|\h T|>|\h T'|$ and the triangle inequality gives
\be\label{triangle}|\h T+\h T'|\geq  |\h T|-|\h T'|>0.\ee
So a vector field $\h T+\h T'$ can only have fixed points on $\M$ if the underlying vector fields $T$ and $T'$ on $\S^3\times \S^1$ have equal length.   In that case,
there actually are fixed points, as analyzed in \cite{BH} and discussed in section \ref{notso}.

Finally,  consider a linear combination $\h U=u\h T+v \h T' +w \h V$.   The formulas  $\h T =-\frac{1}{2}\sum_{a=1}^3 c_a \h\I_a \h\V$ and  $\h T'=-\frac{1}{2}\sum_{a=1}^3 c_a' \h\I_a' \h\V$
show that $\h T$ and $\h T'$ are both everywhere orthogonal to $\h V$.  Therefore the inner product of $\h U$ with $\h V$ is the constant $w|\h V|^2$. If $w\not=0$, the nonzero constant value of that inner product implies that $\h U$ has no zeroes.

\subsection{Large $\N=4$ Algebra for One Chirality Only}\label{concluding}

$M=\S^3\times \S^1$  is the unique compact four-manifold such that a sigma-model with target $M$ has two copies of the large $\N=4$ algebra (for left-
and right-movers, respectively).   For this choice of $M$, we have shown that
 a sigma-model with target the corresponding instanton moduli space $\M$ likewise has two copies of the large $\N=4$ algebra.

What if  $M$ is  such that the sigma-model with target $M$ has just one copy of the large $\N=4$ algebra, say for right-movers only? What choices of $M$ are
possible?   $M$ should have a single hypercomplex structure, with an $\SU(2)$ symmetry that rotates the three complex structures, and an additional $\U(1)$ symmetry.  
As shown in  \cite{Boyer},  a compact hypercomplex manifold of dimension four, if not Kahler, is a Hopf surface, which means in particular that it is locally isomorphic to $\S^3\times \S^1$.   A convenient way to exploit
this fact is to observe that such an $M$ has the same universal cover as that of $\S^3\times \S^1$, namely $\S^3\times \R$.  So we can return to the starting point of section \ref{spherex}.
$\S^3\times \R$ with the familiar metric $\d\Omega^2+\d\tau^2$ is equivalent to $\R^4$ minus the origin with the metric
\be\label{decot}\d s^2=\frac{\d\vec Y^2}{\vec Y^2},\ee
as in eqn. (\ref{undef}).   The orientation-preserving isometry group of $\S^3\times \R$ is $H=(\SU(2)_\ell\times \SU(2)_r)/Z_2 \times \R^*$ where $\R^*$ acts by $\vec Y\to\lambda\vec Y$, $\lambda>0$.
Any orientable manifold locally isomorphic to $\S^3\times \S^1$ is the quotient of $\S^3\times \R$ by a discrete subgroup  $\Gamma\subset H$.   However, since we want to preserve one $\SU(2)$ symmetry,
say $\SU(2)_r$, we should pick $\Gamma$ to commute with $\SU(2)_r$, which means that we want  $\Gamma\subset \SU(2)_\ell\times \R^*$.  The quotient $M=(\S^3\times \R)/\Gamma$ 
is a manifold if $\Gamma$ is a discrete subgroup of $\SU(2)_\ell\times \R^*$, and is compact as long as $\Gamma\not\subset \SU(2)_\ell$. Moreover, as long as $\Gamma\not\subset \R^*$,
a sigma-model with target $M$ will support precisely one copy of the large $\N=4$ algebra.

It is then true that if $\M$ is the instanton moduli space on $M$, a sigma-model with target $\M$ also supports precisely one copy of the large $\N=4$ algebra.  We do not need
any essentially new calculation to show this.   The following structures on $\S^3\times \R$ are $\Gamma$-invariant, and therefore descend to $M$:
the $\SU(2)_\ell$-invariant hypercomplex structure studied in this paper and its associated connection $\nabla$,  and the symmetry group\footnote{The group that acts faithfully on $M$
may be  a quotient of $\SU(2)_r\times \R^*$, depending on $\Gamma$.} $\SU(2)_r\times \R^*$, 
where $\SU(2)_r$ rotates the complex structures that make up the hypercomplex structure.  Since the local geometry is the same as it is for
$\S^3\times\S^1$, the computations that we have done apply without change to show that the vector fields generating the $\SU(2)_r\times \R^*$ action are 
covariantly constant for the connection $\h\nabla$ on $\M$ that corresponds to $\nabla$. This extends the small $\N=4$ algebra 
that we would expect based on the analysis in section \ref{hkt} to a large one.

The example studied in section \ref{novel} with $M=\S^3/\Z_n\times \S^1$ corresponds to $\Gamma=\Z_n\times \Z$, with $\Z_n\subset \SU(2)_\ell$ and $\Z\subset \R^*$.
Another simple example, with $\Gamma=\Z$,  can be obtained if we describe  $M$ by complex variables $Z_1, Z_2$ with an equivalence relation
$(Z_1,Z_2)\cong e^T(Z_1,Z_2)$ for some constant $T$ with  ${\rm Re}\,T>0$.   If ${\rm Im}\,T=0$,
this gives  back the original example $M=\S^3\times \S^1$ with a metric of the standard form, leading to two copies of the large $\N=4$ algebra. But taking ${\rm Im}\,T\not=0$ gives a deformation that allows only one copy
of the large $\N=4$ algebra. 

 \section{Conformal Invariance}\label{conformal}

 There is one more issue to consider.   It has long been understood that for two-dimensional sigma-models
 with $\N=4$ supersymmetry,  there is a potential gap between scale invariance and conformal invariance \cite{HullTownsend}.
 In general, the condition for scale invariance is weaker than the condition for conformal invariance.   Scale invariance says that the trace of the stress tensor is 
 the divergence of a current, and conformal invariance says that the trace of the stress tensor vanishes \cite{Polchinski}.    
 
 In a sigma-model with torsion, the one-loop beta functions for the metric and the $B$-field are conveniently combined together to make $\h R_{KL}$,
 the Ricci tensor of the connection $\h\nabla$ with torsion.   Strong HKT geometry then leads to\footnote{If the same metric and $B$-field 
 on the $\sigma$-model target space are compatible with more than one strong HKT structure -- as happens in the case of the instanton moduli space
 on $\S^3\times\S^1$ -- then a formula like this holds for each of them.  In what follows, we will not need an explicit formula for the Lee form $\h\theta$.  It is difficult
 to get a useful explicit expression for $\h\theta$  in the case of instanton moduli space $\M$, because the definition (\ref{leeform}) of the Lee form
 involves the inverse of the metric on $\M$, and although there is a simple formula for this metric,
 it is difficult to give a useful description of its inverse.}
 \be\label{thef} \h R_{KL}=\h \nabla_K\h\theta_L, \ee
 where $\h\theta$ is a 1-form called the Lee form.   For conformal invariance, one wants instead
 \be\label{cnef}\h R_{KL}=-2\h\nabla_K\h\nabla_L \Phi,\ee
 where $\Phi$ is a scalar field (the dilaton) on the  sigma-model target space $\M$.    Unless $\h\theta$ is such that eqn. (\ref{thef}) can actually
 be put in the form (\ref{cnef}), the model is only scale-invariant and not conformally-invariant.

 There are several ways to argue that the sigma-model with target the instanton moduli space $\M$ is actually conformally-invariant.  However, some of the following
 arguments have technical gaps or are limited, in the form presented here,  to lowest order in
 sigma-model perturbation theory. 
 
 First of all, we can simply invoke the general result \cite{Polchinski}  that a scale-invariant quantum field theory in two dimensions with a discrete spectrum of
 operator dimensions is always conformally invariant.  A sigma-model with smooth compact target space will have a discrete spectrum of operator dimensions.
 The instanton moduli space $\M$ is not smooth and compact, because of singularities arising from  small instantons and un-Higgsing.   However, when the theta-angle is
 non-zero (see section \ref{secondbetti}), the model is expected to have a discrete spectrum of operator dimensions (if the charges are relatively prime), and therefore the general theorem applies.
 
 Even without invoking the theta-angle, we can note that the general argument for conformal invariance \cite{Polchinski} only requires that the theory have a discrete
 spectrum of operator dimensions near dimension zero.   The small instanton and un-Higgsing singularities do produce a continuous spectrum of operator dimensions,
 but only above a positive threshold (which moreover is large if $Q_1$ and $Q_5$ are large).   So one would not expect these singularities to cause difficulty in the proof
 of conformal invariance.

 A number of more explicit alternative arguments are available, though  in the form that will be presented here these arguments have technical gaps or are limited to lowest order of
 sigma-model perturbation theory (that is,  lowest order in an expansion in powers of $1/Q'_5$).   Hopefully these arguments can be sharpened and extended.
 
 One approach is a Perelman-style argument, using an auxiliary 
 Schr\"odinger equation.  This approach was explained in \cite{PW}, following \cite{Perelman,Woolgar}.    See also \cite{StreetsOne}, Corollary 6.11, or \cite{StreetsTwo},
 Proposition 2.6, where the same result is proved in essentially the same way.
 For this argument, one 
 considers on $\M$ the Schr\"{o}dinger operator $-4\Delta^2+R-\frac{1}{12}H^2$.
 The proof of conformal invariance requires that this operator should have a unique, positive ground state.   For example, this is true if $\M$ is compact and smooth except for conical
 singularities which are at finite distance and at which the potential $R-\frac{1}{12}H^2$ is bounded below.   (It is also true under some hypotheses if $\M$ has ends at infinite distance.
 In the related context of the $c$-theorem, this case was studied in \cite{Woolgar2}.)  In the case of the instanton moduli space, the small instanton and un-Higgsing singularities
 are indeed conical singularities at finite distance.   Generically these singularities are hyper-Kahler singularities at which $R$ and $H^2$ remain bounded.   However,
 there are a few exceptional cases.    For example, consider the small instanton singularity on a moduli space of instantons with instanton number
 $Q_1>1$.   When a single instanton bubbles or in other words becomes small and shrinks to a point, $\M$ acquires the same universal small instanton singularity as in the case of instantons
on any other four-manifold.  This universal singularity is a conical hyper-Kahler singularity, and is harmless for our present purposes, since both $R$ and $H$ vanish near such a singularity.
 This is true as long as the remaining instanton solution after instanton bubbling remains irreducible: the details of the
 manifold in which the small instanton is embedded do not matter.   However, if  all $Q_1$ instantons become small, so that the remaining
 instanton solution is simply a flat connection on $\S^3\times \S^1$, we get a singularity at which the global structure is relevant.  This case needs more study before claiming to prove conformal invariance based on a Perelman-style argument.  
 
 Another approach to proving conformal invariance uses more detailed facts about the instanton moduli space $\M$ as well as results of \cite{PW}. However, this argument,
 in the form we will present, is only valid to lowest order in sigma-model perturbation theory.  We have constructed
 on the instanton moduli space $\M$ a hypercomplex structure $\h\I,\h\J,\h\K$, covariantly constant with respect to a connection $\h\nabla$, and a second hypercomplex structure
 $\h\I',\h\J',\h\K'$, covariantly constant with respect to a second connection $\h\nabla'$.   Associated to the first hypercomplex structure is a Lee form $\h\theta$ (though this is not
 obvious from the definition  in
eqn. (\ref{leeform}), the Lee form  depends only on the hypercomplex structure and not on the specific
 choice of $\I$, $\J$, or $\K$ in writing the formula).  Similarly there is a Lee form $\h\theta'$  associated to the second hypercomplex structure. The potential obstruction
 to scale invariance can be expressed in terms of the tensor $\h\nabla_K\h\theta_L$, which is also equal to $\h\nabla'_L\h\theta'_K$.   The equality $\h\nabla_K\h\theta_L =\h\nabla'_L\h\theta'_K$
 gives (as in eqn. (3.16) of \cite{PW}) 
 \be\label{zelmee} D_K(\h\theta-\h\theta')_L+D_L(\h\theta-\h\theta')_K=0,~~~~(\d\h\theta+\d\h\theta')_{KL}=\h H^P{}_{KL}(\h\theta-\h\theta')_P, \ee
 where $D$ is the Riemannian connection.
 The obstruction to conformal invariance vanishes if and only if there is a function $\Phi$ on $\M$ (the dilaton) such that
 \be\label{dileq}\h\nabla_K\theta_L=-2\h\nabla_K \partial_L\Phi.\ee
 Let us see why such a $\Phi$ exists in the case of the instanton moduli space $\M$.
 
Isometries of $\S^3\times \S^1$ that preserve the orientation of each factor  rotate $\h\I$, $\h\J$, and $\h\K$
 among themselves.  Since $\h\theta$ is determined by any of these (or any linear combination $a\h\I+b\h\J+c\h\K$ with $a^2+b^2+c^2=1$), it is invariant under such isometries.
 The same argument applies for $\h\theta'$.   However, as discussed in section \ref{freeaction},  we can also define an isometry $\rho$ that
 acts as a reflection on each factor of $\S^3\times \S^1$, reversing the orientation of each factor and preserving the overall orientation.
 Such an isometry exchanges the two hypercomplex structures, so it exchanges $\h\theta$ and $\h\theta'$.
So $\h\theta+\h\theta'$ is even under $\rho$, and $\h\theta-\h\theta'$ is odd.
 Eqn. (\ref{zelmee}) tells us that $\h\theta-\h\theta'$ is the one-form dual to a Killing vector field.   This Killing vector field must be invariant under rotations of either factor of $\S^3\times \S^1$,
 but odd under the joint reflection $\rho$.   The only Killing vector field  on $\M$ with this property is the Killing vector field $\h V$ associated with  the Killing vector
 field $V$ on $\S^3\times \S^1$ that generates a rotation of $\S^1$.  So $\h\theta-\h\theta'=u\lambda$, where $\lambda$ is the 1-form dual to $\h V$ and $u\in\R$.
  In section \ref{large}, we have proved that $\h H_{PKL}\h V^L=0$.   Consequently, the second
 equation in (\ref{zelmee}) reduces to $\d(\h\theta+\h\theta')=0$.   Thus $\h\theta+\h\theta'$ is a closed 1-form on $\M$, defining an element of $H^1(\M;\R)$.   However,
 $\h\theta+\h\theta'$ is even under the joint reflection $\rho$, and we have seen in section \ref{systematic} that (assuming something along the lines of the Atiyah-Jones
 conjecture) $H^1(\M;\R)$ is one-dimensional, generated by the dual of 
 $\h V$, which is odd under $\rho$.  Hence the closed form $\h\theta+\h\theta'$ is actually exact: $\h\theta+\h\theta'=4\d\Phi$ for some function $\Phi$.
 Putting these statements together, we have $\h\theta=\frac{1}{2}\left((\h\theta+\h\theta')+(\h\theta-\h\theta')\right)=2\d\Phi+ \frac{1}{2} u\lambda$.
 Inserting this in $\h\nabla_K\h\theta_L$, the term proportional to $\lambda$ does not contribute, since $\h\nabla\lambda=0$.   So $\h\nabla\h\theta=2\h\nabla\d\Phi$
 and the condition for conformal invariance holds with dilaton $\Phi$.
 
 A noteworthy fact is that each of these arguments for conformal invariance of the sigma-model with target $\M$ makes use of some global information (compactness, relevant 
 for Polchinski's argument,  some
 knowledge of the possible singularities, relevant for analyzing the effective Schr\"odinger equation, or some more detailed information that entered in the discussion of the Lee forms).
 Local considerations alone do not suffice.

 \section{Searching for Interesting Geometries}\label{novel}
 
 \subsection{Preliminaries}
 
 The moduli space $\M$ of instantons on $\S^3\times \S^1$ has an interesting differential geometric structure that, as we have seen,  leads to $(4,4)$ supersymmetry in two dimensions
 (with large superconformal symmetry) 
 despite the presence on $\M$  of a topologically non-trivial $B$-field.   However, $\M$ is not a smooth compact manifold; it has small instanton singularities and also 
un-Higgsing singularities -- singularities that arise at points in $\M$ corresponding to instantons with a non-trivial automorphism group.

 In the case of instantons on $\T^4$ or $\KK$, the instanton moduli spaces are hyper-Kahler, again with singularities associated to small instantons and un-Higgsing.
 In this case, however,  as also discussed in section \ref{secondbetti}, it is possible to resolve the singularities by turning on a Neveu-Schwarz $B$-field on $\T^4$ or $\KK$.  This leads to a noncommutative deformation
 of the instanton equation \cite{NS}, and resolves all of the singularities, provided that $Q_1$ and $Q_5$ are relatively prime.  (If $Q_1$ and $Q_5$ are not relatively
 prime, certain un-Higgsing singularities are unavoidable.)    In this way one can construct smooth,
 compact hyper-Kahler manifolds that are genuinely new -- they cannot be constructed by taking products of tori and $\KK$ surfaces.  In fact, these are the main known 
 examples of compact hyper-Kahler manifolds.
 
 No comparable examples are known of strong HKT manifolds, even if one only asks for a single strong HKT structure, rather than a pair of  strong HKT
 structures, as  the instanton  moduli space on $\S^3\times \S^1$ possesses.   If one asks for a strong HKT manifold that is compact (and smooth), the known examples
 are compact hyper-Kahler manifolds (in which the torsion $H$ vanishes) and homogeneous spaces (which can have strong HKT structures with $H\not=0$; there are many examples
 \cite{Spindel,OP}, of which
 the simplest is  $\S^3\times \S^1$).   It would be interesting to find genuinely new examples of compact  strong HKT manifolds.   The instanton moduli spaces on 
 $\S^3\times \S^1$ do not qualify, since they have unavoidable singularities associated to small instantons and to unbroken gauge symmetries.   The NS $B$-field modes that
 can resolve these singularities in the case of $\KK$ or $\T^4$ have no analog here, as $H^2(\S^3\times \S^1;\R)=0$.   It is true that there is a theta-angle that,
 as discussed in section \ref{secondbetti}, resolves the singularities in a quantum mechanical sense (if $Q_1$ and $Q_5$ are relatively prime), but this does not give
 a classical deformation or resolution of the singularities of the instanton moduli space.
 
   Aiming to find a classical example of a compact (smooth) manifold with the geometry that leads to large $\N=4$  symmetry -- at least for one chirality in the sigma-model -- we will
   replace $\S^3\times \S^1$ with $M=\S^3/\Z_n\times \S^1$ for some integer $n\geq 2$.   Here we pick $\Z_n$ to act on, say, the left on $\S^3\cong \SU(2)$, thus
   breaking the rotation group $\SU(2)_\ell$ to $\U(1)_\ell$ and leaving $\SU(2)_r$ unbroken.  Instanton moduli space on $M$, which we will again denote as $\M$, 
   now carries just a single strong HKT structure, which will lead to a large $\N=4$ superconformal algebra for right-moving modes of the sigma-model.  Left-moving
   modes will see a single KT structure (Kahler with torsion), leading to an ordinary $\N=2$ algebra for left-movers.   
   
   To take advantage of the torsion in the cohomology of $M$, we will take the gauge group to be $\PU(n)=\SU(n)/\Z_n$, the quotient of $\U(n)$ or $\SU(n)$ by its center,
   rather than $\SU(n)$ as we have assumed up to this point.\footnote{More generally, we could  consider $\PSU(n)$ bundles
   over $\S^3/\Z_m\times \S^1$ with $m\not=n$.  The ideas would be similar but some statements would become slightly more complicated as we would have to consider
   the cohomology of $\S^3/\Z_m$ with $\Z_n$ coefficients.}   
   In fact, replacing $\SU(n)$ with $\SU(n)/\Z_n$ would have  little effect\footnote{The instanton moduli space on $\S^3\times \S^1$ for $\SU(n)/\Z_n$ is obtained by dividing
   the $\SU(n)$ moduli space by a $\Z_n$ symmetry that multiples the holonomy around the second factor of $\S^3\times \S^1$ by an element of the center of $\SU(n)$.}  on the differential geometry of the
   instanton moduli space on $\S^3\times \S^1$.   The reason that $\S^3/\Z_n\times \S^1$ is different is that the there is torsion in the cohomology of $\S^3/\Z_n\times \S^1$
   and as a result there are more choices in the topology of a $\PU(n)$ bundle.
   
   An $\SU(n)$ bundle over  a  four-manifold $E$ is  classified topologically by its second Chern class $c_2(E)$, which we usually think of as the second Chern class of the
   vector bundle associated to $E$ in the fundamental representation.  $\SU(n)$ also has, of course, an adjoint representation.   We will denote the associated vector bundle
   in the adjoint representation as $\ad(E)$.  As a real
    vector bundle, it has  a first Pontryagin class $p_1(\ad(E))$.   The relation between the two invariants is
    \be\label{reltwo}p_1(\ad(E)) = 2n c_2(E).\ee
    Thus, for an $\SU(n)$ bundle, $p_1(\ad(E))$ is always an integer multiple of $2n$.
    
The classification of a $\PU(n)$ bundle $E\to M$ is not quite so simple.  
As $\PU(n)$ has no  analog of the $n$-dimensional fundamental representation of $\SU(n)$, there is no rank $n$ vector bundle
    associated to $E$.  But $\PU(n)$ does, of course, have an adjoint representation, and we can again consider the adjoint bundle $\ad(E)$ associated to $E$.
    It has an invariant $p_1(\ad(E))$ which is still an integer-valued invariant, in the sense that it is valued in $H^4(M;\Z)$ and gives an integer upon integration over $M$.
    However, in general it is no longer divisible by $2n$.
    
However, this is not the whole story. A $\PU(n)$ bundle has an additional invariant $\zeta(E)$   taking values in
    $H^2(M;\Z_n)$.   For $n=2$, $\PU(2)$ is the same as $\SO(3)$, and $\zeta$ is the same as the second Stieffel-Whitney class $w_2(E)$.
    As in that example, $\zeta$ is the obstruction to lifting a $\PU(n)$ bundle to a bundle with structure group $\SU(n)$, the universal cover of $\PU(n)$.
    
    A $\PU(n)$ bundle  $E$ over a four-manifold $M$ is determined topologically by $p_1 (\ad(E))$ and $\zeta(E)$, but these cannot be specified independently.  They are subject to one
    relation, which was determined in \cite{Woodward}:
    \be\label{fido} p_1(\ad(E))=\begin{cases} (n+1)C(\zeta)~{\mathrm{mod}}~2n &~ n ~{\mathrm {even}}  \cr
                                                                       \frac{n+1}{2} C(\zeta) ~{\mathrm{mod}}~2n&~ n~{\mathrm{odd}}.\end{cases}\ee
    Here $C$ is a cohomology operation that generalizes the Pontryagin square (which was first considered in the physics of gauge theory in \cite{Lines}).
    For $n=2$, with $\zeta=w_2(E)$, $C$ actually is the Pontyagin square and the formula is
    \be\label{ponsq}p_1(\ad(E)) =C(\zeta)~{\mathrm{mod}}~4.\ee
       For odd $n$, $C(\zeta)=2\zeta^2$ according to \cite{Woodward} so
    in that case
    \be\label{gido} p_1(\ad(E)) =(n+1)\zeta(E)^2 ~{\mathrm{mod}} ~2n. \ee
    The details for even $n>2$ are slightly more complicated and we omit them.
    
    One has\footnote{See for example \cite{Hatcher} or Prop 3.8 in \cite{Lens}.} $H^i(\S^3/\Z_n;\Z_n)=\Z_n$ for $i=1,2,3$, with generators $x\in H^1(\S^3/\Z_n),$ $y\in H^2(\S^3/\Z_n;\Z_n)$, $xy\in H^3(\S^3/\Z_n;\Z_n)$.   One also has $H^1(\S^1;\Z_n)\cong \Z_n$, say with generator $z$.  The cohomology ring of $\S^3/\Z_n\times \S^1$ with $\Z_n$ coefficients is generated by $x,y,z$ with the
    following relations: $y^2=z^2=0$ for dimensional reasons; moreover, for odd $n$, $x^2=0$
    and for even $n$, $x^2=     \frac{n}{2}y$.   
    
  In particular, since $H^2(\S^3/\Z_n\times \S^1;\Z_n)=\Z_n\times \Z_n$ with generators $y$ and $xz$, we have in general  $\zeta= a y + b xz$, with $a,b\in\Z_n$.    
    We can then make eqn. (\ref{gido}) explicit for odd $n$:
    \be\label{mido} p_1(\ad(E)) = 2 ab~{\mathrm{mod}}~2n. \ee
    Thus in general, for odd $n$, $p_1(\ad(E))$ can take any even value mod $2n$.
    
    We will also make eqn. (\ref{fido}) explicit for $n=2$, $\PU(2)=\SO(3)$.   First of all, for $n=2$, the cohomology of $\S^3/\Z_n$ simplifies slightly as $y=x^2$, and accordingly
    $x^3z$  generates $H^4(\S^3/\Z_2\times \S^1;\Z_2)=\Z_2$.     The group $H^2(\S^3/\Z_2\times \S^1;\Z_2)=\Z_2\times \Z_2$
    has three nonzero elements, namely $x^2$, $xz$, and $x^2+xz$.   Two of these possible nonzero values of $\zeta$ can be realized by flat $\SO(3)$ bundles.   Indeed, since a real line bundle $\varepsilon\to M$ is classified by
    $w_1(\varepsilon)\in H^1(M,\Z_2)$, over  $M=\S^3/\Z_2\times \S^1$, there exists a real line bundle $\varepsilon\to M$ with $w_1(\varepsilon)=x$, and another real line bundle $\varepsilon'\to M$
    with $w_1(\varepsilon')=z$.   Then with $\mathcal O$ representing a trivial real line bundle, there is a flat $\SO(3)$ bundle over $M$ 
    with $\ad(E)=\O\oplus \varepsilon\oplus\varepsilon$ and hence
    $w_2(E)=x^2$, and another flat $\SO(3)$ bundle over $M$ with $\ad(E')=\varepsilon\oplus \varepsilon'\oplus \varepsilon\otimes \varepsilon'$ and  $w_2(E)=x^2+xz$.
    As these bundles are flat, we have $p_1(\ad(E))=p_1(\ad(E'))=0$.   Therefore, by virtue of eqn. (\ref{gido}), in general an $\SO(3)$
     bundle $E\to \S^3/\Z_2\times\S^1$ with $\zeta=x^2$ or $x^2+xz$
    has $p_1(\ad(E)) $ divisible by 4, implying via eqn.  (\ref{ponsq}) that $C(x^2)=C(x^2+xz)=0$.
      To see what happens for $\zeta=xz$, we use the general relation for the Pontryagin square: $C(u+v)=C(u)+C(v)+2uv $ mod 4.
    In the present case, taking $u=x^2$, $v=xz$, this tells us that $C(xz)= 2x^3 z~{\mathrm{mod}}~4=2$ mod 4.   Hence eqn. (\ref{fido}) tells us that in general an $\SO(3)$ bundle with $\zeta=xz$
    will have $p_1(\ad(E))=2$ mod 4.  
    
    We can see this more explicitly by constructing an elementary example.   We begin with a simple construction of an $\SU(2)$ bundle  of instanton number 1 over $\S^3\times \S^1$.   We start with a trivial $\SU(2)$ bundle over the cylinder $\S^3\times I$, where $I$ is an interval.   To build an $\SU(2)$ bundle $E\to \S^3\times \S^1$,
    we glue the top of the cylinder to the bottom after making a gauge transformation by a map $\Phi:\S^3\to \SU(2)$.   To get a bundle of instanton number 1,
    we can take $\Phi$ to be the identity map from $\S^3=\SU(2)$ to itself.    This gives a bundle with $c_2(E)=1$ and hence $p_1(\ad(E))=4$.    To get an $\SO(3)$
    bundle over $\S^3/\Z_2\times \S^1$, we make the same construction starting with a trivial bundle over the cylinder $\S^3/\Z_2\times I$.   Then we make an $\SO(3)$ bundle
    over $\S^3/\Z_2\times \S^1$ by gluing the top of the cylinder
    to the bottom after making a gauge transformation $\Phi:\S^3/\Z_2\to SO(3)$.   For $\Phi$ we choose the identity map from $\S^3/\Z_2=\SO(3)$ to itself.   This
    gives an $\SO(3)$ bundle $E\to \S^3/\Z_2\times \S^1$.  A double cover of this construction gives the previous one, so the curvature integral is half of what it was before
    and $p_1(\ad (E))=2$.   As for the invariant $\zeta(E)$, since the bundle $E$ is trivial when restricted to $\S^3/\Z_2\times p$ for a point $p\in \S^1$, there is no $x^2$
    contribution in $\zeta(E)$.  On the other hand, $\zeta(E)$ must be nonzero since $p_1(\ad(E))$ is not a multiple of 4.  So  $\zeta(E)=xz$.      
  
  \subsection{Instanton Moduli Spaces}\label{instmod}
  
  Now we will begin a more detailed discussion of instanton moduli spaces on $M=\S^3/\Z_n\times \S^1$.   The first question to ask is whether these moduli spaces are nonempty.
  The usual existence theorem for instantons \cite{Taubes2} applies for any simple and simply-connected gauge group $\GG$ and any compact oriented smooth four-manifold $M$.  The
  theorem  shows that the moduli space is non-empty if the instanton number is large enough; if $M$ is simply-connected, it must be
  at least $b_2^+(M)-b_1 +1$ ($b_2^+$ is the dimension of the space of selfdual harmonic two-forms on $M$ and $b_1$ is the first Betti number).
  The proof is made by starting with a trivial flat $\GG$-bundle over $M$.  Then, after picking a suitable embedding of $\SU(2)$ in $\GG$, one
  glues in some number $k$ of small $\SU(2)$ instantons to get an approximate instanton solution over $M$, and one proves that if $k\geq b_2^+(M)-b_1+1$, the approximate solution
  can be corrected to get an exact solution.
  
  This proof is not sufficient for our purposes.   We want to consider instantons with structure group the non-simply-connected group $\PU(n)$ and with $\zeta(E)\not=0$.
  This means that $E$ cannot be a trivial flat bundle.   Actually, the proof in \cite{Taubes2} can easily be modified to start with any flat bundle, not necessarily trivial.   
  But for generic $\zeta$, eqn. (\ref{fido}) shows that $p_1(\ad(E))$ cannot vanish and therefore $E$ cannot be flat.
  
  Luckily, there is another existence theorem for instantons that applies for arbitrary $\zeta$ \cite{Taubes}.    This theorem says that for any compact simple $\GG$ and any
  $\zeta$, the instanton moduli space is non-empty if the instanton number is sufficiently large.   In our context, this means that  for any $\zeta(E)$ the instanton moduli space is nonempty
  for any sufficiently large $p_1(\ad(E))$ that is consistent with eqn. (\ref{fido}).   (Actually this statement is a special case of a more general theorem
  that was already cited in section \ref{systematic}.)  The proof proceeds roughly as follows.   Starting with any
  $\GG$-bundle $E\to M$ with the desired value of $\zeta(E)$, and any connection on this bundle,  one glues in many small instantons.  One shows that the gluing parameters can be chosen
  so as to reduce the $L^2$ norm of the selfdual part of the Yang-Mills field strength, making the field  strength more nearly anti-selfdual.   Once one gets close enough
  to anti-selfduality, one then shows (as in the original proof \cite{Taubes2}) that the connection can be modified to achieve full anti-selfduality.
  
  A drawback of this proof is that it does not tell us what is, for given $\zeta(E)$, the smallest value of $p_1(\ad(E))$ at which the instanton equation can be solved.   We only
  learn that the moduli spaces are non-empty for all sufficiently large values of $p_1(\ad(E))$.   (In fact, there are no ``gaps'': if the moduli space is non-empty for one value
  of $p_1(\ad(E))$, it remains non-empty at larger values consistent with eqn. (\ref{fido}).) 
  
  In short, for all $\zeta$ and all sufficiently large $p_1(\ad(E))$, we do get an instanton moduli space $\M$ that will have the differential geometric properties that lead to
  a single copy of  large $\N=4$ superconformal symmetry.   Our goal now is to argue that with some further choices, $\M$ will be smooth and compact.   If so, these moduli spaces
  are candidates as the first examples of smooth, compact strong HKT manifolds that are not products of hyper-Kahler  manifolds and homogeneous spaces.
  
  For $\M$ to be smooth and compact, we have to avoid singularities associated to bubbling of small instantons, and also un-Higgsing singularities associated to instanton solutions
  with non-trivial automorphism groups. First let us discuss how to eliminate the small instanton singularity.  There is a  natural
  way to do this.   Pick a value of $\zeta$ such that for a $\PU(n)$ bundle $E\to M$ with the given $\zeta$, eqn. (\ref{fido}) implies that $p_1(\ad(E))$ is not a multiple of $2n$.
  If $p_1(\ad(E))$ is negative, then the instanton moduli space is empty.  But  if $p_1(\ad(E))$ is sufficiently positive, then the instanton
  moduli space is non-empty according to the proof that we just sketched.  Therefore, there is a smallest value of $p_1(\ad(E))$ with the given $\zeta$ such that the instanton moduli
  space $\M$ is nonempty.   Since $p_1(\ad(E))$ is not an integer multiple of $2n$, it does not vanish at the minimum.    So the component $\M$ with the minimum value
  of $p_1(\ad(E))$ parametrizes instanton solutions that are not merely flat bundles.  Let us call this the minimal component of instanton moduli space for the given $\zeta$.
  This minimal component
  has no small instanton singularity,
  since such a singularity would connect $\M$ to another component with a smaller value of $p_1(\ad(E))$ and the same $\zeta$.

   The question now arises of whether it is also true, perhaps after putting a further restriction on $\zeta$, that the minimal component of instanton moduli space has
   no un-Higgsing singularity.     
Such a singularity  occurs at a point $p\in \M$ that corresponds to an instanton solution whose structure group is a proper subgroup of $\PU(n)$ -- and more specifically, 
a proper subgroup that has a nontrivial commutant in $\PU(n)$.
   For $n=2$, it is clear that the minimal component can have no such singularity.   A proper subgroup of $\PU(2)=\SO(3)$ is a either a  finite group or is isomorphic to $\SO(2)$
   or to $\OO(2)$,
   and no such group is the structure group of a non-flat instanton solution on $\S^3/\Z_2\times \S^1$.  
   It is not clear what is the dimension of the minimal component for $\PU(2)$, and it is not clear  whether the minimal component might be the product of a hyper-Kahler manifold and a homogeneous strong 
   HKT manifold.   However, the minimal component for $\PU(2)$ is a candidate as a genuinely new strong HKT manifold that is compact and smooth.
   
   To make a similar argument for $n=3$, we observe that a connected proper  subgroup of $\PU(3)$ that is the structure group of a non-flat instanton bundle on $\S^3/\Z_3\times \S^1$ is
   either $\U(2)$ or its subgroup $\SU(2)$ or else $\SO(3)$  (here $\U(2)$ is embedded in $\U(3)$ so that the fundamental representation of $\U(3)$ transforms as ${\mathbf 2}\oplus
   {\mathbf 1}$, and is then projected to $\PU(3)$, and similarly $\SO(3)$ is embedded in $\U(3)$ so that the fundamental representation remains irreducible  and is then projected
   to $\PU(3)$).   A routine check shows that any $\PU(3)$ bundle $E$ whose structure group reduces to $\U(2)$ or $\SO(3)$ has $p_1(\ad(E))$ a multiple of 6, and therefore
   such reductions are not possible for any $\zeta$ such that $p_1(\ad(E))$ is not such a multiple.   (In any event, a reduction of structure group to $\SO(3)$ would
   not produce a singularity of the moduli space, since the stabilizer of $\SO(3)$ in $\PU(3)$ is trivial.)

   For general $n$,  consider a component of the moduli space $\M$ characterized by some given values of $p_1(\ad(E))$ and $\zeta(E)$ at a point in $\M$
   at which the structure group $E$ reduces from $\PU(n)$ to a proper subgroup $K$.  $K$ is not necessarily connected, and it may contain $\U(1)$ factors.
   Gauge fields of a finite group or an abelian group or a product or  extension of these can contribute to $\zeta(E)$ or  $p_1(\ad(E))$
   (for a group containing factors of $\U(1)$, this statement depends upon the fact that $b_2(M)=0$).   So $p_1(\ad(E))$ is unchanged if we simply replace $E$ by another
   instanton bundle whose structure group reduces to a maximal connected semi-simple subgroup $H\subset K$.    What are the possible values of $p_1(\ad(E))$ for
   a bundle with such a reduction?  The embedding
    $H\subset \PU(n)$ gives a homomorphism $\varphi:\pi_1(H)\to \pi_1(\PU(n))=\Z_n$.   Suppose first for simplicity that $n$ is prime.  In that case, the only subgroups
    of $\Z_n$ are  $\Z_n$ itself and the trivial group containing only the identity.   For $H$ a proper subgroup of $\PU(n)$ and $n$ prime, $\pi_1(H)$ does not have an element
    of order\footnote{Indeed, $\PU(n)$ itself is the only semisimple Lie group of rank $\leq n-1$ whose fundamental group contains an element of order $n$.  This follows from the classification of simply-connected simple Lie groups and the explicit description of their centers.}    $n$ and therefore the image of $\varphi$ is trivial.   This means that, after restricting the structure group of $E\to M$ from $\PU(n)$ to $H$, it can be lifted to a bundle
    with structure group $\SU(n)$.  Hence $p_1(\ad(E))$ is divisible by $2n$.   Therefore, for  
    $n$ prime, if $\zeta(E)$ is such that $p_1(\ad(E))$ is not a multiple of $2n$, the instanton
    moduli space has no un-Higgsing singularity related to reduction of structure group. 
    Hence the minimal component of instanton moduli space for the given value of $\zeta(E)$
    is compact and smooth.  
    
     If $n$ is not prime, it is possible to find a proper subgroup  $H\subset \PU(n)$ such that the homomorphism $\varphi$ is surjective
    (for example, take $n=pq$ with relatively prime $p,q$, and choose $H={\mathrm P}(\U(p)\times \U(q))$, with $\U(p)\times \U(q)$ embedded in $\U(n)$ such that the fundamental representation of $\U(n)$ is
    the tensor product of the fundamental representations of $\U(p)$ and $\U(q)$; this can be generalized if $n$ is the product of any number of relatively prime factors).  
    In this case, however, the commutant of $H$ in $\PU(n)$ is trivial, so
    reduction of structure group to $H$ produces no singularity in $\M$.
    If the image of $\varphi$ is a proper subgroup $\Z_k\subset \Z_n$ with $k<n$, then a $\PU(n)$ bundle over $M$ whose structure group restricts to $H$ can
    be lifted to an $\SU(n)/\Z_k$ bundle, which implies that $p_1(\ad(E))$ is an integer multiple of $2n/k$.    If $\zeta(E)$ is chosen so that this is not the case for any proper
    divisor $k$ of $n$, then there is no un-Higgsing singularity and the  minimal component of instanton moduli space  is compact and smooth.
    For example, if $n$ is odd, then according to eqn. (\ref{mido}), for $a=b=1$, $p_1(\ad(E))=2 $ mod 2n, and is therefore not a multiple of $2n/k$ for any $k<n$.
    A similar choice is possible for even $n$.

    \subsection{String Theory Interpretation}\label{stringint}
    
    The construction that we have described actually has a string theory interpretation.
    
    Assuming that the underlying branes considered are D-branes,   the two-form field $B$ and three-form curvature $H=\d B$ 
   in the $\AdS_3\times \S^3\times \S^3\times \S^1$ or $\AdS_3\times \S^3\times \S^3/\Z_n\times \S^1$  geometry are of Ramond-Ramond type.   
    However, we can also turn on the Neveu-Schwarz  two-form field $B_{\rm NS}$.   Topologically, a two-form field $B_{\rm NS}$ on a spacetime $Y$ is classified
    by a characteristic class $\xi$ valued in $H^3(Y;\Z)$.    At the level of differential forms, $\xi$ is represented by $\frac{H_{\rm NS}}{2\pi}$, but here it will be important
    to consider $\xi$ as an integral cohomology class.  
    
    For our present purposes, the interesting case is that $\xi$ is a torsion class, which can be represented by a topologically non-trivial $B_{\rm NS}$ field with $H_{\rm NS}=0$.
    The reason that this is the interesting case is that if we assume that $H_{\rm NS}$ is nonzero, and impose the global symmetries and supersymmetries
    of the $\AdS_3\times \S^3\times \S^3\times \S^1$ or $\AdS_3\times \S^3\times \S_3/\Z_n\times \S^1$  geometry, we will just end up with the same spacetime geometry
    that we have already studied, up to an $S$-duality rotation that replaces $H_{\rm RR}$ with a linear combination of $H_{\rm RR}$ and $H_{\rm NS}$.
   
  But we get something essentially new if $B_{\rm NS}$ is flat but topologically nontrivial.
    The reason that this is possible is that there is torsion in the three-dimensional cohomology of $\AdS_3\times \S^3\times \S_3/\Z_n\times \S^1$.
    This torsion is pulled back from $H^3(\S^3/\Z_n\times \S^1;\Z)=\Z\oplus\Z_n$.   
    
    We have encountered in the preceding analysis $\PSU(n)$ bundles over $\S^3/\Z_n\times \S^1$ that have non-integer values of the instanton number and therefore
    cannot be lifted to $\SU(n)$ or $\U(n)$ bundles over $\S^3/\Z_n\times \S^1$.    In general, a $\PSU(n)$ bundle $E$ over any space $Y$ can be lifted to a $\U(n)$ bundle $E'\to Y$
    if and only if the characteristic class $\zeta(E)\in H^2(Y;\Z_n)$ that has been important in our analysis
    can be lifted to a class  $\zeta'\in H^2(Y;\Z)$ (which will then be the first Chern class $c_1(E')$).  The obstruction to
    this lifting can be understood by considering the long exact cohomology sequence associated to the short exact sequence $0\to \Z\overset{n}{\rightarrow}\Z\overset{r}{\rightarrow}\Z_n
    \to 0$, where the first map is multiplication by $n$ and the second is reduction mod $n$.  The associated long exact cohomology sequence reads in part 
            \be\label{terfo}\cdots H^2(\S^3/\Z_n\times \S^1;\Z)\overset{r}{\rightarrow} H^2(\S^3/\Z_n\times \S^1;\Z_n)\overset{\beta}{\rightarrow} H^3(\S^3/\Z_n\times \S^1;\Z)\cdots,\ee
            where $\beta$ is called the Bockstein map.  Thus $\zeta\in H^2(\S^3/\Z_n\times \S^1;\Z_n)$  is not the mod $n$ reduction of an integer class $\zeta'$ -- and so is not in the
            image of $r$ -- if and only if $\beta(\zeta)$ is a nonzero element of $H^3(\S^3/\Z_n\times \S^1;\Z_n)$.

           If $\beta(\zeta)\not=0$, it is not possible to lift the $\PSU(n)$ bundle $E\to \S^3/\Z_n\times \S^1$ to a $\U(n)$ bundle  or in other words to a rank $n$ vector
           bundle.   But it can be lifted to what is called a twisted vector bundle, twisted by the class $\beta(\zeta)$.   Such a twisted vector bundle, rather than an ordinary one,
           is precisely what one gets in D-brane physics in the presence of a background field $B_{\rm NS}$ whose characteristic class $\xi$ is a torsion class.
           Indeed,  a system of $n$ D-branes interacting with a background  $B_{\rm NS}$ field such that $\xi=\beta(\zeta)$ supports not an ordinary vector bundle but a twisted vector bundle
           associated to a $\PSU(n)$ bundle with characteristic class $\zeta$ \cite{WittenK}.   Thus in the presence of a suitable $B_{\rm NS}$ field, the slightly exotic
           instanton moduli spaces studied in this section do actually appear in D-brane physics.

\vskip1cm
 \noindent {\it {Acknowledgements}}   I thank O. Aharony, L. Eberhardt, J. Maldacena, G. W. Moore, R. Moraru, S. Murthy,  N. Nekrasov,
 G. Papadopoulos, A. Tomasiello, and M. Verbitsky for discussions.
  Research supported in part by NSF Grant PHY-2207584.
  
 \bibliographystyle{unsrt}

\end{document}